\documentclass[a4paper,11pt]{article}
\pdfoutput=1

\usepackage{jcappub}
\usepackage[T1]{fontenc}

\usepackage{hyperref}
\usepackage{verbatim}
\usepackage{amsmath}
\usepackage{amssymb}
\usepackage{slashed}
\usepackage[dvipsnames]{xcolor}
\usepackage{booktabs}
\usepackage{fontawesome}

\newcommand\myshade{80}
\colorlet{mylinkcolor}{ForestGreen}
\colorlet{mycitecolor}{Red}
\colorlet{myurlcolor}{violet}
\hypersetup{
  linkcolor  = mylinkcolor!\myshade!black,
  citecolor  = mycitecolor!\myshade!black,
  urlcolor   = myurlcolor!\myshade!black,
  colorlinks = true
}

\newcommand{\revised}[1]{#1}                   

\usepackage[frozencache,cachedir=.]{minted}
\usepackage{mdframed}  
\definecolor{bg}{rgb}{0.94,0.94,0.94}
\surroundwithmdframed[backgroundcolor=bg,linewidth=0.1]{minted}
\newcommand{\mil}[1]{\mintinline{python}{#1}}

\newcommand{\zenodoLink}{\href{https://doi.org/10.5281/zenodo.3347114}{\faTags}}

\newcommand{\nbLink}[1]{%
    \href{https://github.com/LoganAMorrison/Hazma/tree/master/notebooks/hazma_paper/#1.ipynb}{\faBook}%
}
\newcommand{\scriptLink}[1]{%
    \href{https://github.com/LoganAMorrison/Hazma/tree/master/notebooks/hazma_paper/#1.py}{\faFileCodeO}%
}

\newcommand{\us}[1]{{~\mathrm{#1}}}
\renewcommand\u[1]{{\mathrm{#1}}}
\newcommand{\ee}[1]{\times10^{{#1}}}
\renewcommand\L{\mathcal{L}}
\renewcommand\d{\partial}
\newcommand{\rvertchi}[1]{\left. {#1} \right\rvert_{\bar{\chi}\chi}}
\newcommand{\feff}{{f_\u{eff}^\chi}}

\newcommand{\mdm}{{m_\chi}}
\newcommand{\ev}[1]{\langle {#1} \rangle}
\newcommand{\sv}{\ev{\sigma v}}
\newcommand{\svann}{\ev{\sigma v}_{\bar{\chi} \chi}}
\newcommand{\svcmb}{\ev{\sigma v}_{\bar{\chi} \chi, \u{CMB}}}

\title{Hazma: A Python Toolkit for Studying Indirect Detection of Sub-GeV Dark Matter}

\author[a1]{Adam Coogan}
\author[b,c]{Logan Morrison}
\author[b,c]{Stefano Profumo}

\affiliation[a]{GRAPPA, Institute of Physics, University of Amsterdam, 1098 XH Amsterdam, The Netherlands}
\affiliation[b]{Department of Physics, 1156 High St., University of California Santa Cruz, Santa Cruz, CA 95064, USA}
\affiliation[c]{Santa Cruz Institute for Particle Physics, 1156 High St., Santa Cruz, CA 95064, USA}

\emailAdd{a.m.coogan@uva.nl}
\emailAdd{loanmorr@ucsc.edu}
\emailAdd{profumo@ucsc.edu}

\abstract{With several proposed MeV gamma-ray telescopes on the horizon, it is of paramount importance to perform accurate calculations of gamma-ray spectra expected from sub-GeV dark matter annihilation and decay. We present \mil{hazma}, a python package for reliably computing these spectra, determining the resulting constraints from existing gamma-ray data, and prospects for upcoming telescopes. For high-level analyses, \mil{hazma} comes with several built-in dark matter models where the interactions between dark matter and hadrons have been determined in detail using chiral perturbation theory. Additionally, \mil{hazma} provides tools for computing spectra from individual final states with arbitrary numbers of light leptons and mesons, and for analyzing custom dark matter models. \mil{hazma} can also produce electron and positron spectra from dark matter annihilation, enabling precise derivation of constraints from the cosmic microwave background.
\href{https://github.com/LoganAMorrison/Hazma}{\faGithub} \zenodoLink}

\begin{document}
\maketitle
\flushbottom

\section{Introduction}
\label{sec:introduction}

The search for particle debris from dark matter (DM) annihilation or decay has thus far largely centered on \revised{DM masses in} the GeV-TeV scale, for a variety of reasons. First, if the DM shares electroweak interactions with the Standard Model, as in the weakly-interacting massive particle (WIMP) scenario, \revised{then, by the Lee-Weinberg limit, its mass is expected to be more than a few GeV}\cite{Lee:1977ua}\footnote{Note that exceptions exist to the Lee-Weinberg limit, see e.g. \cite{Profumo:2008yg}}. Second, the general expectation for the scale of new physics based on the ``small hierarchy problem'' is that new physics, and thus new massive particles possibly including the particle making up the cosmological DM, should appear around the electroweak scale. Finally, the GeV scale is testable with an array of currently-operating gamma-ray and cosmic-ray observatories, including, but not limited to, the Fermi Large Area Telescope (LAT) \cite{fermilat}, Cherenkov Telescope Arrays such as MAGIC \cite{magic}, HESS \cite{hess}, \revised{VERITAS} \cite{holder2006first}, and the Alpha-Magnetic Spectrometer (AMS-02) \cite{AMS}.

On the theory front, the calculation of the detailed expected {\em particle spectrum} of the debris resulting from DM annihilation or decay has thus focused on the GeV-TeV regime. State-of-the-art codes utilize simulations describing the results of high-energy collisions of elementary particles yielding jets and leptons, which in turn decay and produce stable final-state particles. Many such codes, such as DarkSUSY~\cite{Gondolo:2004sc}, micrOMEGAs~\cite{Belanger:2013oya} and PPPC4DM~\cite{PPPC} utilize tabulated results from PYTHIA~\cite{pythia}, one of the most widely-used programs for performing these simulations. Such results are reliable at center-of-mass energy scales at or above roughly 5 GeV~\cite{pythia}, but not at lower energies, where, for instance, strongly-interacting particles form hadronic bound states and are no longer described by parton showers, fragmentation and decay. It is well known that the resulting spectra of gamma rays, electrons and positrons and antiprotons, are dramatically different in that case.

For a variety of reasons it is now quite timely to offer the community a reliable computational package that provides the spectra of particles resulting from lighter, sub-GeV DM annihilation or decay. First and foremost, MeV astronomy will soon be revolutionized with a new generation of telescopes such e-ASTROGAM~\cite{eastrogam,DeAngelis:2017gra}, \revised{AMEGO}~\cite{mcenery2019all} and others, including concept telescopes such as the Advanced Energetic Pair Telescope (AdEPT)~\cite{adept}, the PAir-productioN Gamma-ray Unit (PANGU)~\cite{pangu}, and the Gamma-Ray Imaging, Polarimetry and Spectroscopy (``GRIPS'')~\cite{grips}. Second, the persistent absence of any conclusive astrophysical signal from DM in the GeV-TeV range has furthered theoretical and phenomenological interest in the mass range below the GeV, providing additional motivation to investigate the details of DM decay or annihilation processes. Lastly, at present no code exists that allows users to readily study gamma-ray and cosmic-ray production from DM particles annihilating dominantly into hadronic bound states. 

With these motivations, we here introduce a Python toolkit, \mil{hazma}\footnote{Hazma is a small rounded Pokemon, with light green spikes running down its back and tail, making it appear somewhat dinosaurian. Its body resembles a yellow hazardous materials suit, with a face resembling a respirator or gas mask, and a zipper-like marking running down its stomach. It has two stubby legs, the feet of which are green. Hazma is one of the few stable Nuclear types, the others being Nucleon and Urayne. Its leaded skin makes it immune to nuclear radiation. In the aftermath of a nuclear accident, groups of Hazma will appear and feed on the radioactive gas, eventually cleaning the air of the area over time.~\cite{hazma}}, that computes spectra of gamma rays and cosmic-ray electrons and positrons from the decay of muons and pions, calculates constraints from gamma-ray observations and the cosmic microwave background, and allows users to compute composite spectra for selected built-in models of DM-parton interactions.

From a field theoretic standpoint, the description of the interactions of fundamental fields with hadrons is performed in the context of chiral perturbation theory (ChPT, see e.g. Ref.~\cite{Scherer2003} for a review). A full account of mapping a fundamental, parton-level Lagrangian onto its ChPT counterpart will be given elsewhere \cite{companion}; here, however, we do provide selected examples of how models where the DM interacts with a {\em mediator} of specified spin and parity produces ChPT vertices.

As for any effective theory, ChPT possesses a certain range of validity which depends upon the size of some dimensionless parameter, here the ratio of the meson momentum to a scale $\Lambda_{\rm ChPT}\equiv 4\pi f_\pi\sim 1$ GeV. Below we will describe the range of dark matter and mediator masses for which our EFT framework can be reliably used to compute annihilation cross sections and mediator decay rates. The mass ranges dictate which combination of mesons we include in the computational package we hereby present.

In the light DM mass limit, we also found that the standard approach for studying radiative emission from leptonic final states is problematic. In short, utilizing the Altarelli-Parisi splitting function to calculate the final state radiation spectrum assumes that radiating particle's center-of-mass frame energy is much larger than its mass. For $\mathcal{O}(100\us{MeV})$ dark matter annihilating into muons, this is not the case. As a result we compute the \emph{exact} spectrum for a few model cases. We also provide spectra for the final state radiation off of charged pions, and account for radiative decays of all relevant particles (e.g. $\pi^+\to \mu^+\nu\gamma$, $\pi^+\to e^+\nu\gamma$ etc).

A general issue with light, sub-GeV DM models is that constraints from perturbations to the cosmic microwave background (CMB) are generically very strong if the DM freezes out as a thermal relic. Of course there exist a broad variety of workarounds and caveats (see e.g.~\cite{DEramo:2018khz}), but any light DM model is prone to CMB constraints. \mil{hazma} implements such constraints by including functionality for computing electron and positron spectra from dark matter annihilation.

The remainder of this paper is structured as follows. Sec.~(\ref{sec:structure_workflow}) offers a high-level overview of the \mil{hazma} code and introduces the \mil{Theory} class, the main user-facing component of \mil{hazma}. Section~(\ref{sec:framework}) describes the effective field theory framework used to study sub-GeV dark matter, provides details of the scalar mediator (sec.~(\ref{sec:models_scalar_mediator}) and vector mediator (sec.~(\ref{sec:models_vector_mediator})) models, and describes the particle physics outputs of \mil{hazma}, including cross section, decay widths, and branching fractions. Section~(\ref{sec:computing_gamma_ray_spectra}) explains how to calculate gamma-ray spectra from individual annihilation final states, while Sec.~(\ref{sec:dm_gamma_ray_spectra}) combines the latter with the scalar- and vector-mediator models to obtain the {\em overall} gamma-ray spectra from DM annihilation. Section~(\ref{sec:dm_positron_spectra}) describes the calculation of the positron spectra, and, finally, Sec.~(\ref{sec:gr_limits}) and Sec.~(\ref{sec:cmb_limits}) describe, respectively, gamma-ray and CMB limits. Section~(\ref{sec:conclusion}) concludes. Examples of how to use \mil{hazma} are woven throughout the text. Appendix~(\ref{sec:installation}) describes the installation process for \mil{hazma}, App.~(\ref{sec:basic_usage}) review the basics of using \mil{hazma}, and App.~(\ref{sec:usage}) gives examples of more advanced applications of the code, such as incorporating new models. \revised{Appendices~(\ref{sec:rambo}-\ref{sec:gamma_ray}) describe modules provided with \mil{hazma} for convenience which can perform arbitrary phase-space integrations and compute photon spectrum for $\geq 3$-body final states.}

Our code is available on GitHub at \href{https://github.com/LoganAMorrison/Hazma}{https://github.com/LoganAMorrison/Hazma}, and the manual is located at \href{https://hazma.readthedocs.io/}{https://hazma.readthedocs.io/}. The icons \faBook{} and \faFileCodeO{} provide the jupyter notebook and python script used to make each figure, which are paired with \href{https://github.com/mwouts/jupytext}{\mil{jupytext}}. The code snippets appearing throughout this work are collected in a jupyter notebook: \href{https://github.com/LoganAMorrison/Hazma/blob/master/notebooks/hazma_paper/snippets.ipynb}{\faBook}. An archived version of \mil{hazma} is available on Zenodo \zenodoLink. \revised{Finally, we have created a Mathematica package, \mil{HazmaTools}, to reproduce all calculations of cross sections, mediator decay widths, and final state radiation spectra used in Hazma. The package includes FeynRules~\cite{Alloul:2013bka} model files for the scalar and vector mediator models, and utilizes FeynArts~\cite{Hahn:2000kx} and FeynCalc~\cite{Shtabovenko:2016sxi}. It is available for download at \href{https://github.com/LoganAMorrison/HazmaTools/}{https://github.com/LoganAMorrison/HazmaTools/}.}

\paragraph{Conventions:} throughout this paper, unless otherwise noted, the units used are MeV for energies, masses and decay widths and $\u{cm}^3 / \u{s}$ for the thermally-averaged DM self-annihilation cross section $\sv$. The \mil{numpy} package~\cite{numpy} is referred to in some of the code snippets as \mil{np}. Lines in code blocks beginning with \mil{>>>} indicate the python command prompt. We have sometimes rounded or formatted the output from \mil{hazma} to make the code snippets more readable.

\section{Structure of \texorpdfstring{\mil{hazma}}{hazma}}
\label{sec:structure_workflow}

\begin{figure}[thb!]
    \centering
    
    \includegraphics[width=0.75\textwidth]{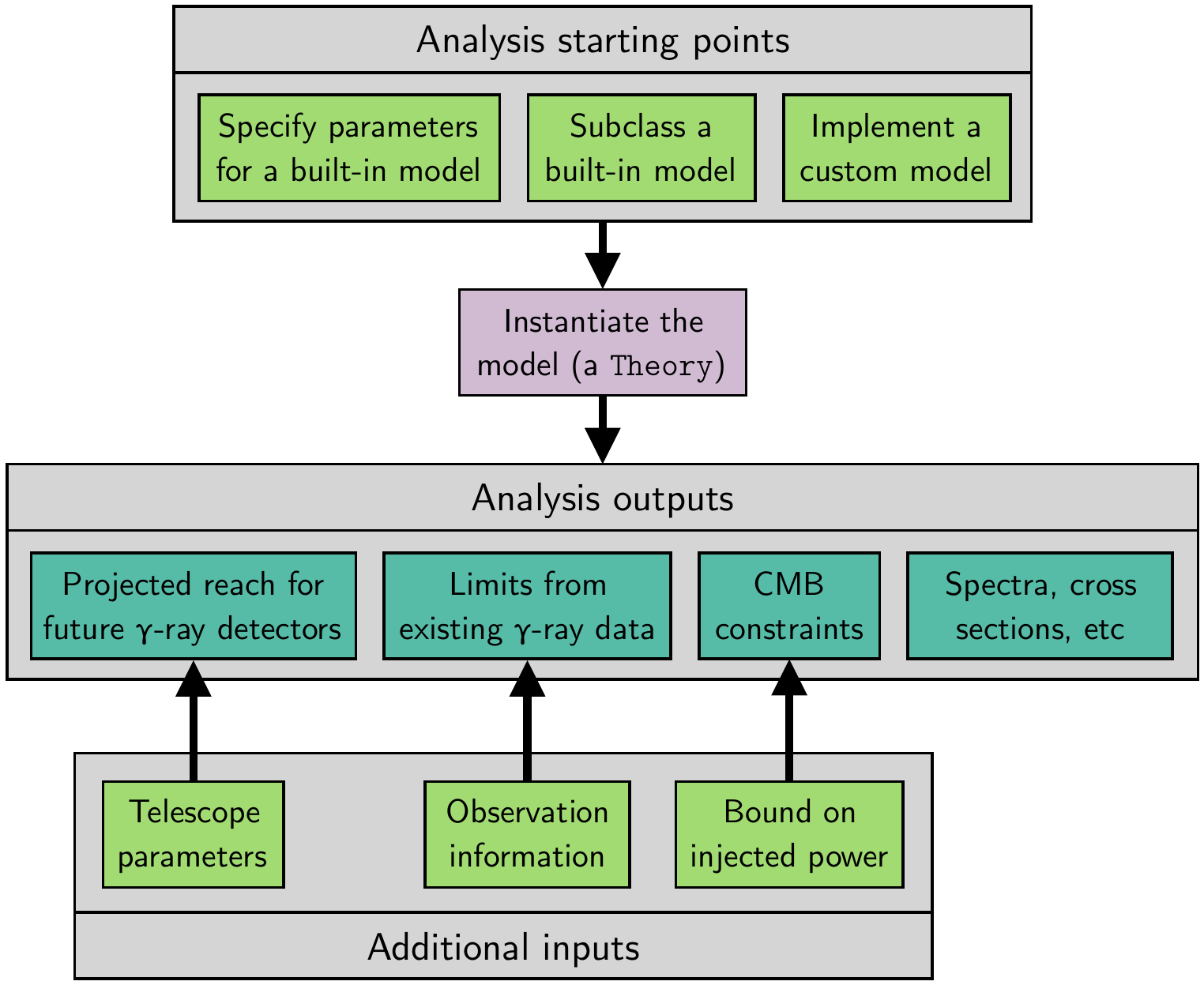}
    \caption{\textbf{Overview of the \mil{hazma} workflow, showing different starting points for analyzing sub-GeV dark matter models and possible outputs.} The light green boxes \revised{at the top} show different types of models the user can analyze. After a model has been instantiated (purple box), various functions can be called to compute the outputs in the dark green boxes. \revised{The inputs in the lower light green boxes are required to compute the three corresponding indirect detection constraints.}}
    \label{fig:user_workflow}
\end{figure}

In this section, we describe the structure of the \mil{hazma} codebase and the intended workflow. The general workflow for a user of \mil{hazma} is shown in Fig.~(\ref{fig:user_workflow}). The light green boxes at the top indicate the types of physics models a user can study, the dark green boxes show possible analysis outputs, and the light green boxes at the bottom denoting inputs required for these outputs.

The user has several options for tapping into the resources provided by \mil{hazma}. The easiest is to use one of the built-in models, where a user only needs to specify the parameters of the model (see below and Sec.~(\ref{sec:models_scalar_mediator}) and Sec.~(\ref{sec:models_vector_mediator}) for details on the built-in models and their corresponding parameters). Alternatively, if the user is working with a model which is a specific version one of the included models (e.g., where the mediator's couplings to Standard Model particles are interrelated), they can define their own subclass of that model. By using inheritance, one retains all of the functionality of the built-in model (such as functions for computing final state radiation spectra, cross sections, mediator decay widths, etc.) while supplying the user with a simpler, more specialized interface to the underlying models. For a detailed explanations of how to set model parameters and make subclasses of built-in models, see App.~(\ref{sec:basic_usage}). Another option is for the user to define their own model. To do this, they need to define a class which contains functions for the gamma-ray and positron spectra, as well as the annihilation cross sections and branching fractions. In App.~(\ref{sec:usage}) we provide a detailed example of how to do this for a toy model\revised{, utilizing various helper modules provided by \mil{hazma}.}

\revised{After choosing a model, the user represents it in \mil{hazma} by creating an instance of the \mil{Theory} class (purple box in Fig.~(\ref{fig:user_workflow})). Various particle physics quantities (the DM self-annihilation cross section, gamma-ray spectra per DM annihilation, etc) can be computed using the values of the masses and couplings in the \mil{Theory}. Computing constraints on the DM self-annihilation cross section requires additional inputs (green boxes at bottom of Fig.~\ref{fig:user_workflow}). To make it straightforward for the user to constrain models using existing data, \mil{hazma} comes with flux measurements from several gamma-ray telescopes, along with Planck's upper bound on how much energy DM annihilations could inject into the CMB. For projecting the reach of future gamma-ray telescopes, the user must provide information characterizing the detector.}

\begin{figure}[tbh!]
    \centering
    \includegraphics[width=0.75\textwidth]{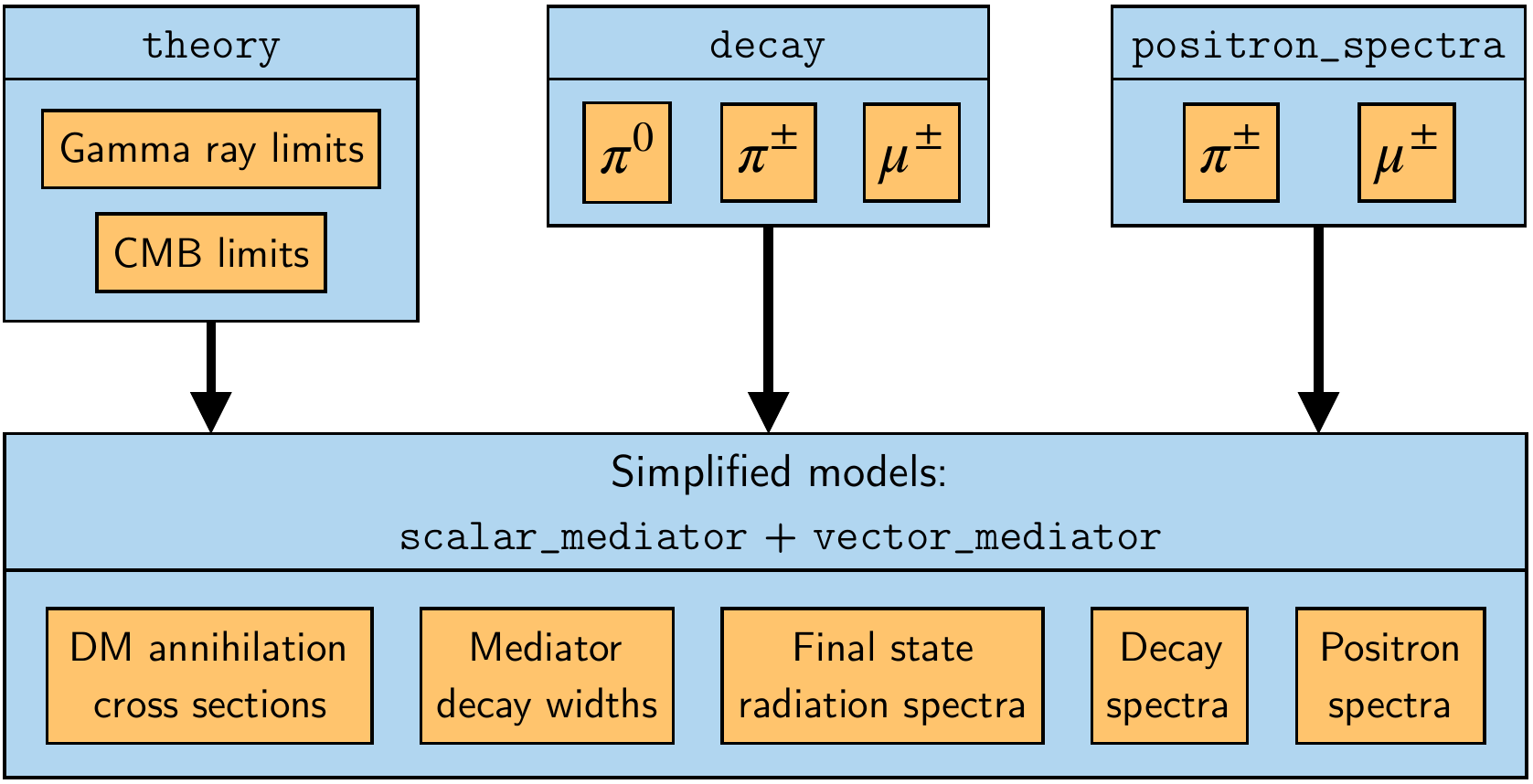}
    \caption{\textbf{Structure of \mil{hazma}.} At the core of \mil{hazma} are modules for computing gamma-ray and positron spectra: \mil{decay}, \mil{positron_spectra} and \mil{gamma_ray}\revised{. The \mil{theory} module contains the \mil{Theory} class, which possesses methods for computing limits on models MeV gamma-ray telescopes.} \mil{hazma} comes with predefined models located in the \mil{scalar_mediator} and \mil{vector_mediator} modules. \revised{These have pre-implemented functions for computing the decay, FSR, and positron spectra, mediator branching fractions, and DM self-annihilation cross sections.}}
    \label{fig:hazma_structure}
\end{figure}{}

Fig.~(\ref{fig:hazma_structure}) displays the modules contained in \mil{hazma} as well as the general structure of the code, which are explained in detail below.

\paragraph*{\underline{\mil{theory}}}
The primary goal of \mil{hazma} is to provide users with a simple interface for setting indirect detection constraints on $\langle \sigma v \rangle$ as a function of DM mass for sub-GeV dark matter models. This is done using the \mil{Theory} abstract base class, which every dark matter model in \mil{hazma} should inherit from. At a high level, it contains three methods for determining these constraints: one that uses existing gamma-ray data (\mil{Theory.binned_limit}), one for computing the discovery reach for planned gamma-ray detectors (\mil{Theory.unbinned_limit}), and one for deriving CMB constraints (\mil{Theory.cmb_limit}). The class also provides methods for computing particle physics quantities such as annihilation cross sections (with \mil{annihilation_cross_sections}), mediator decay widths (\mil{partial_widths}), the continuum and monochromatic gamma-ray spectra from DM annihilation (\mil{total_spectrum}, \mil{gamma_ray_lines}), and the same for positron spectra (\mil{total_positron_spectrum}, \mil{positron_lines}), and more.

Custom models must implement functions for computing gamma-ray spectra, positron spectra and branching fractions to gain use of the methods in \mil{Theory}. \revised{As mentioned above,} Appendix~(\ref{sec:usage}) shows how to implement a simple custom model. \revised{Two other modules may be useful for complex custom models: \mil{rambo}, which performs $N$-body phase space integrations using a Monte Carlo algorithm, and \mil{gamma_ray}, which computes gamma ray spectra for $N$-body final states. These are not used by the simplified models that ship with \mil{hazma}, and are described in detail in Appendices~(\ref{sec:rambo}-\ref{sec:gamma_ray}).}

\paragraph*{\underline{\mil{decay}}}
This contains high-performance functions for computing the decay spectra from $\pi^{\pm}, \pi^{0}$ and $\mu^{\pm}$. Details of how these are computed are found in Sec.~(\ref{sub:radiative_decay_spectra}). The functions in this module allow the user to compute the decay spectra for particles decaying with arbitrary lab-frame energies. This requires computing the decay spectra in the rest frame of the parent-particle and performing a Lorentz boost, amounting to performing a change-of-variables along with a convolution integral. To achieve high computational performance, we perform all integrations in \mil{c} using \mil{cython}~\revised{\cite{behnel2011cython}} and build extension modules to interface with python.

\paragraph*{\underline{\mil{positron_spectra}}}
This module computes the electron/positron spectra from decays of $\pi^{\pm}$ and $\mu^{\pm}$, which are critical inputs for constraining dark matter models using CMB observations. See Sec.~(\ref{sec:dm_positron_spectra}) for details on how these spectra are computed. As in the \mil{decay} module, the \mil{positron_spectra} module allows users to compute the electron/positron spectra for arbitrary energies of the parent-particle. The procedure for computing the spectra for arbitrary parent-particle energies is identical to the procedure used for \mil{decay}.\\

\noindent \mil{hazma} also ships with various particle physics models of sub-GeV DM. These models are located in the modules \mil{scalar_mediator} and \mil{vector_mediator}, and contain all the relevant annihilation cross sections, branching fractions, decay spectra, FSR spectra and positron spectra. They can be used by instantiating the appropriate class (for example, the \mil{HiggsPortal} class from \mil{scalar_mediator} for the Higgs-portal model.) The user only needs to specify the parameters of the model. The particle physics frameworks used to construct these models, the specialized subclasses of these models included with \mil{hazma}, and accessing functions to compute their annihilation cross sections and other particle physics quantities is the topic of the following section.

\section{Particle physics framework}
\label{sec:framework}

Each of the models distributed with \mil{hazma} contain two BSM particles: a dark matter particle $\chi$ and a mediator $M$ that interacts with the DM as well as Standard Model particles. The Lagrangian can be expressed as
\begin{align}
    \L & = \L_\u{SM} + \L_\u{DM} + \L_\u{M} + \L_{\u{Int}(M)},
\end{align}
which consists of the SM Lagrangian, the free Lagrangians for the Dark Matter (DM) and mediator (M), and the mediator's interactions with the DM and SM fields, collectively included in the term $\L_{\u{Int}(M)}$. For the models currently available in \mil{hazma} the DM is taken to be a Dirac fermion, so that
\begin{align}
    \L_\u{DM} & = \bar{\chi} (i \slashed \d - m_\chi) \chi.
\end{align}
Both the dark matter and the mediator are taken to be uncharged under the Standard Model gauge group. The Lagrangian is defined in terms of the microscopic degrees of freedom of the Standard Model \revised{(quarks, leptons and gauge bosons)}. However, at the energy scale of interest for self-annihilations of non-relativistic MeV dark matter ($\sqrt{s}\sim 1-100~\mathrm{MeV}$), quarks and gluons are not the correct strongly-interacting degrees of freedom. \revised{At these energies, QCD confines, and quarks and gluons group into bound states called mesons and baryons. Given that the DM interacts with quarks and gluons, at these energies interactions with mesons and baryons are induced. Since QCD is non-perturbative for energies $\sqrt{s} \lesssim \mathrm{GeV}$, we must use an effective field theory to describe the interactions of DM with mesons and baryons.} We thus match our microscopic Lagrangian onto the effective Lagrangian for pions and other mesons using the techniques of chiral perturbation theory (ChPT)~\cite{WEINBERG1979327,Gasser:1983yg,GASSER1985465,ecker1995chiral,Scherer2003}. The models currently implemented in \mil{hazma} utilize leading-order ChPT.

Before describing these models, we review the range of validity of ChPT, which is limited since it is an effective theory. ChPT is organized as an expansion in a small parameter, the meson momentum squared $p^2$ divided by the squared mass scale associated with loop diagrams, $\Lambda_\u{ChPT} \sim 4 \pi f_\pi \approx 1.2\us{GeV}$, where $f_\pi = 92.2\us{MeV}$ is the pion decay constant. Derivatives of the meson fields and factors of the meson masses are counted as $\mathcal{O}(p)$. The expansion parameter $p^2$ is used in two ways. First, the chiral Lagrangian is organized as a sum of terms of decreasing importance
\begin{align}
    \mathcal{L}_\u{ChPT} = \mathcal{L}^{(2)} + \mathcal{L}^{(4)} + \mathcal{L}^{(6)} + \dots,
\end{align}
where the superscript indicates the number of derivatives or powers of meson masses. Second, the expansion parameter can be used to assess the relative size of an individual Feynman diagram's contribution in the calculation of a given observable. This is quantified using the \emph{chiral dimension}, which is lower for more important diagrams. Generally the tree-level diagrams from $\mathcal{L}^{(2)}$ have the lowest chiral dimension and provide the largest contributions to observables, followed by tree-level diagrams from $\mathcal{L}^{(4)}$ and one-loop diagrams from $\mathcal{L}^{(2)}$, and so on. As $p^2 \to \Lambda_\u{ChPT}^2$, this scheme for organizing calculations in terms of the chiral dimension breaks down and ChPT becomes unreliable.

Assessing the precise range of validity of leading-order ChPT is complicated. A simple estimate suggests that for processes with meson energies of e.g. $500\us{MeV}$ the leading-order observables receive $\sim (500\us{MeV} / \Lambda_\u{ChPT})^2 \sim 20\%$ corrections from next-to-leading-order (NLO) contributions~\cite{Pich:1995bw}. In reality, the magnitude of NLO corrections is larger for annihilation final states with the same quantum numbers as resonances such as the $\rho$~\cite{Meissner:1993ah} and $f_0(500)$~\cite{f0500_review}, since the final state particles can rescatter. While resonances can be accounted for with unitarization techniques~\cite{PhysRevLett.61.2526,f0500_review} or as explicit resonance fields~\cite{ECKER1989311,Soto:2011ap}, we instead use leading-order ChPT, restricting to DM masses below 250 MeV in the case of annihilation into Standard-Model particles and mediator masses below 500 MeV for analysis of their decays to avoid resonances. An important consequence is that annihilation into kaons and heavier mesons is not considered herein since that would require DM masses far above this range. User-defined models (see App.~(\ref{app:user_defined_models})) need not use leading-order ChPT, and this is an interesting topic for future work.

While the preceding discussion is quite general, we now specialize to the two models that come with \mil{hazma}: \mil{scalar_mediator}, which contains a real scalar mediator $S$, and \mil{vector_mediator}, where the mediator is a vector $V$~\footnote{\revised{In future work, we plan to add pseudoscalar and axial-vector mediators.}}. These are subclasses of the abstract class \mil{Theory}, and thus implement a variety of functions for computing physical quantities. In the following two subsections, we present $\L_{\u{Int}(S)}$ and $\L_{\u{Int}(V)}$ at the level of quarks and gluons as well as the Lagrangians obtained by performing the ChPT matching.\footnote{The interaction Lagrangians, matching procedure and a review of the chiral Lagrangian are explained in detail in a forthcoming companion paper~\cite{companion}.} Snippets are provided to demonstrate how to construct each model and change its parameters. The third subsection shows how to access various particle physics quantities in \mil{hazma}.

\subsection{Scalar Mediator}
\label{sec:models_scalar_mediator}

The free Lagrangian for a real scalar is
\begin{align}
    \L_S & = \frac{1}{2} (\d_\mu S) (\d^\mu S) - \frac{1}{2} m_S^2 S^2,
\end{align}
where $m_S$ is the scalar's mass. The interactions with the light fundamental SM degrees of freedom read
\begin{align}
    \L_{\u{Int}(S)} & = -S \left( g_{S\chi} + g_{Sf} \sum_f \frac{y_f}{\sqrt{2}} \bar{f} f \right)                                                                                     \\
                    & \hspace{2cm} + \frac{S}{\Lambda} \left( g_{SG} \frac{\alpha_\u{EM}}{4\pi} F_{\mu\nu} F^{\mu\nu} + g_{SF} \frac{\alpha_s}{4\pi} G_{\mu\nu}^a G^{a \mu\nu} \right).\notag
\end{align}
The sum runs over fermions with mass below the GeV scale ($f = e, \mu, u, d, s$). Note that the coupling $g_{Sf}$ is outside the sum. The Yukawa couplings are defined to be $y_f = \sqrt{2} m_f / v_h$, with the Higgs vacuum expectation value (vev) defined as $v_h = 246\us{GeV}$. The parameter $\Lambda$ is the mass scale at which $S$ acquires (non-renormalizable) interactions with the photon and gluon.

After performing the matching onto the chiral Lagrangian and expanding to leading order in the pion fields, the resulting interaction Lagrangian is
\begin{align}
    \label{eq:LagIntS}
    \L_{\mathrm{Int}(S)} & = \frac{2 g_{SG}}{9 \Lambda} S \left[ (\d_\mu \pi^0) (\d^\mu \pi^0) + 2 (\d_\mu \pi^+) (\d^\mu \pi^-) \right]                                                                        \\
                         & \hspace{1cm} + \frac{4 i e g_{SG}}{9 \Lambda} S A^\mu \left[ \pi^- (\d_\mu \pi^+) - \pi^+ (\d_\mu \pi^-) \right]\notag                                                                     \\
                         & \hspace{1cm} - \frac{B (m_u + m_d)}{6} \left( \frac{3 g_{Sf}}{v_h} + \frac{2 g_{SG}}{3 \Lambda} \right) S \left[ (\pi^0)^2 + 2 \pi^+ \pi^- \right]\notag                                   \\
                         & \hspace{1cm} + \frac{B (m_u + m_d) g_{SG}}{81 \Lambda} \left( \frac{2 g_{SG}}{\Lambda} - \frac{9 g_{Sf}}{v_h} \right) S^2 \left[ (\pi^0)^2 + 2 \pi^+ \pi^- \right]\notag                   \\
                         & \hspace{1cm} + \frac{4 e^2 g_{SF}}{9\Lambda} S \pi^+ \pi^- A_\mu A^\mu\notag                                                                                                               \\
                         & \hspace{1cm} - g_{S \chi} S \bar{\chi} \chi - g_{Sf} S \sum_{\ell=e,\mu} \frac{y_\ell}{\sqrt{2}} \bar{\ell} \ell.\notag
\end{align}
where $B = m_{\pi^{\pm}}^2/(m_{u} + m_{d}) \approx 2800 $ MeV \cite{ecker1995chiral}.

The parameters for the scalar model are attributes of the \mil{scalar_mediator} class. Their names in \mil{hazma} are
\begin{align*}
    (m_\chi, m_S, g_{S\chi}, g_{Sf}, g_{SG}, g_{SF}, \Lambda) \leftrightarrow (\mathtt{mx}, \mathtt{ms}, \mathtt{gsxx}, \mathtt{gsff}, \mathtt{gsGG}, \mathtt{gsFF}, \mathtt{lam}).
\end{align*}
The following snippet shows how to instantiate \mil{scalar_mediator}, change the value of a parameter, and print its new value:
\begin{minted}{python}
>>> from hazma.scalar_mediator import ScalarMediator
>>> sm = ScalarMediator(mx=150., ms=1e3, gsxx=1., gsff=0.1,
...                     gsGG=0.1, gsFF=0.1, lam=2e5)
>>> sm.gsff
0.1
>>> sm.gsff = 0.5
>>> sm.gsff
0.5
\end{minted}
In addition to the general \mil{scalar_mediator} model, \mil{hazma} also comes with two subclasses of \mil{scalar_mediator}, which represent UV completions: a Higgs-portal model~\cite{Krnjaic:2015mbs} (\mil{HiggsPortal}) and a model containing a heavy quark (\mil{HeavyQuark}.) The Higgs-portal model assumes the scalar mediator interacts with the Higgs through gauge-invariant interactions resulting in the scalar mediator mixing with the Higgs. After diagonalizing the scalar-mediator/Higgs mass matrix, the scalar mediator and Higgs are replaced with:
\begin{align}
    h &\to h\cos\theta - S\sin\theta, 
    & S &\to h\sin\theta + S\cos\theta
\end{align}
where $\theta$ is the scalar-mediator/Higgs mixing angle. This replacement results in the following cutoff scale and couplings of the scalar mediator with the Standard Model fermions, gluons and photon, the later of which require integrating out the $\tau$, $c$, $b$, $t$, $W$ and $Z$~\cite{Marciano:2011gm}:
\begin{align}
    g_{Sf} &= \sin\theta, & g_{SG} &= 3\sin\theta, & g_{SF}& = -\frac{5}{6}\sin\theta, & \Lambda &= v_{h}.
\end{align}
where $v_{h} = 246 \ \mathrm{GeV}$ is the Higgs vacuum expectation value. The parameters for the \mil{HiggsPortal} class are:
\begin{align}
    (m_\chi, m_S, g_{S\chi}, \sin\theta) \leftrightarrow (\mathtt{mx}, \mathtt{ms}, \mathtt{gsxx}, \mathtt{stheta}).
\end{align}

The heavy-quark model assumes the existence a new heavy, colored and charged fermion which enters the Lagrangian as:
\begin{align}
    \mathcal{L} \supset i\bar{Q}\slashed{D}Q - m_{Q}\bar{Q}Q - \frac{m_{Q}}{v_{h}}h\bar{Q}Q - g_{SQ}S\bar{Q}Q
\end{align}
with $D_{\mu} = \partial_{\mu} -ieQ_{Q}A_{\mu} -ig_{s}G_{\mu}^{a}\lambda^{a}/2$ where $Q_{Q}$ is the $\operatorname{U}_\u{EM}(1)$ charge of the heavy-quark. Integrating out the heavy quark induces a cutoff scale and effective couplings of the scalar mediator with gluons and photons:
\begin{align}
    g_{SG} &= g_{SQ}, & g_{SF} &= 2Q_{Q}^2 g_{SQ}, & \Lambda &= m_{Q}.
\end{align}
Note that integrating out the heavy-quark also induced effective couplings to the SM fermions, but at two-loop order. We do not include these interactions.  The parameters for the \mil{HeavyQuark} class are:
\begin{align}
    (m_\chi, m_S, g_{S\chi}, g_{SQ}, m_Q, Q_Q) \leftrightarrow (\mathtt{mx}, \mathtt{ms}, \mathtt{gsxx}, \mathtt{gsQ}, \mathtt{mQ}, \mathtt{QQ}).
\end{align}
Both of these models can be imported and used in a similar fashion to the generic \mil{ScalarMediator} class. The following snippet displays how to initialize these subclasses:
\begin{minted}{python}
>>> from hazma.scalar_mediator import HiggsPortal, HeavyQuark
>>> hp = HiggsPortal(mx=150., ms=1e3, gsxx=1., stheta=1e-3)
>>> hq = HeavyQuark(mx=150., ms=1e3, gsxx=1., gsQ=1.0, mQ=1e6, QQ=1.)
\end{minted}
Trying to set the underlying \mil{ScalarMediator} model's attributes will result in an error when these are fully determined by the attributes of the subclass:
\begin{minted}{python}
>>> hp.gsff        # can only be accessed
0.001
>>> hp.gsff = 0.1  # fully determined by hp.stheta
AttributeError: Cannot set gsff
\end{minted}

\subsection{Vector Mediator}
\label{sec:models_vector_mediator}
For the vector mediator the free part of the Lagrangian is
\begin{align}
    \L_V & = -\frac{1}{4} V_{\mu\nu} V^{\mu\nu} + \frac{1}{2} m_V^2 V_\mu V^\mu,
\end{align}
where $m_V$ is the mass of the vector. The interactions considered are
\begin{align}
    \L_{\u{Int}(V)} & = V_\mu \left( g_{V\chi} \bar{\chi} \gamma^\mu \chi + \sum_f g_{Vf} \bar{f} \gamma^\mu f \right).
\end{align}
The sum again runs over the light fermions ($f = e, \mu, u, d, s$), and $V$ may have different couplings to each of these.

A kinetic mixing term between the mediator and photon is also possible:
\begin{align}
    \L_{\u{Int}(V)} &\supset - \frac{\epsilon}{2} V^{\mu\nu} F_{\mu\nu}.
\end{align}
However, this can be eliminated by redefining the photon field using $A_\mu \to A_\mu - \epsilon V_\mu$. Upon this field redefinition $V$ acquires an $\epsilon$-suppressed interaction with the SM fermions, which is captured by changing the fermion couplings
\begin{align}
    g_{Vf} \to g_{Vf} - \epsilon e Q_f,
\end{align}
where $Q_f$ is the electric charge of the fermion $f$ and $e > 0$ is the electron's charge. Because of this, we do not include $\epsilon$ as a coupling in the vector model, and instead provide a subclass \mil{KineticMixing} to handle this case.

Matching onto the chiral Lagrangian and isolating the terms contributing at leading order to the quantities computed in \mil{hazma} gives
\begin{align}
    \label{eq:LagIntV}
    \L_{\mathrm{Int}(V)} & = -i (g_{Vu} - g_{Vd}) V^\mu \left( \pi^+ \d_\mu \pi^- - \pi^- \d_\mu \pi^+ \right) \\
                         & \hspace{1cm} + ( g_{Vu} - g_{Vd} )^2 V_\mu V^\mu \pi^+ \pi^- \notag \\
                         & \hspace{1cm} + 2 e (Q_u - Q_d) (g_{Vu} - g_{Vd}) A_\mu V^\mu \pi^+ \pi^-\notag\\
                         & \hspace{1cm} + V_{\mu}\left(g_{Ve}\bar{e}\gamma^{\mu}e + g_{V\mu}\bar{\mu}\gamma^{\mu}\mu\right)\notag\\
                         & \hspace{1cm} + \frac{1}{8\pi^2 f_\pi} \epsilon^{\mu\nu\rho\sigma} (\d_\mu \pi^0)\notag\\
                         & \hspace{2cm} \times \left\{ e (2 g_{Vu} + g_{Vd}) \left[ (\d_\nu A_\rho) V_\sigma + (\d_\nu V_\rho) A_\sigma \right] \right.\notag \\
                         & \hspace{4cm} \left. + 3 (g_{Vu}^2 - g_{Vd}^2) (\d_\nu V_\rho) V_\sigma \right\}\notag.
\end{align}
The contributions in the last three lines come from matching onto the anomalous terms in the chiral Lagrangian~\cite{Witten:1983tw,Wess:1971yu,Scherer2003,Gasser:1983yg}, which is explained in detail in our companion paper~\cite{companion}. While these terms come from the NLO chiral Lagrangian, the resulting final states contribute significantly to the gamma-ray spectra, so we include them here. In particular, in the case where $V$ couples only to quarks, the \emph{only} final state accessible below the two-pion threshold is $\pi^0\gamma$. In contrast, other terms from the NLO chiral Lagrangian only lead to corrections for scattering rates into final states accessible at tree-level, and are thus not included. 

The correspondence between the microscopic parameters for the vector model and attributes of the base \mil{vector_mediator} class is\footnote{Note that $\mil{gvss}$ is not currently used since \mil{vector_mediator} is based on leading-order ChPT.}
\begin{align*}
      & (m_\chi, m_V, g_{V\chi}, g_{Vu}, g_{Vd}, g_{Vs}, g_{Ve}, g_{V\mu})                                                           \\
      & \hspace{1cm} \leftrightarrow (\mil{mx}, \mil{mv}, \mil{gvxx}, \mil{gvuu}, \mil{gvdd}, \mil{gvss}, \mil{gvee}, \mil{gvmumu}).
\end{align*}
The following snippet shows how to instantiate \mil{vector_mediator}:
\begin{minted}{python}
>>> from hazma.vector_mediator import VectorMediator
>>> vm = VectorMediator(mx=150., mv=1e3, gvxx=1., gvuu=0.1,
...                     gvdd=0.2, gvss=0.0, gvee=0.4,
...                     gvmumu=0.5)
\end{minted}

For the subclass \mil{KineticMixing} which handles the important case of a kinetically-mixed mediator, the parameters are
\begin{align*}
    (m_\chi, m_V, g_{V\chi}, \epsilon) \leftrightarrow (\mil{mx}, \mil{mv}, \mil{gvxx}, \mil{eps}).
\end{align*}
While the underlying parameters \mil{gvuu}, \dots, \mil{gvmumu} can be accessed by instances of \newline\mil{KineticMixing}, they cannot be set directly since they are fully determined by \mil{eps}:
\begin{minted}{python}
>>> from hazma.vector_mediator import KineticMixing
>>> km = KineticMixing(mx=150., mv=1e3, gvxx=1., eps=0.1)
>>> km.gvuu
-0.020187846690459792  # = -0.1 * 2/3 * sqrt(4 pi / 137)
>>> km.gvuu = 0.1
AttributeError: Cannot set gvuu
\end{minted}

Finally, the subclass \mil{QuarksOnly} is provided for analyzing the case where the mediator is hadrophilic and has only couplings to quarks. The parameters are those for the full vector-mediator model, with the leptonic couplings set to zero.

\subsection{Computing Particle physics quantities}
\label{sub:particle_physics_quantities}

\begin{figure}[tbh!]
    \centering
    \includegraphics[width=\textwidth]{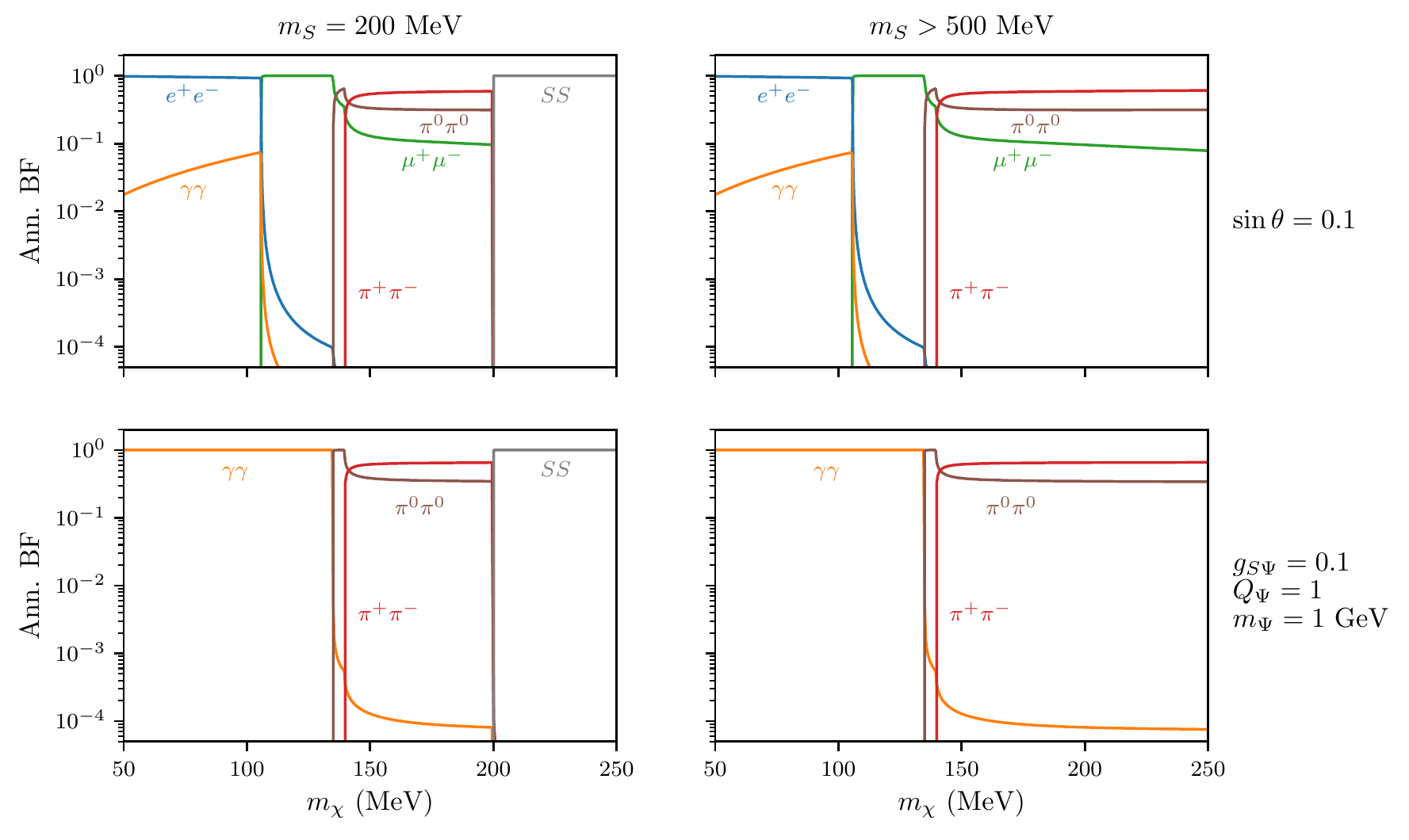}
    \caption{\textbf{Annihilation branching fractions for the scalar mediator model.} The coupling $g_{S\chi\chi}$ between the DM and mediator is set to one. The rows correspond to the Higgs portal and heavy quark UV completions and the columns are for a light (left) and heavy (right) mediator.
    \nbLink{scalar_ann_bfs}
    \scriptLink{scalar_ann_bfs}}
    \label{fig:scalar_ann_bfs}
\end{figure}

\begin{figure}[tbh!]
    \centering
    \includegraphics[width=\textwidth]{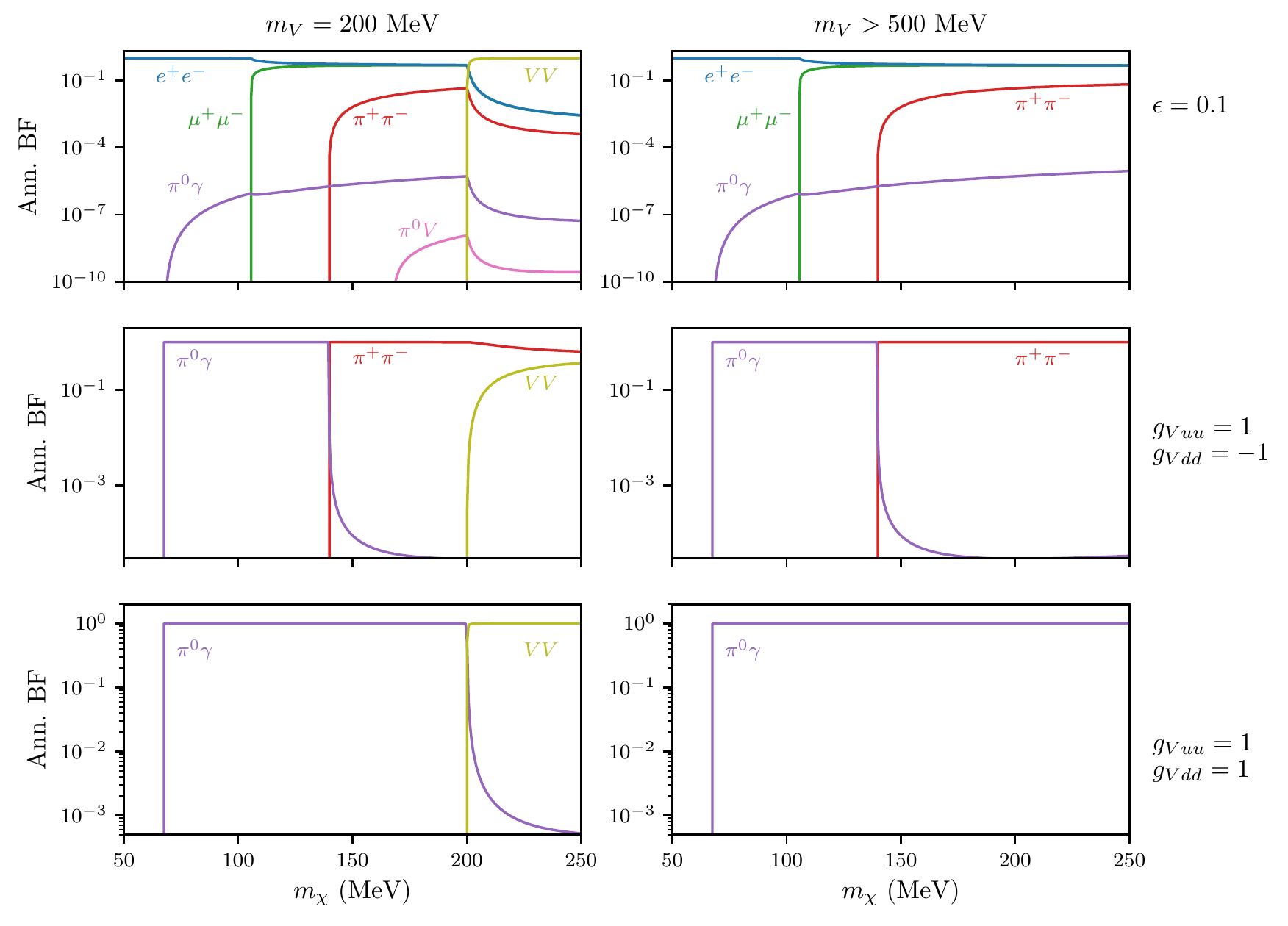}
    \caption{\textbf{Annihilation branching fractions for the vector mediator model.} The coupling $g_{S\chi\chi}$ between the DM and mediator is set to one. The rows correspond to a kinetically mixed mediator and a mediator with couplings only to quarks, and the columns are for a light (left) and heavy (right) mediator.
    \nbLink{vector_ann_bfs}
    \scriptLink{vector_ann_bfs}}
    \label{fig:vector_ann_bfs}
\end{figure}

Every model in \mil{hazma} is required to implement functions listing the available annihilation final states and providing corresponding functions for computing annihilation cross sections. For example, for the models described above we have:
\begin{minted}{python}
>>> ScalarMediator.list_annihilation_final_states()
['mu mu', 'e e', 'g g', 'pi0 pi0', 'pi pi', 's s']
>>> VectorMediator.list_annihilation_final_states()
['mu mu', 'e e', 'pi pi', 'pi0 g', 'pi0 v', 'v v']
\end{minted}
These lists \emph{exclude} final states where the annihilation is extremely suppressed by couplings or phase-space factors (e.g., $\pi^+ \pi^- \gamma \gamma$, $S \pi^0 \pi^0$, $S \pi^+ \pi^-$ for the scalar model). They also exclude final states arising from radiative processes (such as $e^+ e^- \gamma$). Depending on the couplings' values and the center of mass energy of the DM, some of these final state may be inaccessible.

The corresponding annihilation cross sections and branching fractions can be accessed using the \mil{annihilation_cross_sections} function:
\begin{minted}{python}
>>> from hazma.scalar_mediator import ScalarMediator
>>> sm = ScalarMediator(mx=180., ms=190., gsxx=1., gsff=0.1,
...                     gsGG=0.1, gsFF=0.1, lam=2e5)
>>> e_cm = 400.
>>> sm.annihilation_cross_sections(e_cm)  # MeV^-2
{'mu mu': 1.018678054222354e-16,
 'e e': 3.892649866478948e-21,
 'g g': 2.4381469306302895e-21,
 'pi0 pi0': 7.975042237472552e-17,
 'pi pi': 1.5169347389281449e-16,
 's s': 1.78924656025887e-07,
 'total': 1.7892465635920504e-07}
\end{minted}
Note that in this example $m_\chi < m_S$, but the center of mass energy is large enough to allow the process $\bar{\chi}\chi \to S S$. The function \mil{Theory.annihilation_branching_fractions()} is invoked the same way and returns a \mil{dict} of branching fractions for annihilation into each final state $X$:
\begin{align}
    \operatorname{BF}(\bar{\chi}\chi\to X) = \frac{\sigma_{\bar{\chi}\chi\to X}}{\sum_{Y} \sigma_{\bar{\chi}\chi\to Y}}
\end{align}

Fig.~(\ref{fig:scalar_ann_bfs}) shows the annihilation branching fractions for the scalar model. The different rows exhibit results for the Higgs portal and heavy quark UV completions discussed above, which will be referred to throughout this work. In the first case, the branching fraction is largest into whichever final state is closest to threshold since $S$ couples to Standard Model states roughly proportionally to their Yukawas. The branching fraction into $\gamma\gamma$ is small for non-relativistic DM annihilations due to the derivative coupling between $S$ and the photon field. For the heavy quark coupling pattern only hadronic final states are accessible since the mediator couplings exclusively to photons and gluons.

Fig.~(\ref{fig:vector_ann_bfs}) collects the branching fractions for the vector model, assuming it kinetically mixes with the photon (top row) or couples only to quarks (middle and bottom rows). The coupling-dependence of the vector's cross sections can be understood without performing detailed calculations. Since the vector model can be recast as a $\operatorname{U}(1)$ gauge theory (by using the Stuckelberg mechanism to generate the mass term and decoupling the Stuckelberg field), the vector's couplings to pions are the sums of its couplings to their constituents: $g_{V\pi^\pm\pi^\pm} = g_{Vuu} - g_{Vdd}$, $g_{V\pi^0\pi^0} = 0$. The cross section $\sigma_{\bar{\chi}\chi \to \pi^+ \pi^-}$ is thus proportional to $(g_{Vuu} - g_{Vdd})^2$, explaining why that final state has a branching fraction of zero in the bottom row of the figure. The vector's coupling to $\pi^0\gamma$ and $\pi^0 V$ come from the anomalous part of the chiral Lagrangian. Since the anomaly is exact at one loop, the relevant diagrams in the former case can be evaluated with a proton running in the loop, making it clear that $\sigma_{\bar{\chi}\chi\to\pi^0\gamma} \propto (2 g_{Vuu} + g_{Vdd})^2$. This cross section is suppressed due to the loop factors, but provides an important contribution to the gamma-ray spectrum. For the coupling choices used for the plots in this work, the spectrum contribution from the $\pi^0 V$ final state is either negligible or identically zero.

Finally, the mediator decay partial widths are easily obtained with the \mil{partial_widths} function:
\begin{minted}{python}
>>> from hazma.scalar_mediator import ScalarMediator
>>> sm = ScalarMediator(mx=120., ms=280., gsxx=1., gsff=0.1,
...                     gsGG=0.1, gsFF=0.1, lam=2e5)
>>> sigmas = sm.partial_widths()
>>> sigmas  # MeV
{'g g': 1.472617003459079e-13,
 'pi0 pi0': 3.261686997076484e-09,
 'pi pi': 1.8664869506864194e-09,
 'x x': 1.522436123428156,
 'e e': 4.798160511838726e-13,
 'mu mu': 5.792897882600423e-09,
 'total': 1.522436134349855}
\end{minted}

\section{Building Blocks of MeV Gamma-Ray Spectra}
\label{sec:computing_gamma_ray_spectra}

The total gamma-ray spectrum for dark matter annihilation consists of three parts:
\begin{align}
  \label{eq:total_ann_spec}
  \left. \frac{dN}{dE_\gamma} \right\rvert_{\bar{\chi}\chi} &= \left. \frac{dN}{dE_\gamma} \right\rvert_{\bar{\chi}\chi,\text{line}} + \left. \frac{dN}{dE_\gamma} \right\rvert_{\bar{\chi}\chi,\text{dec.}} + \left. \frac{dN}{dE_\gamma} \right\rvert_{\bar{\chi}\chi,\text{FSR}}.
\end{align}
The first term accounts for photons from final states containing one or more monochromatic gamma rays, which result in spectral lines. The second term is the spectrum of photons produced through the radiative decays of final state particles. Decays can produce dramatic spectral features such as the neutral pion ``box'' as well as $E_\gamma^{-1}$ spectra. \revised{The third term accounts for photons produced through final state radiation (FSR) from annihilation into electromagnetically charged final states, and possibly internal bremmstrahlung (IB).} FSR generically produces spectra that scale as $dN/dE_\gamma \propto E_\gamma^{-1}$ at low energies, in accordance with Low's theorem~\cite{Low1958,BurnettKroll1968}, which could be distinguished from the softer $dN/dE_\gamma \propto E_\gamma^{-\Gamma}$, with $\Gamma\sim2$ for typical astrophysical background.

The calculations required to account for these contributions are described in detail in this section, beginning with the trivial case of spectral lines and the associated \mil{hazma} functions. The model-dependent FSR spectra are explained in detail since we compute them exactly for leptonic and hadronic final states. With regard to radiative decays, we present a comprehensive overview of the contributions to the $\pi^\pm$ decay spectrum, which is currently absent from the literature and thus has not been considered in prior work on sub-GeV DM. Code snippets for computing different decay spectra are included throughout.

The section concludes by showing how to compute FSR and radiative decay spectra in \mil{hazma} for the built-in models. A key feature of \mil{hazma} is that the spectrum functions can be employed to analyze user-defined models, as demonstrated in a detailed example in App.~(\ref{app:user_defined_models}).

\subsection{Monochromatic Gamma Rays}
\label{sub:lines}

The only final states containing monochromatic photons that are relevant for this work are $\pi^0 \gamma$ and $\gamma \gamma$:
\begin{align}
  \label{eq:line_ann_spec}
  \left. \frac{dN}{dE_\gamma} \right\rvert_{\bar{\chi}\chi,\text{line}} (E_\gamma) & = \operatorname{Br}(\bar{\chi}\chi \to \pi^0 \gamma) ~ \delta(E_{\pi^0 \gamma} - E_\gamma) + 2 \operatorname{Br}(\bar{\chi}\chi \to \gamma\gamma) ~ \delta(E_{\text{CM}} / 2 - E_\gamma),
\end{align}
where the energy of the line from the $\pi^0 \gamma$ final state is $E_{\pi^0 \gamma} \equiv (E_{\text{CM}}^2 - m_{\pi^0}^2) / (2 E_{\text{CM}})$ and $E_{\text{CM}}$ is the center of mass energy. This contribution to the DM annihilation spectrum is thus obtained by computing the branching fractions appearing in this expression.

The \mil{Theory.gamma_ray_lines()} method returns a \mil{dict} with information about monochromatic gamma-ray lines produced in dark matter annihilations. For example, the scalar mediator model has a $2\gamma$ line and the vector model has a line from the $\pi^0\gamma$ final state. Assuming heavy quark-type couplings for the former and quark-only couplings for the later, the line energies and branching fractions are found to be
\begin{minted}{python}
>>> from hazma.scalar_mediator import HeavyQuark
>>> sm = HeavyQuark(mx=140., ms=1e3, gsxx=1., gsQ=0.1, mQ=1e3, QQ=0.1)
>>> sm.gamma_ray_lines(e_cm=300.)
{'g g': {'energy': 150.0, 'bf': 1.285653242415087e-08}}
>>> from hazma.vector_mediator import QuarksOnly
>>> vm = QuarksOnly(mx=140., mv=1e3, gvxx=1., gvuu=0.1, gvdd=0.1, gvss=0.)
>>> vm.gamma_ray_lines(e_cm=300.)
{'pi0 g': {'energy': 119.63552908740002, 'bf': 1.0}}
\end{minted}

\subsection{Final State Radiation}
\label{sub:fsr}
Dark matter annihilating into charged SM particles generically produces photons via final state radiation (FSR), which is accounted for by computing
\begin{align}
  \left. \frac{dN}{dE_\gamma} \right\rvert_{\bar{\chi}\chi,\text{FSR/IB}} &= \frac{1}{\sigma_{\bar{\chi}\chi}} \sum_A \frac{d\sigma_{\bar{\chi}\chi \to A \gamma}}{dE_\gamma}(E_\gamma)\\
                                                                       &\approx \sum_A \left[ \frac{\sigma_{\bar{\chi}\chi \to A}}{\sigma_{\bar{\chi}\chi}} \right] \left[ \frac{1}{\sigma_{\bar{\chi}\chi \to A}} \frac{d\sigma_{\bar{\chi}\chi \to A \gamma}}{dE_\gamma}(E_\gamma) \right]\\
                                                                       &\equiv \sum_A \operatorname{Br}(\bar{\chi}\chi \to A) \left. \frac{dN}{dE_\gamma} \right\rvert_{\bar{\chi}\chi \to A} (E_\gamma),
  \label{eq:fsr_ann_spec}
\end{align}
where in the second line we assumed $\sigma_{\bar{\chi}\chi \to A\gamma} \ll \sigma_{\bar{\chi}\chi \to A}$ and the sum is over final states containing electromagnetically-charged particles. The required branching fractions were computed in Sec.~(\ref{sub:particle_physics_quantities}).

\revised{It is well-known that the FSR physics factorizes from the hard process in the limit where the mass of the radiating particle is zero and the radiation is collinear with the radiating particle\cite{collins1989factorization}. This leads to a universal spectrum for FSR for DM annihilating into charged particle-antiparticle pairs $\bar{X} X$, which, at leading order in $\alpha$ is: \footnote{Note that the approximation takes several forms in the literature, often omitting the $-1$ term, which are equivalent in the small-$\mu$ limit.}
\begin{align}
  \label{eq:altarelli_parisi}
  \left. \frac{dN}{dE_\gamma} \right\rvert_{\bar{\chi}\chi\to \bar{X} X} &\approx \frac{2 \alpha}{\pi Q} \cdot P_{\gamma X}(x) \cdot \left[ \log\left( \frac{1 - x}{\mu_{X}^2} \right) - 1 \right].
\end{align}
with $\mu_{X} = m_{X}/Q$ is the ratio of the mass of the radiating particle divided by the center of mass energy. The form factor $P_{\gamma X}(x)$ is called the splitting function, and depends only on the spin of the radiating particle~\cite{Chen:2016wkt} and the fraction of energy it takes from the splitting particle, $x$: \footnote{Note that the scalar splitting function in Ref.~\cite{Birkedal:2005ep} is a factor of two smaller than in Ref.~\cite{Chen:2016wkt}. We have rederived the splitting function and agree with the result in the later reference.}
\begin{align}
    P_{\gamma X}(x) = \begin{cases}
        \frac{2(1 - x)}{x} & \text{for } X=\u{scalar}\\
        \frac{1 + \left( 1 - x \right)^2}{x} & \text{for } X=\u{fermion}
    \end{cases},
\end{align}
where $x \equiv 2 E_\gamma / Q$.}

\revised{This spectrum, which we call the Altarelli-Parisi (AP) approximation, has been employed in prior studies of indirect detection of sub-GeV DM~\cite{Bartels2017,PhysRevLett.94.171301,cascadia,1126-6708-2008-01-049,PhysRevD.80.023506}.} However, the AP approximation insufficient for our analysis \revised{since it is only valid near the mass singularity, i.e. for $\mu_{X}\ll1$}. This does not hold for MeV DM in the Milky Way halo annihilating non-relativistically into $\mu^+ \mu^-$ (or $\pi^+ \pi^-$), since in this case $Q \sim m_\mu \sim \mathcal{O}(100~\mathrm{MeV})$. As a result of this assumption, the AP spectrum cuts off at $x = 1- \exp(1)\mu^2$, which is different from the exact kinematic threshold $x = 1-4\mu^2$.

In this section we instead exactly compute the model-dependent FSR spectra for the $e^+ e^-$ and $\mu^+ \mu^-$ final states, as well as the $\pi^+ \pi^-$ final state. The resulting spectra for leptonic final states are (see chapter 20 of \cite{schwartz_2017} for details of how these calculations can be performed)
\begin{align}
    \label{eq:lepton_fsr_spectra}
    \left. \frac{dN}{dE_\gamma} \right\rvert_{\bar{\chi}\chi \to S^* \to \bar{\ell} \ell} &= \frac{\alpha}{E_\gamma \pi (1 - 4 \mu^2)^{3/2}} \left[ \left( 2 (1 - x - 6 \mu^2) + (x + 4 \mu^2)^2 \right) \log \frac{1 + \sqrt{1 - \frac{4 \mu^2}{1-x}}}{1 - \sqrt{1 - \frac{4 \mu^2}{1-x}}} \right. \notag \\
    & \hspace{2.5cm} \left. - 2 (1-4\mu^2)(1-x) \sqrt{1 - \frac{4 \mu^2}{1-x}} \right]\\
    \left. \frac{dN}{dE_\gamma} \right\rvert_{\bar{\chi}\chi \to V^* \to \bar{\ell} \ell}
    &= \frac{\alpha}{E_\gamma \pi \sqrt{1 - 4\mu^2} (1 + 2 \mu^2)} \left[ (1 + (1-x)^2 - 4\mu^2 (x + 2\mu^2)) \log \frac{1 + \sqrt{1 - \frac{4 \mu^2}{1-x}}}{1 - \sqrt{1 - \frac{4 \mu^2}{1-x}}} \right. \notag \\
    & \hspace{2.5cm} \left. - (1 + (1 - x)^2 + 4\mu^2 (1-x)) \sqrt{1 - \frac{4 \mu^2}{1-x}} \right].
\end{align}
The FSR spectrum for the $\pi^+ \pi^-$ final state can be computed using the interaction Lagrangians derived in our companion paper~\cite{companion} and reproduced in Eqn.~(\ref{eq:LagIntS}) and~(\ref{eq:LagIntV}). The final expressions for the spectra are
\begin{align}\label{eq:pion_fsr_spectra}
    \left.\frac{dN}{dE_\gamma} \right\rvert_{\bar{\chi}\chi \to S^* \to \pi^{+}\pi^{-}}
    &= \frac{2 \alpha}{E_\gamma \pi \sqrt{1 - 4 \mu^2}} \left[ (1 - x - 2\mu^2) \log \frac{1 + \sqrt{1 - \frac{4 \mu^2}{1-x}}}{1 - \sqrt{1 - \frac{4 \mu^2}{1-x}}} - (1 - x) \sqrt{1 - \frac{4 \mu^2}{1-x}} \right]\\
    \left. \frac{dN}{dE_\gamma} \right\rvert_{\bar{\chi}\chi \to V^* \to \pi^{+}\pi^{-}}
    &= \frac{2 \alpha}{E_\gamma \pi (1 - 4\mu^2)^{3/2}} \left[ (1 - x - 2\mu^2)(1 - 4\mu^2) \log \frac{1 + \sqrt{1 - \frac{4 \mu^2}{1-x}}}{1 - \sqrt{1 - \frac{4 \mu^2}{1-x}}} \right. \notag \\
    & \hspace{4.0cm} \left. - \left( (1 - x)(1 - 4 \mu^2) - x^2 \right) \sqrt{1 - \frac{4 \mu^2}{1-x}} \right].
\end{align}
\revised{In the limit $\mu \ll 1$, these spectra reduce to the AP approximation, up to the additional, non-collinear terms: $\alpha x / \pi$ and $2 \alpha x / \pi$ for the processes $\bar{\chi}\chi \to S^* \to \bar{\ell} \ell \gamma$ and $\bar{\chi}\chi \to V^* \to \pi^{+}\pi^{-} \gamma$ respectively. The spectra all scale as $E_\gamma^{-1}$ in accordance with Low's theorem at low photon energies.

Computing the FSR spectra for annihilation into $\pi^+ \pi^-$ requires including internal bremmstrahlung diagrams with the 4-point vertices $S,V - \pi^+ - \pi^- - \gamma$. In the scalar mediator case, this diagram is only present when $g_{SG} \neq 0$. The FSR spectrum is in accord with the AP approximation in the small $\mu$ limit. In the vector case, this diagram cannot be eliminated since it is required by gauge invariance.}

\begin{figure}[tbh!]
  \centering
  \includegraphics[width=\textwidth]{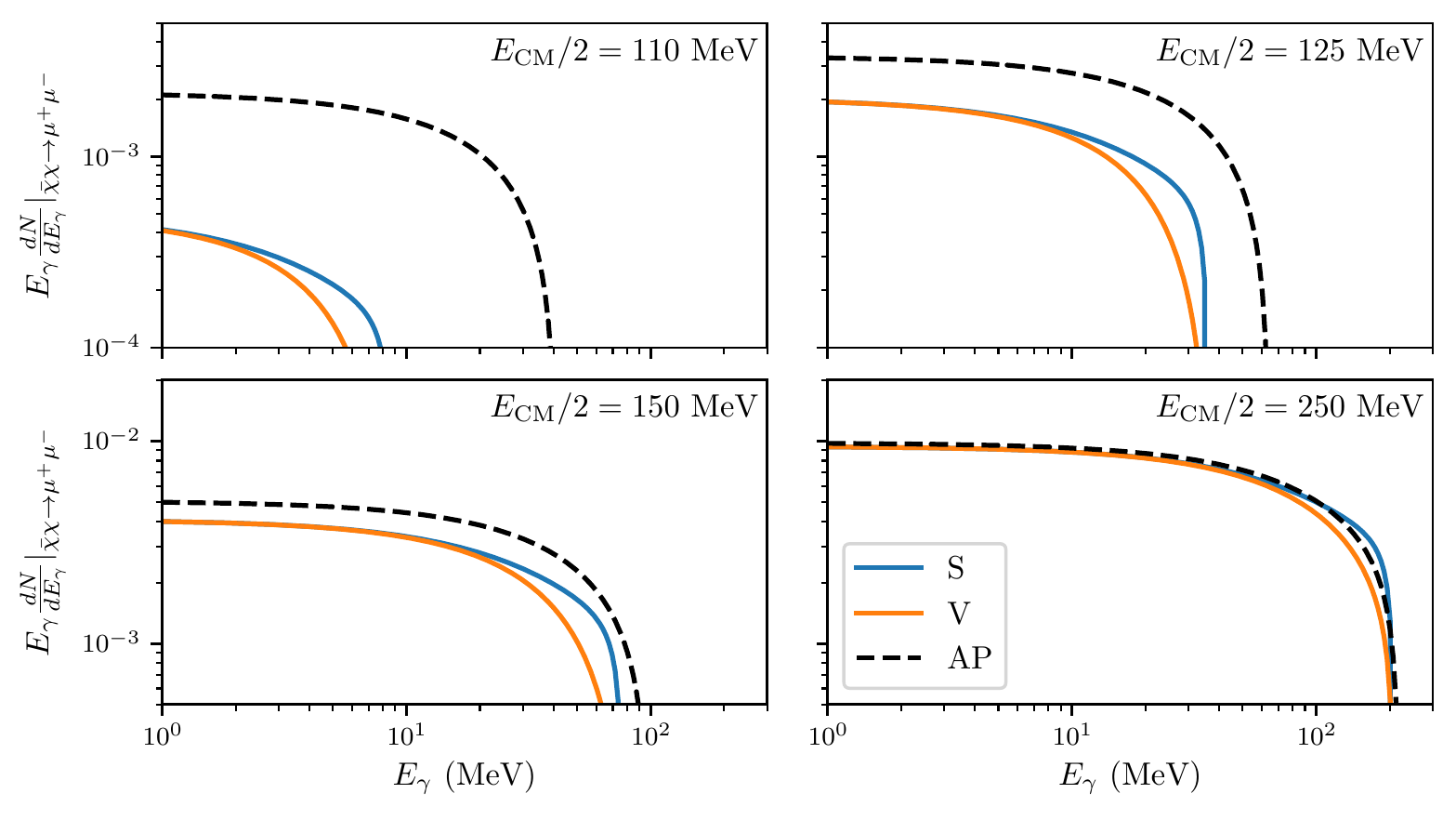}
  \caption{\textbf{Final state radiation spectrum for dark matter annihilating into $\mu^+ \mu^-$.} The curves correspond to the scalar (blue curve) and vector (orange curve) models, with the indicated center of mass energies. The Altarelli-Parisi spectrum from Eqn.~(\ref{eq:altarelli_parisi}) is also shown (dashed black curve), illustrating the limiting behavior of the spectra as $m_\ell \ll Q$. \revised{As seen in Eqn.~(\ref{eq:lepton_fsr_spectra}), the spectra depend only on the center of mass energy of the interaction and not the mediator or DM mass, since these cancel in the cross section ratio in Eqn.~(\ref{eq:fsr_ann_spec}).}
  \nbLink{muon_fsr}
  \scriptLink{muon_fsr}}
  \label{fig:lepton_fsr}
\end{figure}

\begin{figure}[tbh!]
  \centering
  \includegraphics[width=\textwidth]{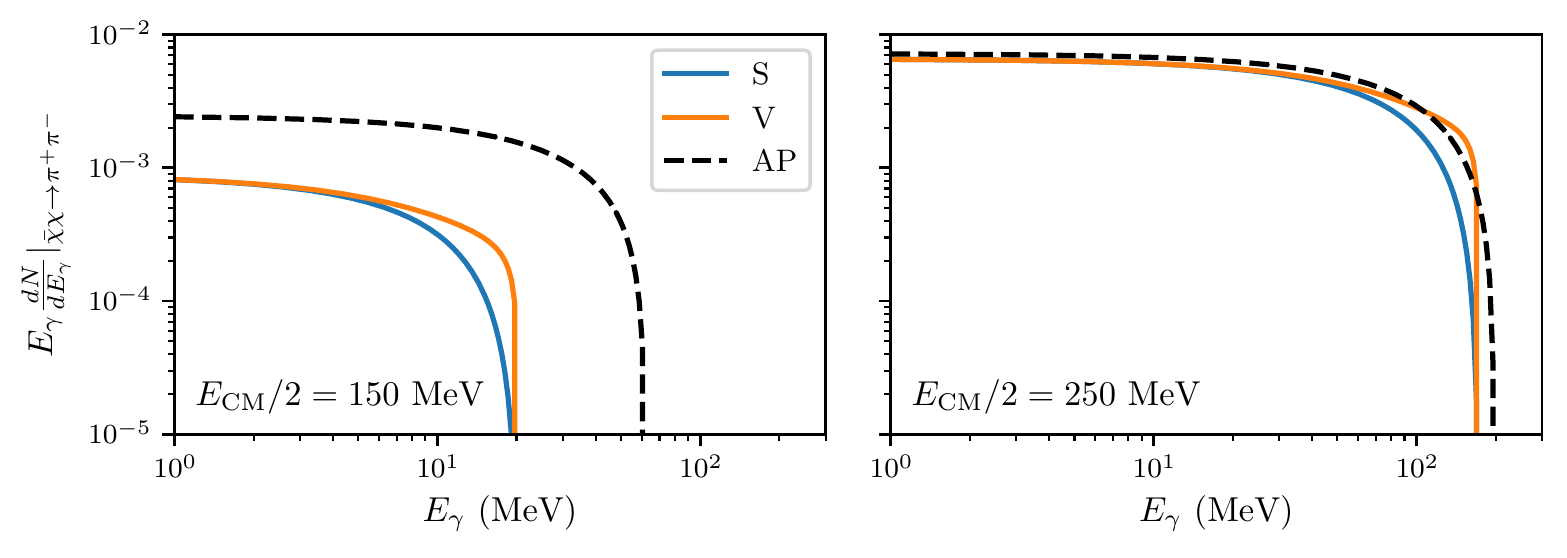}
  \caption{\textbf{Final state radiation spectrum for dark matter annihilating into $\pi^+ \pi^-$.} The blue curve corresponds to the scalar-mediator model and the orange to the vector-mediator case. The dashed black curve shows the Altarelli-Parisi spectrum. The panels are labeled with the annihilation's center of mass energy.
  \nbLink{charged_pion_fsr}
  \scriptLink{charged_pion_fsr}}
  \label{fig:pion_fsr}
\end{figure}

In Fig.~(\ref{fig:lepton_fsr}) we plot the spectra for FSR in the scalar and vector models as well as the AP spectrum (blue, red and yellow curves respectively). As can be seen from the above equations, the specific values of the couplings, mediator masses and DM mass do not impact the shape of the spectrum since they cancel when dividing by $\sigma_{\bar{\chi} \chi \to \bar{\ell} \ell}$. As expected, the AP approximation performs poorly near the dimuon threshold (upper left panel). The photon energy at which the spectrum cuts off is an order of magnitude too large, and while the normalization is within a factor of two for the scalar model's spectrum, it is an order of magnitude larger than the vector model's one. The situation improves at larger center of mass energies, and just above the $\pi \pi$ threshold the spectra all agree to within a factor of two. The pion FSR spectra are shown in Fig.~(\ref{fig:pion_fsr}).

We demonstrate how to compute these spectra in \mil{hazma} at the end of this section.

\subsection{Radiative Decay Spectra}
\label{sub:radiative_decay_spectra}

Photons are produced when dark matter annihilates into a state $A$ that subsequently undergoes radiative decay $A \to B \gamma$:
\begin{align}
  \label{eq:dec_ann_spec}
  \left. \frac{dN}{dE_\gamma} \right\rvert_{\bar{\chi}\chi,\text{dec.}} (E_\gamma) &= \sum_A \operatorname{Br}(\bar{\chi}\chi \to A) \left. \frac{dN}{dE_\gamma} \right\rvert_A (E_\gamma).
\end{align}
In addition to collecting well-known expressions for the neutral pion and muon radiative decay spectra, we carefully account for the three dominant contributions to the charge pion's radiative decay spectrum. Functions for computing radiative decay spectra are contained in the \mil{decay} module, and described below. These can be utilized to compute the decay contribution to the annihilation spectrum for arbitrary final states.

\subsubsection{Decay Spectra in Different Frames}

Before computing the radiative decay spectra for the $\mu^\pm$, $\pi^0$ and $\pi^\pm$, we review how to boost a spectrum $dN/dE_R$ from the parent particle's rest frame to obtain the ``lab frame'' spectrum $dN/dE_L$, where the particle has boost $\gamma = E / m$ along the $z$-axis. Let the photon phase space coordinates be $(E_R, \cos\theta_R)$ in the rest frame and $(E_L, \cos\theta_L)$ in the lab frame, where $\theta_R$ and $\theta_L$ are the angles with respect to the $z$ axis and we have assumed azimuthal symmetry. Since the number of photons in a small patch of phase space does not depend on which coordinates we choose, we have
\begin{align}\label{eq:frame_trans_spec}
    dN &= f(E_R, \cos \theta_R)\ dE_R\ d\cos\theta_R\\
       &= \begin{vmatrix}
            \frac{dE_R}{dE_L} & \frac{dE_R}{d\cos\theta_L}\\
            \frac{d\cos\theta_R}{dE_L} & \frac{d\cos\theta_R}{d\cos\theta_L}
        \end{vmatrix} f(E_R(E_L, \cos\theta_L), \cos\theta_R(E_L, \cos\theta_L)) dE_L\ d\cos\theta_L,
\end{align}
where we recognize the partial phase space density as $f(E, \cos\theta) \equiv dN/(dE\ d\cos\theta)$ and the term with the two vertical lines is the Jacobian factor for the change of variables. If \(f(E_R, \cos\theta_R)\) is independent of \(c\theta_R\) (which is the case for all the spectra we will consider), we can write:
\begin{align}
  f(E_{R}, c\theta_{R}) = \dfrac{1}{2}\dfrac{dN}{dE_{R}}.
\end{align}
The variables \(E_{1}, E_{2}, c\theta_{1}\) and \(c\theta_{2}\) are related via a Lorentz boost. The relationships are
\begin{align}
  E_R &= \gamma E_L (1 - \beta c\theta_L), &
  \cos\theta_R &= \dfrac{\beta - \cos\theta_L}{1 - \beta \cos\theta_L},
\end{align}
where \(\beta = \sqrt{1 - 1 / \gamma^2}\) is the particle's velocity in natural units. The spectrum in the lab frame is then obtained by integrating Eqn.~(\ref{eq:frame_trans_spec}) over \(\cos\theta_L\):
\begin{align}\label{eq:spec_boost}
  \dfrac{dN}{dE_L} & = \int \cos\theta_L\dfrac{1}{2\gamma(\beta \cos\theta_L - 1)}\dfrac{dN}{dE_R}.
\end{align}

\subsubsection{Neutral Pions}
The dominant decay mode for neutral pions is \(\pi^{0}\to\gamma\gamma\) with a branching fraction of about \(99\%\). Due to this decay modes' large branching fraction and the fact that it has two photons in the final state, we ignore the $\pi^0$'s other decay modes. In the pion's rest frame, the gamma-ray spectrum is trivial:
\begin{align}
    \left. \dfrac{dN}{dE_\gamma} \right\rvert_{\pi^0}(E_{\pi^0} = m_{\pi^0}) & = 2\times\delta\left(E_{R} - \dfrac{m_{\pi}}{2}\right)
\end{align}
where the factor of \(2\) comes from the fact that there are two photons in the final state.

Applying Eqn.~(\ref{eq:spec_boost}) gives that the gamma ray spectrum in the laboratory frame is
\begin{align}
    \left. \dfrac{dN}{dE_\gamma} \right\rvert_{\pi^0}(E_{\pi^0}) &= \int \cos\theta_{L}\dfrac{1}{2\gamma(\beta \cos\theta_{L} - 1)} 2\times\delta\left(E_{R} - \dfrac{m_{\pi^0}}{2}\right)\\
                                                  &= \dfrac{2}{\gamma\beta m_{\pi^0}} \left[\theta(E_{\gamma} - E_{-})- \theta(E_{\gamma} -E_{+})\right]
\end{align}
with \(E_{\pm} = m_{\pi^0} / 2\gamma(1\mp\beta)\), which is the characteristic box spectrum centered at $m_{\pi^0} / 2$.

The boosted spectrum can be computed in \mil{hazma} as follows:
\begin{minted}{python}
>>> from hazma.decay import neutral_pion as dnde_pi0
>>> e_gams = np.array([100., 125., 150])  # photon energies
>>> e_pi0 = 180.                          # pion energy
>>> dnde_pi0(e_gams, e_pi0)
array([0.0165965, 0.0165965, 0.       ])
\end{minted}

\subsubsection{Muons}

The primary muon decay channel is $\mu^- \to e^- \bar{\nu}_e \nu_\mu$. The photon in the corresponding radiative decay $\mu^- \to e^- \bar{\nu}_e \nu_\mu \gamma$ has the following spectrum in the muon's rest frame~\cite{RevModPhys.73.151}:
\begin{align}
    \dfrac{dN}{dE_\gamma} \bigg{|}_{\mu^\pm}(E_\mu = m_\mu) &= \frac{\alpha (1-x)}{36 \pi E_\gamma} \left[ 12 \left(3 - 2 x {(1-x)}^2\right) \log \left(\frac{1-x}{r}\right) \right. \notag\\
    &\hspace{5cm} \left. + x (1-x) (46 - 55x) - 102 \right],
\label{eq:muraddec}
\end{align}
where $r \equiv {(m_{e}/m_{\mu})}^2$, $x \equiv 2 E_\gamma / m_\mu$ and the kinematic bounds on $E_\gamma$ translate into $0 \leq x \leq (1 - r)$. To obtain the spectrum in the lab frame we substitute the above expression into Eqn.~(\ref{eq:spec_boost}) and evaluate the integral numerically. 

The lab-frame decay spectrum can be compute for arbitrary muon energies using:
\begin{minted}{python}
>>> from hazma.decay import muon as dnde_mu
>>> e_gams = np.array([1., 10., 100.])  # photon energies
>>> e_mu = 130.                         # muon energy
>>> dnde_mu(e_gams, e_mu)
array([1.76076858e-02, 1.34063877e-03, 4.64775301e-08])
\end{minted}

\subsubsection{Charged Pions}

The dominant pion decay is $\pi^+ \to \ell^+ \nu_\ell$, where $\ell = e, \mu$ and $\ell$ is produced approximately on shell, thanks to the narrow width approximation when $\ell = \mu$. Contributions to the pion's radiative decay spectrum come from initial state radiation from the photon or FSR from the lepton, as well as emission from virtual hadronic states (the ``structure-dependent'' component). An expression for $d^2 \Gamma_{\pi^+ \to \ell^+ \nu \gamma} / dx dy$ is given in Eqn. (72) of~\cite{bryman1982pi} and Eq. (68.3) in the Review of Particle Physics~\cite{PhysRevD.98.030001}~\footnote{Note that the Particle Data Group uses the convention $f_\pi = 130~\mathrm{MeV}$, which is a factor of $\sqrt{2}$ larger than the value used in this work.}, where $x \equiv 2 E_\gamma / m_{\pi^+}$ and $y \equiv 2 E_\ell / m_{\pi^+}$. After changing to the Mandelstam variables
\begin{align*}
    s &\equiv (p_\pi - p_\gamma)^2 = m_{\pi^+}^2 (1-x),\\
    t &\equiv (p_\pi - p_\ell)^2 = m_{\pi^+}^2 \left( 1 - y + \frac{m_\ell^2}{m_{\pi^+}^2} \right),
\end{align*}
integrating over $0 \leq t \leq (m_{\pi^+}^2 - s) (s - m_\ell^2) / s$, and dividing by the total $\pi^+$ decay width, we obtain an analytic expression for this channel's contribution to the $\pi^+$ decay spectrum:
\begin{align}
    \label{eq:pdg_contrib_to_pi_dec_spec}
    \operatorname{Br}(\pi^+ \to \ell^+ \nu_\ell) \cdot \left. \frac{dN}{dE_\gamma} \right\rvert_{\pi^+ \to \ell^+ \nu_\ell} (E_{\pi^+} = m_{\pi^+}) &= \frac{\Gamma_{\pi^+ \to \ell^+ \nu_\ell}}{\Gamma_{\pi^+}} \cdot \frac{\alpha (f(x) + g(x))}{24 \pi m_{\pi^+} f_\pi^2 (r-1)^2 (x-1)^2 r x},
\end{align}
where $r \equiv m_\ell^2 / m_{\pi^+}^2$ and
\begin{align}
    f(x) & = (r+x-1) \left[m_{\pi^+}^2 x^4 \left(F_A^2+F_V^2 \right) \left(r^2-r x+r-2 (x-1)^2\right) \right. \notag\\[1em]
    &\hspace{1.3in} \left. - 12 \sqrt{2} f_\pi m_{\pi^+} r (x-1) x^2 (F_A (r-2 x+1)+x F_V) \right. \notag\\[1em]
    &\hspace{1.3in} \left. - 24 f_\pi^2 r (x-1) \left(4 r (x-1)+(x-2)^2\right)\right], \\[1em]
    g(x) & = 12 \sqrt{2} f_\pi r (x-1)^2 \log\left(\frac{r}{1-x}\right) \left[ m_{\pi^+} x^2 (F_A (x-2 r) - x F_V) \right. \notag\\
    &\hspace{2.5in} \left. + \sqrt{2} f_\pi \left(2 r^2-2 r x-x^2+2x-2\right) \right].
\end{align}
The vector form factor is $F_V(q^2) = F_V(0) (1 + a q^2)$ with $F_V(0) = 0.0254$, slope parameter $a = 0.10$ and $q^2 = 1 - x$; the axial form factor is $F_A = 0.0119$~\cite{PhysRevD.98.030001}.

When $\ell = \mu$, the muon's subsequent radiative decay also contributes significantly to the pion's decay spectrum.  Since we can take the muon to be on shell by the narrow width approximation, this decay path's contribution is simply the product of the branching fraction for $\pi^+ \to \mu^+ \nu_\mu$ and the muon decay spectrum Eqn.~(\ref{eq:muraddec}) evaluated at the muon's energy. Our final expression for the charged pion decay spectrum is thus
\begin{align}
    \label{eq:pi_dec_spec}
    \dfrac{dN}{dE_\gamma} \bigg{|}_{\pi^+}(E_{\pi^+} = m_{\pi^+}) &= \sum_{\ell = e, \mu} \operatorname{Br}(\pi^+ \to \ell^+ \nu_\ell) \cdot \left. \frac{dN}{dE_\gamma} \right\rvert_{\pi^+ \to \ell^+ \nu_\ell} (E_{\pi^+} = m_{\pi^+})\\
    &\hspace{0.3in} + \operatorname{Br}(\pi^+ \to \mu^+ \nu_\mu) \cdot \dfrac{dN}{dE_\gamma} \bigg{|}_{\mu^\pm} \left( E_\mu = \frac{m_{\pi^+}^2 + m_\mu^2}{2 m_{\pi^+}} \right). \notag
\end{align}
This rest frame spectrum can be substituted into Eqn.~(\ref{eq:spec_boost}) and numerically integrated to obtain the total charged pion radiative decay spectrum. 

\begin{figure}[tbh!]
    \centering
    \includegraphics[scale=1]{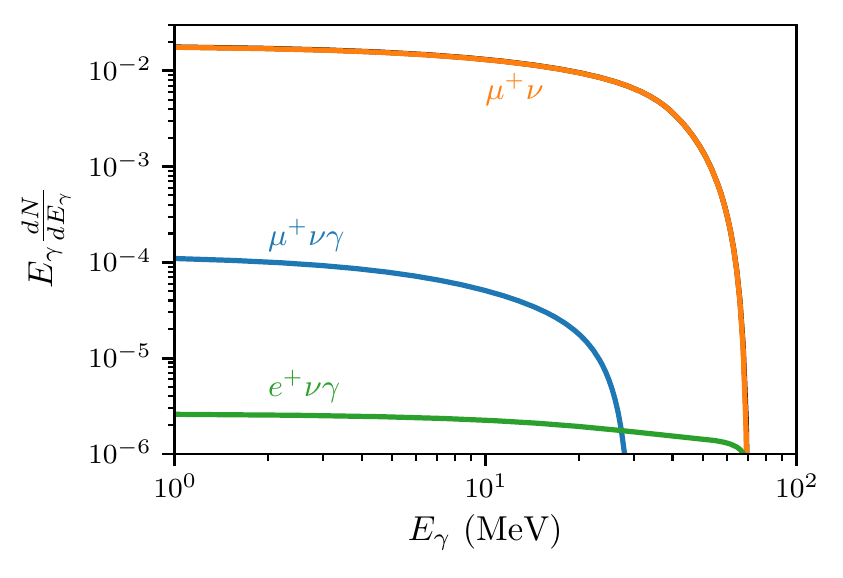}
    \caption{\textbf{Contributions to the charged pion radiative decay spectrum in the pion's rest frame.} The decay $\pi^+ \to \mu^+ \nu_\mu$ followed by radiative muon decay is plotted in orange. The other curves show the spectra for $\pi^+ \to \ell^+ \nu_\ell$, where the photon is produced via initial/final state radiation from the lepton or emission from virtual hadronic states. The total spectrum completely overlaps with the orange curve.
    \nbLink{charged_pion_decay}
    \scriptLink{charged_pion_decay}}
    \label{fig:charged_pion_decay_spec}
\end{figure}

Fig.~(\ref{fig:charged_pion_decay_spec}) shows the contributions of each term and the total spectrum in the pion's rest frame. The muon's radiative decay is by far the most important component due to its large branching fraction. There is also more phase space available for the photon in comparison with $\pi^+ \to \mu^+ \nu_\mu \gamma$ since the other final state particles are massless: the former spectrum cuts off at $\sim 69\us{MeV}$ and the latter at $(m_{\pi^+}^2 - m_\mu^2) / (2 m_{\pi^+}) \approx 27\us{MeV}$. The contribution from $\pi^+ \to e^+ \nu \gamma$ is smaller than the others due to helicity suppression.

The charged pion decay spectrum is included in \mil{hazma.decay}:
\begin{minted}{python}
>>> from hazma.decay import charged_pion as dnde_pi
>>> e_gams = np.array([1., 10., 100.])  # photon energies
>>> e_pi = 150.                         # pion energy
>>> dnde_pi(e_gams, e_pi)
array([1.76949944e-02, 1.32675207e-03, 1.16607174e-09])
\end{minted}
The individual contributions can also be computed by setting the \mil{mode} argument. The orange, blue and green curves are obtained using \mil{mode="munu"}, \mil{"munug"} and \mil{"enug"} respectively.

\subsection{Continuum Gamma-Ray Spectra in \texorpdfstring{\mil{hazma}}{hazma}}
\label{sub:continuum_spectra_hazma}

Models in \mil{hazma} provide a few methods for computing gamma-ray spectra at different levels of detail\revised{: accessing the total gamma-ray spectrum including all available final states, computing spectrum from specific final states or even decomposing the spectra into FSR or decay for individual channels and processes.} The \mil{total_spectrum} method gives the total \emph{continuum} gamma-ray spectrum at specified photon energies and fixed center-of-mass energy:
\begin{minted}{python}
>>> from hazma.scalar_mediator import HiggsPortal
>>> e_cm = 305.                         # DM center of mass energy
>>> e_gams = np.array([1., 10., 100.])  # photon energies
>>> hp = HiggsPortal(mx=150., ms=1e3, gsxx=0.7, stheta=0.1)
>>> hp.total_spectrum(e_gams, e_cm)
array([0.02484114, 0.00186874, 0.01827828])
\end{minted}
Underlying this are methods for computing the gamma-ray spectra for individual final states. These are accessible through the \mil{spectra} method, which also accepts a list of photon energies and a center of mass energies. It returns the total continuum spectrum and the contribution from each final state (ie, $dN/dE_\gamma$ multiplied by the final state's branching fraction):
\begin{minted}{python}
>>> hp.spectra(e_gams, e_cm)
{'mu mu': array([4.87977497e-03, 3.83154301e-04, 1.22977086e-06]),
 'e e': array([4.19129970e-07, 3.93151015e-08, 2.22413275e-09]),
 'pi0 pi0': array([0.        , 0.        , 0.01827702]),
 'pi pi': array([1.99609455e-02, 1.48554303e-03, 2.28115016e-08]),
 's s': array([0., 0., 0.]),
 'total': array([0.02484114, 0.00186874, 0.01827828])}
 \end{minted}
 The \mil{spectra} and \mil{total_spectrum} methods are provided by the \mil{Theory} class, and are driven by the method \mil{spectrum_funcs} which classes inheriting from \mil{Theory} are required to implement. This returns a \mil{dict} whose keys are strings corresponding to annihilation final states and whose values are methods returning the continuum gamma-ray spectrum for annihilations into that final state (ie, omitting the branching fraction factor).

The models built into \mil{hazma} provide more fine-grained methods for studying different channels' spectra. Each of these has a name corresponding to the final state and takes an additional string argument specifying the spectrum type (\mil{"decay"}, \mil{"fsr"} or \mil{"all"}, the default). The return value depends on the spectrum type argument. In the first case, the returned spectrum only accounts for the final state particles' radiative decays, which are model-independent. In the second case, the method returns the radiative decay spectrum, which is model-dependent. In the third, default case, the method returns the total continuum spectrum.

For example, in the scalar mediator model, the annihilation final states $e^+ e^-$, $\mu^+ \mu^-$, $\pi^0\pi^0$, $\pi^+ \pi^-$ and $S S$ contribute to the continuum annihilation spectrum. The following snippet shows how to call the spectrum methods for the instance of \mil{HiggsPortal} created above. The methods all follow the same naming conventions and return $dN/dE_\gamma$ in $\u{MeV}^{-1}$:
\begin{minted}{python}
>>> from hazma.scalar_mediator import HiggsPortal
>>> e_cm = 305.                         # DM center of mass energy
>>> e_gams = np.array([1., 10., 100.])  # photon energies
>>> hp = HiggsPortal(mx=150., ms=1e3, gsxx=0.7, stheta=0.1)
>>> hp.dnde_ee(e_gams, e_cm, spectrum_type="fsr")
>>> array([0.05435176, 0.00509829, 0.00028842])
>>> hp.dnde_ee(e_gams, e_cm, spectrum_type="decay")
>>> array([0., 0., 0.])  # electrons don't decay
>>> hp.dnde_pipi(e_gams, e_cm, spectrum_type="fsr")
>>> array([1.01492577e-03, 5.00245201e-05, 0.00000000e+00])
>>> hp.dnde_pipi(e_gams, e_cm, spectrum_type="decay")
>>> array([3.53972084e-02, 2.65985674e-03, 4.16120296e-08])
>>> hp.dnde_pipi(e_gams, e_cm)
>>> array([3.64121342e-02, 2.70988126e-03, 4.16120296e-08])
\end{minted}

\section{Gamma Ray Spectra from DM annihilation}%
\label{sec:dm_gamma_ray_spectra}

\begin{figure}[tbh!]
    \centering
    \includegraphics[width=1\linewidth]{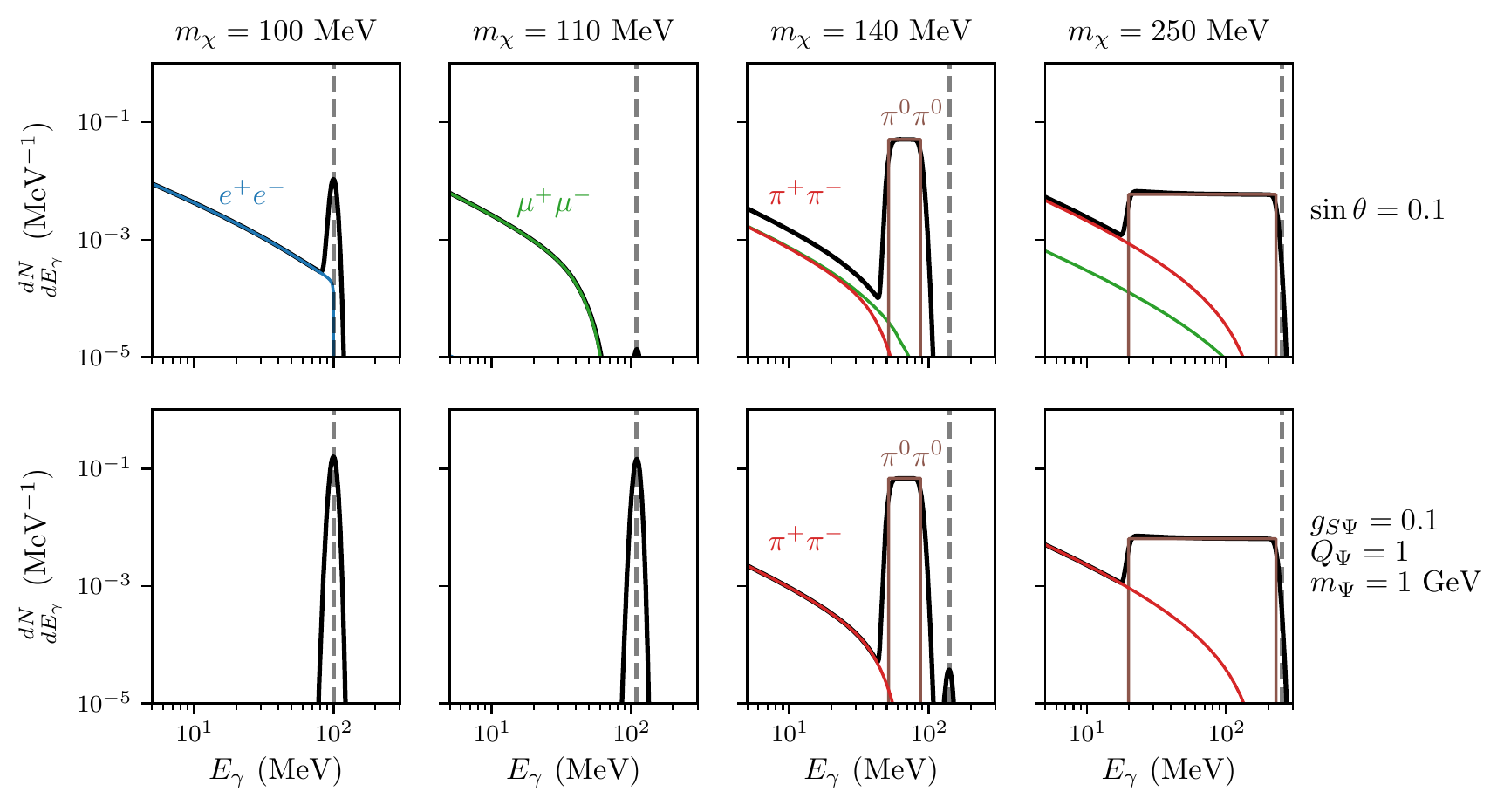}
    \caption{\textbf{Spectra from DM annihilation into Standard Model particles for scalar-mediator models.} The two-mediator final state is kinematically forbidden. The black curve is the total spectrum as seen by an instrument with 5\% energy resolution. The labeled colored curves are the spectra for individual final states, and the vertical dashed line marks the photons' energies in the $\gamma\gamma$ final state. The rows correspond to Higgs portal and heavy quark coupling patterns. The DM mass is fixed to the indicated value in each column.
    \nbLink{scalar_spectra_ann_to_sm}
    \scriptLink{scalar_spectra_ann_to_sm}}
    \label{fig:scalar_spectra_ann_to_sm}
\end{figure}

\begin{figure}[tbh!]
    \centering
    \includegraphics[width=1\linewidth]{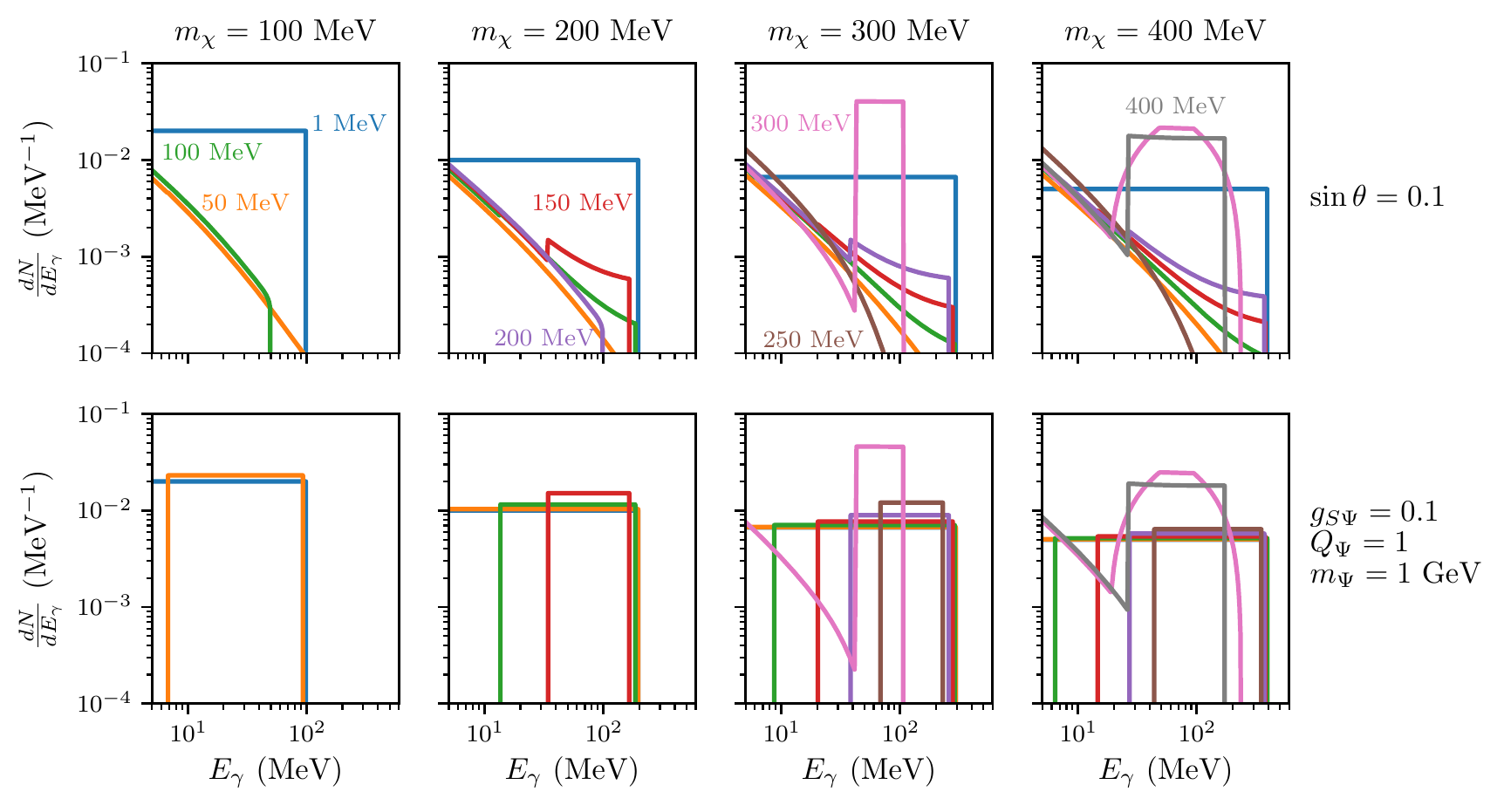}
    \caption{\textbf{Spectra from DM annihilation into mediators for scalar-mediator models.} Curves are only shown when $E_\u{CM} \geq 2 m_S$, and are labeled with the corresponding mediator mass. The spectra have not been convolved with an energy resolution function since the $S S$ final state does not produce gamma-ray lines.
    \nbLink{scalar_spectra_ann_to_sm}
    \scriptLink{scalar_spectra_ann_to_sm}}
    \label{fig:scalar_spectra_ann_to_med}
\end{figure}

\begin{figure}[tbh!]
    \centering
    \includegraphics[width=1\linewidth]{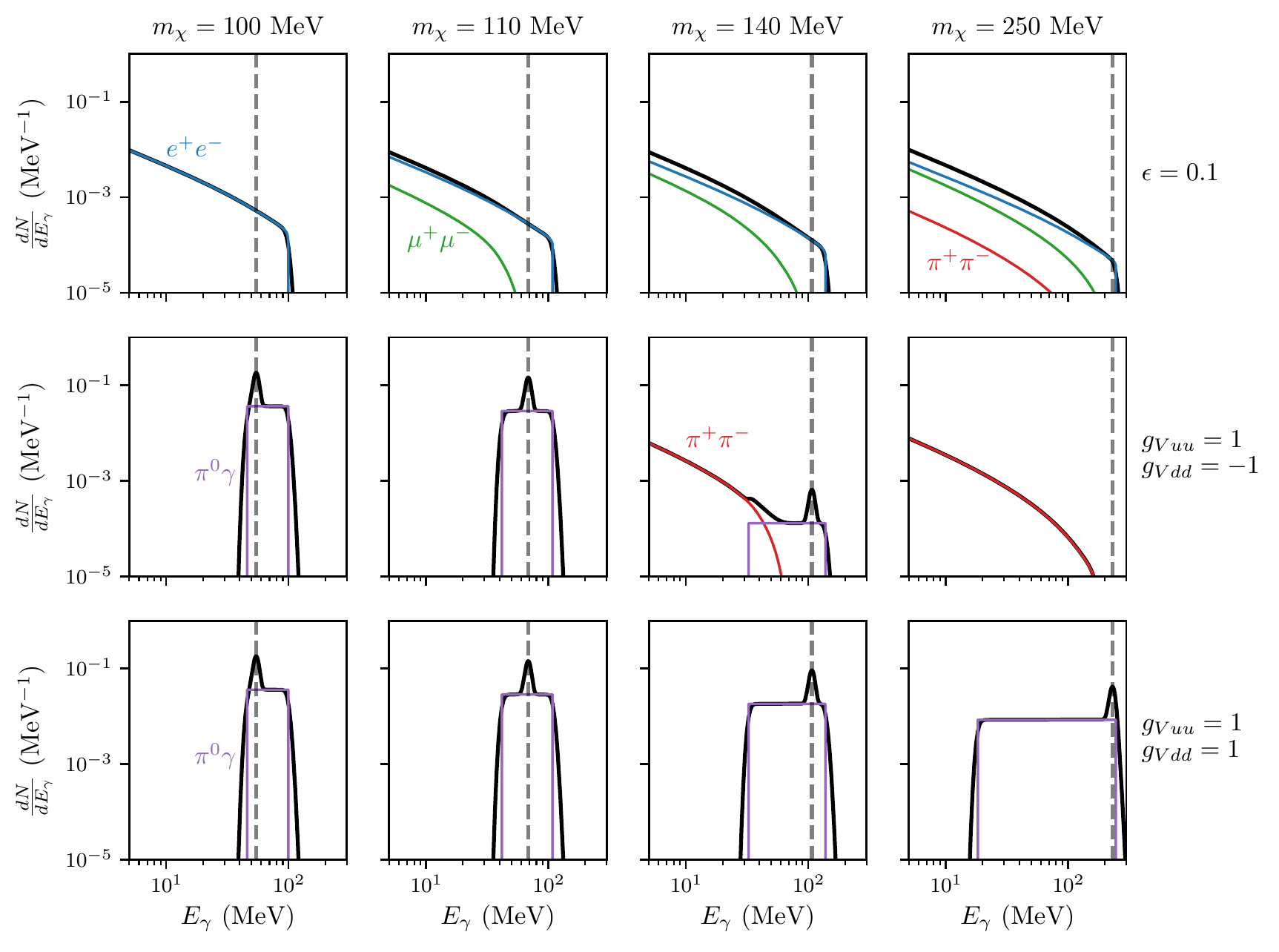}
    \caption{\textbf{Spectra from DM annihilation into Standard Model particles for vector-mediator models.} The two-mediator final state is taken to be kinematically forbidden. The black curve is the total spectrum as seen by an instrument with 5\% energy resolution. The rows correspond to the case where the mediator kinetically mixes with the photon and where it couples only to quarks. The vertical dashed line indicates the energy of the photon in the $\pi^0 \gamma$ final state.
    \nbLink{vector_spectra_ann_to_sm}
    \scriptLink{vector_spectra_ann_to_sm}}
    \label{fig:vector_spectra_ann_to_sm}
\end{figure}

\begin{figure}[tbh!]
    \centering
    \includegraphics[width=1\linewidth]{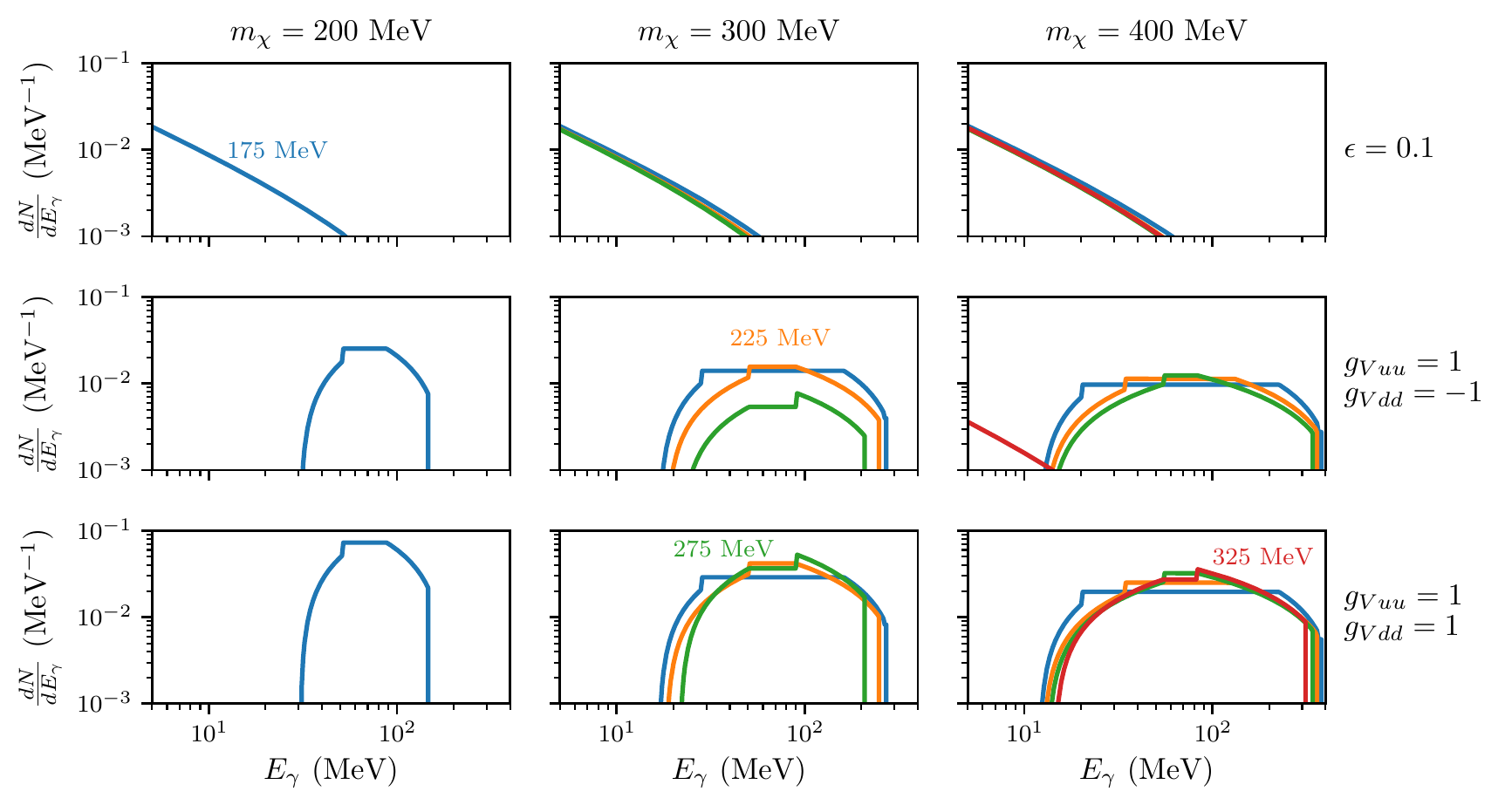}
    \caption{\textbf{Spectra from DM annihilation into mediators for vector-mediator models.} The labels and colors indicate the value of $m_V$ for each curve.
    \nbLink{vector_spectra_ann_to_med}
    \scriptLink{vector_spectra_ann_to_med}}
    \label{fig:vector_spectra_ann_to_med}
\end{figure}

Using the ingredients described in the previous section, we now compute the photon spectra from DM annihilation for the scalar and vector models with the same couplings as in Fig.~(\ref{fig:scalar_ann_bfs}) and Fig.~(\ref{fig:vector_ann_bfs}). In what follows, the center of mass energy is set to the non-relativistic approximation $E_{\mathrm{CM}} = 2 m_\chi \left( 1 + \frac{1}{2} v_{\mathrm{MW}}^2 \right)$, where $v_{\mathrm{MW}} = 10^{-3} c$ is the fiducial velocity dispersion for the Milky Way.

Fig.~(\ref{fig:scalar_spectra_ann_to_sm}) displays the spectrum for the scalar model when annihilation into the mediator is kinematically prohibited ($m_S \gtrsim m_\chi$). The dark matter mass is fixed for each column and the model parameters in each row. Individual final states' spectra are indicated by the colored curves. To visualize the di-photon final state's contribution, the black curve shows the total spectrum convolved with a 5\% energy resolution function (the case for COMPTEL~\cite{Kappadath:1993}), with a vertical dashed line highlighting the photons' energies ($E_{\mathrm{CM}} / 2$).

For low DM masses, the electron and di-photon line are responsible for the spectrum. Above the muon threshold the Yukawa-suppressed electron contribution is negligible. Since the branching fractions into different pion species are identical (up to isospin-breaking corrections proportional to $m_{\pi^\pm} - m_{\pi^0}$), the charged pion spectrum is always sub-dominant to the neutral pion box for $E_{\mathrm{CM}} > 2 m_\pi$. Depending on the ratio of the microscopic coupling to photons ($g_{SF}$) with the couplings to gluons and fermions ($g_{SG}$ and $g_{Sf}$), the neutral pion box or di-photon line are generally the most distinctive spectral features for large DM masses, which can also be seen in the figure. Note that if $g_{Sff}$ and $g_{SGG}$ are considered as independent parameters, they can take values such that the cross section for annihilation into pions is zero, though this scenario is very fine-tuned.

Spectra for annihilation into \emph{mediators} are shown in Fig.~(\ref{fig:scalar_spectra_ann_to_med}). The panels are labeled with the DM mass and the different colored curves are the spectra computed with the scalar's mass set to the indicated values. For different mediator masses, the following processes dictate the shape of the spectrum:
\begin{itemize}
    \item $m_S < 2 m_e$: the mediator can only decay to two photons, so the annihilation spectrum is a box centered at $m_S/2$, just like for the neutral pion.
    \item $2 m_e < m_S < 2 m_\mu$: the spectrum is a superposition of the box and the electron FSR spectra, and when the scalar can decay into two muons this dominates the spectrum.
    \item $m_S > 2 m_\pi$: the spectrum consists of a peak from boosting the $\pi^0$ decay spectrum and a softer component from FSR and decays from the other final states into which $S$ decays. The peak is more rounded if the scalar is highly boosted ($m_S \ll E_{\mathrm{CM}} / 2$) and boxy for $m_S \sim E_{\mathrm{CM}} / 2$. The difference in the case of heavy quark microscopic couplings is that the muon and electron final states are not present.
\end{itemize}

The components of the spectrum for DM annihilation into SM final states in the vector model are illustrated in Fig.~(\ref{fig:vector_spectra_ann_to_sm}). In the kinetic mixing case the spectrum consists purely of a continuum, since the branching fraction for annihilation into $\pi^0 \gamma$ is highly suppressed. In contrast, if the vector couples only to quarks the spectrum is composed of a monochromatic gamma ray line and box from the $\pi^0$ decay at low energies. If not forbidden by the vector's couplings (as is the case for the bottom row), as the center-of-mass energy grows above the two-pion threshold the continuum of photons from the $\pi^+ \pi^-$ final state overwhelm those from $\pi^0 \gamma$.

Fig.~(\ref{fig:vector_spectra_ann_to_med}) exhibits spectra from DM annihilating into $V V$ followed by the vectors' decays for a kinetically mixed vector (top row) and couplings to quarks only. As in the case of annihilations directly into SM particles, a power law spectrum is obtained in the kinetically mixed case. For the spectrum for quark-only couplings (middle and bottom rows) is sourced exclusively by the vector's decay to $\pi^0 \gamma$. The resulting spectrum is comprised of a box (from the boosting the monochromatic photon) and a rounded peak (from boosting the neutral pion's decay spectrum). \revised{The box component is offset towards higher energies than the rounded peak component, leading to the sawtooth pattern in the total spectrum.} This feature would only by observable at a detector with very good energy resolution.

The colored curves in the Fig.~(\ref{fig:scalar_spectra_ann_to_sm})-(\ref{fig:vector_spectra_ann_to_med}) can be reproduced using the \mil{Theory.spectra()} function described in Sec.~(\ref{sub:continuum_spectra_hazma}). The total spectrum convolved with an energy resolution function can be obtained using \mil{Theory.total_conv_spectrum_fn()}. The arguments specify the range of photon energies over which to perform the convolved spectrum, the center of mass energy of the DM annihilation, the energy resolution function, and the number of points to use when computing the convolution:
\begin{minted}{python}
>>> from hazma.scalar_mediator import HiggsPortal
>>> e_cm = 305.                  # DM center of mass energy
>>> e_min, e_max = 1., 100.      # define energy range
>>> energy_res = lambda e: 0.05  # 5% energy resolution function
>>> hp = HiggsPortal(mx=150., ms=1e3, gsxx=0.7, stheta=0.1)
>>> dnde_conv = hp.total_conv_spectrum_fn(
...     e_min, e_max, e_cm, energy_res, n_pts=1000)
\end{minted}
Decreasing \mil{n_pts} from its default value of 1000 increases speed at the cost of accuracy. The return value (\mil{dnde_conv} above) is an interpolator of type \mil{InterpolatedUnivariateSpline} from the \mil{scipy.interpolate} module. This can be used to compute the spectrum at a particular photon energy:
\begin{minted}{python}
>>> dnde_conv(25.)
array(0.00048767)
\end{minted}
This spectrum is can easily and efficiently be integrated over a range of photon energies between \mil{e_min} and \mil{e_max}:
\begin{minted}{python}
>>> dnde_conv.integral(25., 85.)
0.815998464406668
\end{minted}

\section{Electron and Positron Spectra from DM annihilation}%
\label{sec:dm_positron_spectra}

Dark matter annihilation in the mass range under consideration produces relativistic electrons and positrons. Such particles could be in principle directly detected as part of the cosmic radiation, albeit at energies well below the GeV solar modulation strongly affects cosmic-ray electron spectra. Additionally, relativistic electrons and positrons can be indirectly detected through (1) effects on photons of the cosmic microwave background and (2) the secondary emission of radiation via e.g. up-scattering of background photons (inverse Compton processes), synchrotron radiation, and bremsstrahlung. The calculation of the electron-positron spectra for MeV dark matter is therefore needed to reliably compute CMB limits and the spectrum of secondary radiation.

In this section we compute the positron spectrum for MeV dark matter, which follows along the lines of the gamma-ray ones.\footnote{The electron spectrum looks identical to the positron one since our dark matter particles do not carry lepton number.} The spectrum consists of a line-like spectrum from $\bar{\chi}\chi\to e^+ e^-$ and a continuum piece from the decays of unstable particles:
\begin{align}
    \label{eq:total_ann_pos_spec}
    \rvertchi{\frac{dN}{dE_{e^+}}}(E_{e^+}) & = \operatorname{Br}(\bar{\chi}\chi \to e^+ e^-) ~ \delta(E_{\text{CM}}/2 - E_{e^+}) + \left. \frac{dN}{dE_{e^+}} \right\rvert_{\bar{\chi}\chi,\text{dec}}(E_{e^+})
\end{align}
For the scalar and vector model, the decay piece receives contributions from the $\mu^+\mu^-$ and $\pi^+\pi^-$ final states.

The muon decays nearly $100\%$ of the time through $\mu^+\to e^+\nu_{e}\bar{\nu}_{\mu}$. In the muon's rest frame, the positron's energy spectrum is~\cite{ku} 

\begin{align}
    \label{eq:muon_positron_spectrum}
    \frac{dN}{dE_{e^+}} & = -\frac{4 \sqrt{x^2-4 r^2} \left(r^2 (4-3 x)+x (2 x-3)\right)}{m_{\mu}}
\end{align}
where $r = m_{e} / m_{\mu}$ and $x = 2E_{e^+} / m_\mu$. The spectrum can be boosted to other frames using Eqn.~(\ref{eq:spec_boost}). It is accessible in \mil{hazma} using:
\begin{minted}{python}
>>> from hazma.positron_spectra import muon as dnde_p_mu
>>> e_mu = 150.                      # muon energy
>>> e_p = np.array([1., 10., 100.])  # positron energies
>>> dnde_p_mu(e_p, e_mu)
array([4.86031362e-05, 4.56232320e-03, 4.45753994e-03])
\end{minted}

The charged pion primarily decays through $\pi^+ \to \mu^+ \nu_\mu$, which yields a positron when the muon decays. The energy spectrum is thus obtained by boosting the muon spectrum from Eqn.~(\ref{eq:muon_positron_spectrum}). An additional contribution comes from the monochromatic positron produced through $\pi^+ \to e^+ \nu_e$, which has a helicity suppressed branching fraction $\operatorname{Br}(\pi^+ \to e^+ \nu_e) = 1.23\ee{-4}$. The total spectrum can be computed in \mil{hazma} with
\begin{minted}{python}
>>> from hazma.positron_spectra import charged_pion as dnde_p_pi
>>> e_pi = 150.                      # charged pion energy
>>> e_p = np.array([1., 10., 100.])  # positron energies
>>> dnde_p_pi(e_p, e_pi)
array([3.84163631e-05, 3.85242442e-03, 2.55578895e-05])
\end{minted}

\begin{figure}[tbh!]
  \centering
  \includegraphics[width=\textwidth]{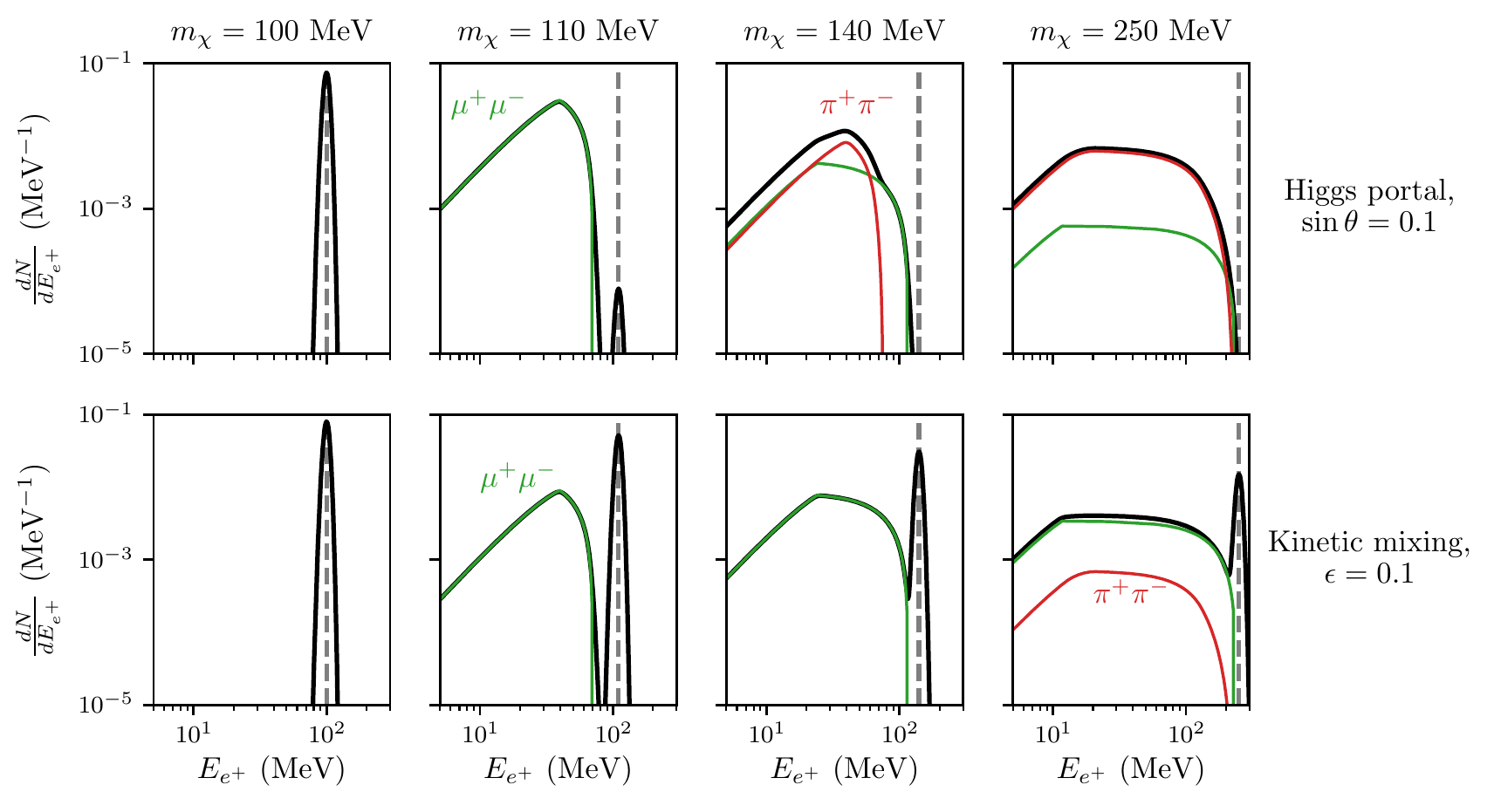}
  \caption{\textbf{Positron spectra for DM annihilating into Standard Model final states.} Results for the scalar model with Higgs portal couplings are shown in the top row and for the vector model with a kinetically-mixed mediator in the bottom. The grey vertical dashed line indicates the location of the monochromatic $\bar{\chi}\chi\to e^+e^-$ final state. The total spectrum (black curve) is convolved with a 5\% energy resolution function.
  \nbLink{positron_spectra_ann_to_sm}
  \scriptLink{positron_spectra_ann_to_sm}}
  \label{fig:positron_spectra_ann_to_sm}
\end{figure}

\begin{figure}[tbh!]
  \centering
  \includegraphics[width=\textwidth]{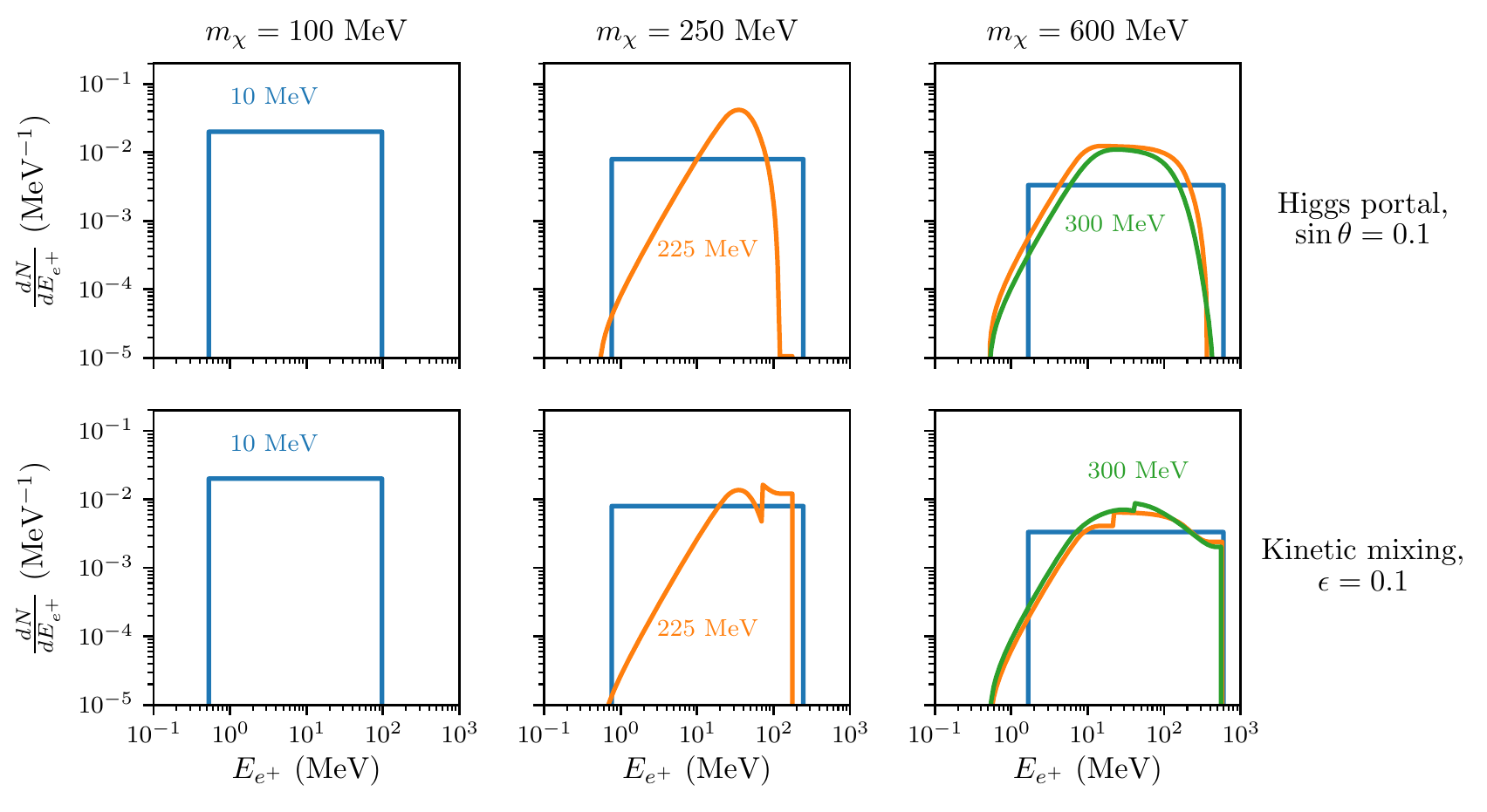}
  \caption{\textbf{Positron spectra for DM annihilating into two mediators.} \revised{The rows correspond to the scalar model Higgs portal couplings and vector model with kinetic mixing couplings. The different colored curves are labeled by the mediator masses. The spectra consist of a box component coming from the mediator decaying into $e^+ e^-$, which dominates at low energies, and a smooth component from mediator decays into muons and pions.}
  \nbLink{positron_spectra_ann_to_med}
  \scriptLink{positron_spectra_ann_to_med}}
  \label{fig:positron_spectra_ann_to_med}
\end{figure}

Fig.~(\ref{fig:positron_spectra_ann_to_sm}) exhibits representative positron annihilation spectra from the scalar model with \revised{Higgs portal couplings and vector model with kinetic mixing couplings (first and second rows)}, and several DM masses (different columns). Changing the couplings to prohibit annihilation into leptons has a straightforward impact on the total spectrum. Here annihilation into mediators is kinematically forbidden. Positron spectra for annihilation into mediators are shown for these models in Fig.~(\ref{fig:positron_spectra_ann_to_med}). When the mediator is light the spectrum is box-shaped since the mediator can only produce electrons by decaying into them directly. At higher energies the spectrum is a superposition of the rounded spectra from mediator decays into muons and pions and this box spectrum. \revised{For the Higgs portal model, the box is subdominant to the rounded part of the spectrum since the mediator decay width into $e^+ e^-$ is Yukawa-suppressed. The top edge of the box is barely visible for the orange curve near $200~\us{MeV}$ in the top middle panel. For the vector model, the box component of the spectrum is visible as a sawtooth edge in the lower middle and right panels.}

The interface for computing positron spectra mirrors the gamma-ray spectrum interface. The \mil{total_positron_spectrum} and \mil{positron_spectra} methods in \mil{Theory} give the total positron spectrum and individual channels' contributions (scaled by branching fraction). The \mil{positron_lines} method gives information about final states producing monochromatic positrons:
\begin{minted}{python}
>>> from hazma.scalar_mediator import HiggsPortal
>>> e_cm = 305.                       # DM center of mass energy
>>> e_ps = np.array([1., 10., 100.])  # positron energies
>>> hp = HiggsPortal(mx=150., ms=1e3, gsxx=0.7, stheta=0.1)
>>> hp.total_positron_spectrum(e_ps, e_cm)
array([2.60677406e-05, 2.58192085e-03, 4.09383294e-04])
>>> hp.positron_spectra(e_ps, e_cm)
{'mu mu': array([6.10588051e-06, 5.70145391e-04, 5.78482356e-04]),
 'pi pi': array([2.14034924e-05, 2.13725359e-03, 9.48371777e-05]),
 's s': array([0., 0., 0.]),
 'total': array([2.75093729e-05, 2.70739898e-03, 6.73319534e-04])}
>>> hp.positron_lines(e_cm)
{'e e': {'energy': 152.5, 'bf': 7.711433862697906e-06}}
\end{minted}
For the built-in models, functions computing the positron spectra for annihilations into particular final states follow a similar naming scheme to the gamma-ray ones, and can be called directly.

A method (more precisely, an \mil{InterpolatedUnivariateSpline}) to compute the total positron spectrum convolved with an energy resolution function is obtained using:
\begin{minted}{python}
>>> e_cm = 305.                  # DM center of mass energy
>>> e_p_min, e_p_max = 1., 100.  # define energy range
>>> energy_res = lambda e: 0.05
>>> dnde_p_conv = hp.total_conv_positron_spectrum_fn(
...     e_p_min, e_p_max, e_cm, energy_res)
>>> dnde_p_conv(20.)
0.00864851382906508
>>> dnde_p_conv.integral(10, 100)  # integrate spectrum
0.6538810882108401
\end{minted}

\section{Gamma Ray Limits}
\label{sec:gr_limits}

In \mil{hazma} we include two routines to set constraints on the DM self-annihilation cross section $\svann$ using current gamma-ray flux measurements and to make projections for planned detectors.

As is well-known, the primary gamma-ray flux from DM annihilating in a region of the sky subtending a solid angle $\Delta \Omega$ is
\begin{align}
    \rvertchi{\frac{d\Phi}{dE_\gamma}}(E) &= \frac{\Delta \Omega}{4\pi} \cdot \frac{\svann}{2 f_\chi m_\chi^2} \cdot J \cdot \rvertchi{\frac{dN}{dE_\gamma}}(E). \label{eq:gamma_ray_flux}
\end{align}
The factor $f_\chi$ is 1 if the DM is self-conjugate and 2 otherwise (2 for the models currently included with \mil{hazma}). The $J$ factor counts the number of pairs of DM particles in the observing region:
\begin{align}
    J \equiv \frac{1}{\Delta\Omega} \int_{\Delta\Omega} d\Omega\, \int_{\u{LOS}} dl\, \rho(r(l, \psi))^2,
\end{align}
where $\rho(r)$ is the DM density profile. The annihilation spectrum $dN/dE_\gamma|_{\bar{\chi}\chi}$ was computed in detail in the previous section. A detector with finite energy resolution observes the smoothed flux
\begin{align}
    \rvertchi{\frac{d\Phi_\epsilon}{dE_\gamma}}(E) &\equiv \int dE'\, R_\epsilon(E | E') \rvertchi{\frac{d\Phi}{dE_\gamma}}(E'),
\end{align}
where the spectral resolution function is the probability that a gamma ray with true energy $E'$ is reconstructed with energy $E$, and can be approximated as a Gaussian~\cite{Bringmann2009}:
\begin{align}
    R_\epsilon(E|E') &\equiv \frac{1}{\sqrt{2\pi}\, \epsilon(E')\, E'} \exp\left[ -\frac{(E - E')^2}{2 (\epsilon(E')\, E')^2} \right].
\end{align}

In the limit-setting procedure for \emph{existing} experiments, the integral of the smoothed spectrum in each energy bin is required not to exceed the observed value plus twice the upper error bar: $\Phi_\epsilon^{(i)} |_{\bar{\chi}\chi} < \Phi^{(i)}|_\u{obs} + 2 \sigma^{(i)}$. While better limits can be obtained by assuming a background model, this introduces significant systematic uncertainties since the model is derived from precisely the observations under consideration and would thus only marginally improve the limits on $\svann$~\cite{Essig:2013goa}.

For assessing the discovery reach by \emph{upcoming} experiments, \mil{hazma} uses an unbinned procedure. A detector can be characterized by an effective area $A_\u{eff}(E)$ and an (energy-independent) observing time $T_\u{obs}$. Over the energy range $[E_\u{min}, E_\u{max}]$, the number of photons observed is then
\begin{align}
    \rvertchi{N_\gamma} &= \int_{E_\u{min}}^{E_\u{max}} dE\, T_\u{obs} \, A_\u{eff}(E) \,  \rvertchi{\frac{d\Phi_\epsilon}{dE_\gamma}}(E).
\end{align}
Assuming Poisson statistics, the minimum-detectable $\svann$ is obtained by requiring the signal-to-noise ratio to be significant at the $n_\sigma = 5$ level: $N_\gamma|_{\bar{\chi}\chi} / \sqrt{N_\gamma|_\u{bg}} = n_\sigma$. Holding the DM mass fixed, the lowest-possible value of $\svann$ is found by optimizing over $E_\u{min}$ and $E_\u{max}$ to maximize $N_\gamma|_{\bar{\chi}\chi} / \sqrt{N_\gamma|_\u{bg}}$. This energy range does not completely align with spectra features. For example, in the case of the $\pi^0$ spectrum, setting $E_\u{min}$ equal to the left boundary of the ``box'' would include too much of the $d\Phi/dE_\gamma|_\u{bg} \propto E_\gamma^{-2}$ spectrum.

The unbinned analysis requires a background model, two of which are included with \mil{hazma}. The simplest was obtained by fitting a power law to COMPTEL and EGRET data from high galactic latitudes over the energy range $0.8\us{MeV} - 10\us{GeV}$~\cite{Boddy2015}. This is suitable for projecting limits on $\svann$ from dwarf galaxy observations. A model derived using \mil{GALPROP} to fit existing MeV gamma-ray data in the region $|b| < 10^\circ$, $|l| < 30^\circ$ is also included to assess the sensitivity to emission from the Galactic Center.

\begin{table}[tbh!]
    \centering
    \begin{tabular}{c c c c}
        \toprule
        Detector & Energy range & $A_\u{eff}$ (cm$^2$) & $\epsilon$\\
        \midrule
        COMPTEL~\cite{Kappadath:1993} & $0.7 - 27\us{MeV}$ & 50 & 5\%\\
        EGRET~\cite{Thompson:1993zz} & $27\us{MeV} - 8.6\us{GeV}$ & $10 - 10^3$ & 18\%\\
        Fermi-LAT~\cite{FermiInstrument} & $150\us{MeV} - 95\us{GeV}$ & $10^3 - 10^4$ & 7.5\%\\
        \midrule
        e-ASTROGAM~\cite{eASTROGAMWhitebook} & $0.3\us{MeV} - 3\us{GeV}$ & $10^2 - 10^3$ & $\substack{\lesssim 2\%,\ E_\gamma < 10\us{MeV} \\ 20-30\%,\ E_\gamma \geq 10\us{MeV}}$\\
        \bottomrule
    \end{tabular}
    \caption{\textbf{Figures of merit for the detectors used in \mil{hazma}.} Note that the effective area is only used in \mil{hazma} for projecting limits for e-ASTROGAM.}
    \label{tab:detector_figures_of_merit_data}
\end{table}

\begin{table}[tbh!]
    \centering
    \begin{tabular}{c c c c}
        \toprule
        Detector & Region & $\Delta\Omega$ (sr) & $J$ (MeV$^2$ cm$^{-5}$ sr$^{-1}$)\\
        \midrule
        COMPTEL~\cite{Kappadath:1993} & $|b| < 20^\circ,\ |l| < 60^\circ$ & 1.433 & $3.725\times10^{28}$\\
        EGRET~\cite{Strong:2004de} & $20^\circ < |b| < 60^\circ,\ |l| < 180^\circ$ & 6.585 & $3.79\times10^{27}$\\
        Fermi-LAT~\cite{Ackermann2012} & $8^\circ < |b| < 90^\circ,\ |l| < 180^\circ$ & 10.817 & $4.698\times10^{27}$\\
        \midrule
        e-ASTROGAM~\cite{eASTROGAMWhitebook} & $|b| < 10^\circ,\ |l| < 10^\circ$ & 0.121 & $1.795\times10^{29}$\\
        \bottomrule
    \end{tabular}
    \caption{\textbf{Target regions and data references (first column) for detectors included in \mil{hazma}.} The $J$-factors for COMPTEL, EGRET and Fermi-LAT are taken from~\cite{Essig2013} and the value for e-ASTROGAM is from ref.~\cite{PPPC}. All assume an NFW profile for the galactic DM distribution~\cite{PPPC}.}
    \label{tab:target_regions}
\end{table}

These two limit-setting procedures can currently be applied in \mil{hazma} to data from three existing experiments (COMPTEL, EGRET and Fermi-LAT) and one proposed detector (e-ASTROGAM). The energy ranges, approximate effective areas and energy resolutions for these are supplied in Tab.~(\ref{tab:detector_figures_of_merit_data}). Each gamma-ray data set is paired with a target region, defined by the galactic coordinate ranges, solid angles and $J$-factors (assuming an NFW distribution for the galactic DM halo) in Tab.~(\ref{tab:target_regions}). These are packaged into \mil{FluxMeasurements} objects in \mil{hazma}.

The limit-setting procedures are methods in the \mil{Theory} class. For example, the following snippet imports a \mil{FluxMeasurement} object containing information about EGRET observations and computes the $\svann$ limits for the Higgs portal model at three DM masses in $\u{cm}^3/\u{s}$:
\begin{minted}{python}
>>> from hazma.scalar_mediator import ScalarMediator
>>> from hazma.gamma_ray_parameters import egret_diffuse, TargetParams
>>> egret_diffuse.target  # target region information
TargetParams(J=3.79e+27, dOmega=6.584844306798711)
>>> sm = ScalarMediator(mx=150., ms=1e4, gsxx=1., gsff=0.1, gsGG=0.5,
...                     gsFF=0, lam=1e5)
>>> sm.binned_limit(egret_diffuse)
7.890021123433229e-27  # cm^3 / s
\end{minted}
It is important to note that \mil{hazma} only uses the model parameters ($m_S$, $g_{S\chi\chi}$, $g_{Sff}$, $g_{SGG}$, $g_{SFF}$ and $\Lambda$) to compute the \emph{shape} of the spectrum and treats $\svann$ as an independent parameter (c.f. \ref{eq:gamma_ray_flux}). To relate the model parameters to $\svann$, the user can employ the method \mil{Theory.annihilation_cross_sections()}. The \mil{FluxMeasurement} objects for COMPTEL and Fermi-LAT observations are named \mil{comptel_diffuse} and \mil{fermi_diffuse} and can be imported similarly. App.~(\ref{app:gamma_ray_limits}) describes how to create new flux measurement objects.

Computing unbinned limits requires specifying several quantities \revised{characterizing the behavior of a gamma-ray detector}:
\begin{itemize}
    \item An effective area function returning $A_\u{eff}(E)$ in $\u{cm}^2$;
    \item An energy resolution function returning $\epsilon(E')$ in \%;
    \item An observing time in seconds;
    \item An object of type \mil{TargetParams} containing the $J$-factor and $\Delta\Omega$ for the target region;
    \item A \mil{BackgroundModel} object, which has a method \mil{dPhi_dEdOmega()} returning the differential gamma-ray flux flux per solid angle for the background model, and an attribute \mil{e_range} specifying the range of $E_\gamma$ values for which the model is valid.
\end{itemize}
Using the same instance of \mil{scalar_mediator} and DM masses as above, we can compute the projected limits for e-ASTROGAM using the $10^\circ\times10^\circ$ window around the Galactic Center from Tab.~(\ref{tab:target_regions}) as the target:
\begin{minted}{python}
>>> from hazma.gamma_ray_parameters import A_eff_e_astrogam
>>> from hazma.gamma_ray_parameters import energy_res_e_astrogam
>>> from hazma.gamma_ray_parameters import T_obs_e_astrogam
>>> from hazma.gamma_ray_parameters import gc_target, gc_bg_model
>>> gc_target.J                               # target J-factor
1.795e+29
>>> gc_target.dOmega                          # target extent
0.12122929761504078
>>> sm.unbinned_limit(A_eff_e_astrogam,       # effective area
...                   energy_res_e_astrogam,  # energy resolution
...                   T_obs_e_astrogam,       # observing time
...                   gc_target,              # target region
...                   gc_bg_model)            # background model
5.6701062876845636e-30  # cm^3 / s
\end{minted}
Modifying these ingredients to study other planned gamma-ray detectors is described in App.~(\ref{app:gamma_ray_limits}.

We do not currently use secondary radiation to set limits on our models. Such limits are highly dependent on (for example) the choice of observational target, the magnetic field intensity and spatial structure~\cite{Colafrancesco:2005ji, Colafrancesco:2006he}; Ref.~(\cite{Bartels2017}) calculates the secondary emission for select MeV dark matter annihilation final states. Since \mil{hazma} can already compute the relevant input spectra for determining secondary spectra, we plan to incorporate this into future versions of the code.

\section{Cosmic Microwave Background limits}
\label{sec:cmb_limits}

Dark matter annihilations at recombination inject ionizing particles into the photon-baryon plasma. The resulting changes in the residual ionization fraction and baryon temperature modify the CMB temperature and polarization power spectra, particularly at small scales, which puts a constraint on $\svcmb$, the DM self-annihilation cross section at recombination~\cite{Chen:2003gz,Padmanabhan:2005es,Galli:2009zc,Slatyer:2009yq,PhysRevD.93.023527,planck2018}. The size of the effect of the changes depend on the energy per unit volume per unit time, which is related to the effective parameter
\begin{align}
    p_\u{ann} \equiv \feff \, \frac{\svcmb}{\mdm}.
\end{align}
The current 95\% upper limit on $p_\u{ann}$ from Planck is $3.5\ee{-31}\us{cm}^3\us{s}^{-1}\us{MeV}^{-1}$~\cite{planck2018}. $\feff$ is the fraction of energy per DM annihilation imparted to the plasma. This depends on how many electrons/positrons and photons are produced per DM annihilation (ie, the spectra computed earlier in this work) and how efficiently these particles re-ionize the CMB (quantified by the factors $f_\u{eff}^\gamma$ and $f_\u{eff}^{e^+ e^-}$)~\cite{PhysRevD.93.023527}:
\begin{align}
    \label{eq:f_eff}
    f_\u{eff}^\chi = \frac{1}{E_\u{CM}} \int_0^{E_\u{CM} / 2} dE\, \left( 2 f_\u{eff}^{e^+e^-} E \, \rvertchi{\frac{dN}{dE_{e^+}}} + f_\u{eff}^\gamma E \, \rvertchi{\frac{dN}{dE_\gamma}} \right).
\end{align}

If the DM self-annihilation cross section is velocity dependent (as is the case for the scalar mediator model, where the annihilation is $p$-wave), the DM velocity $v_{\chi,\u{CMB}}$ at recombination is required to relate $\svcmb$ to the present-day value of $\svann$ in indirect detection targets. This velocity is also required for computing the center of mass energy in DM self-annihilation events at recombination. The velocity depends on the kinetic decoupling temperature, $T_\u{KD}$. Before kinetic decoupling DM initially is coupled to the plasma so that $v_{\chi,\u{CMB}} = \sqrt{3 T_\gamma / \mdm}$, but afterwards it cools more quickly ($T_\chi \propto z^2$). The DM velocity is thus~\cite{Essig:2013goa}
\begin{align}
    v_{\chi,\u{CMB}} = \sqrt{\frac{3 T_\chi}{\mdm}} \approx 2\ee{-4} \left( \frac{T_\gamma}{1\us{eV}} \right) \left( \frac{1\us{MeV}}{\mdm} \right) \left( \frac{10^{-4}}{x_\u{kd}} \right)^{1/2},
\end{align}
where $x_\u{kd} \equiv T_\u{KD} / \mdm$ generally lies in the range $10^{-6} - 10^{-4}$. Since this velocity is much smaller than the characteristic velocity in the Milky Way, the CMB constraints on $\svann$ for models with $p$-wave suppression are in general much weaker than the gamma-ray ones. The CMB constraints in the $s$-wave case are generally much stronger, but can be evaded in models where the DM is partially produced through the decay of a heavier dark-sector degree of freedom after recombination~\cite{DEramo:2018khz}.

\begin{figure}
    \centering
    \includegraphics[width=\textwidth]{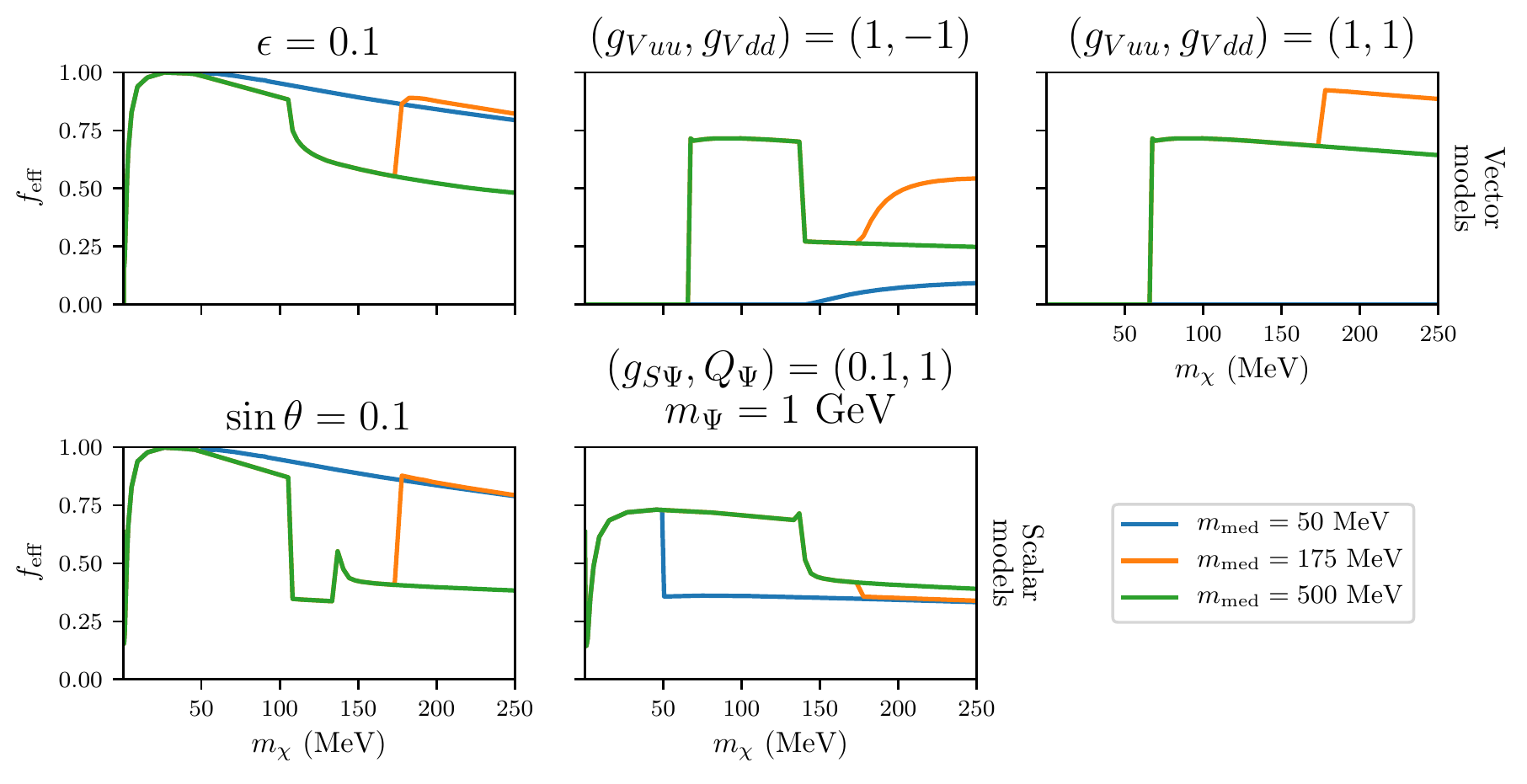}
    \caption{\textbf{Efficiency of energy injection into the CMB by dark matter annihilation.} The top panels are for the vector model with kinetic mixing and two possible values for mediator couplings to quarks only. The bottom row shows $\feff$ for the scalar mediator with Higgs portal and heavy quark-type couplings. The DM-mediator coupling is fixed to one in all panels. Note that the curves for different mediator masses have some overlap.
    \nbLink{f_eff}
    \scriptLink{f_eff}}
    \label{fig:f_eff}
\end{figure}

The quantity $\feff$ as well as the separate $e^\pm$ and $\gamma$ terms in Eqn.~(\ref{eq:f_eff}) can be computed for a given kinetic decoupling temperature:
\begin{minted}{python}
>>> from hazma.scalar_mediator import ScalarMediator
>>> sm = ScalarMediator(mx=150., ms=1e4, gsxx=1., gsff=0.1, gsGG=0.5,
...                     gsFF=0, lam=1e5)
>>> x_kd = 1e-4
>>> sm.f_eff(x_kd), sm.f_eff_ep(x_kd), sm.f_eff_g(x_kd)
(0.4364732027954761, 0.16674887821288106, 0.26972432458259504)
\end{minted}
Fig.~(\ref{fig:f_eff}) shows this factor as a function of DM and mediator mass for the scalar and vector models with the couplings we have been using as benchmarks. There are sharp variations in $\feff$ arising from SM or two-mediator final states becoming accessible as the DM mass changes. Deriving the kinetic decoupling temperature for particular models rather than leaving it as a free parameter is left for future work (or the user).

The \mil{Theory.cmb_limit()} function computes the constraint on $\svcmb$ given a value of $x_\u{kd}$ and upper limit on $p_\u{ann}$:
\begin{minted}{python}
>>> from hazma.cmb import p_ann_planck_temp_pol as p_ann
>>> p_ann
3.5e-31  # cm^3 / MeV / s
>>> x_kd = 1e-4
>>> sm = HiggsPortal(mx=150., ms=1e4, gsxx=1., stheta=0.01)
>>> sm.cmb_limit(x_kd, p_ann)
1.2398330863024796e-28  # cm^3 / s
\end{minted}
As with the gamma-ray limits described in the previous section, the parameters of the theory (here $m_S$, $g_{S\chi\chi}$ and $s_\theta$) determine the \emph{shapes} of the photon and $e^\pm$ spectra, and \mil{Theory.cmb_limit()} treats $\svcmb$ as a free parameter. The user can find which model parameters reproduce this cross section using \mil{Theory.annihilation_cross_sections()}.

\section{Conclusion}
\label{sec:conclusion}

We have introduced \mil{hazma}, a code for the calculation of gamma-ray and electron-positron spectra as well as constraints and projected reach for future telescopes for sub-GeV dark matter models. \mil{hazma} is timely both because of forthcoming new observational capabilities at sub-GeV gamma-ray energies, and because of renewed interest in dark matter models beyond the WIMP paradigm and at lower-than-usually-considered masses. In future work we will use \mil{hazma} to study the complementarity between indirect and other probes of sub-GeV dark matter.

\mil{hazma} allows users to compute both spectra of gamma rays, electrons and positrons for individual annihilation final states containing arbitrarily many particles, as well as for user-implemented or a set of built-in dark matter models. In the latter case, the code computes the relevant branching fractions and the ensuing decay chains leading to photons, electrons and positrons, and calculates model-dependent emission processes such as e.g. internal bremsstrahlung. The interactions between the mediator particles in these models and the final state mesons is implemented using chiral perturbation theory techniques.

\mil{hazma} contains several updates and refinements over how gamma-ray spectra have been calculated in the recent literature. These include several radiative decay spectra, such as from charged pion decay or from muon decay, final state radiation calculations going beyond the Altarelli-Parisi splitting function approximation for the built-in models, and the ability to include arbitrarily many mesons and leptons in the final states.

We presented numerous snippets of code, and all code used to produce the figures shown herein manuscript is available through clickable links. We also showed a few advanced usage features, such as adding new gamma-ray experiments for deriving constraints from existing data or assessing projected reach for planned detectors, and implementing user-defined models.

\acknowledgments
LM and SP are partly supported by the U.S.\ Department of Energy grant number de-sc0010107. We acknowledge with gratitude early collaboration on this project with Francesco D'Eramo. Francesco D'Eramo was the first to put forward the idea of using chiral perturbation theory in the context of accurately calculating the annihilation and decay products of MeV dark matter. He was also responsible for several of the early calculations, part of the writing of the text, and for overall educating us all on numerous topics in chiral perturbation theory. We thank Michael Peskin for important input and feedback, and Regina Caputo for discussions about MeV gamma-ray observatories. \revised{We thank Graham White as well as Nicholas Rodd for pointing out typos in an earlier version of the manuscript.} Finally, we thank Eulogio Oset and Jose Antonio Oller for clarifying some points about unitarized chiral perturbation theory.

\appendix

\section{Installation}
\label{sec:installation}

\mil{hazma} was developed for python3. Before installing \mil{hazma}, the user needs to install several well-established python packages: \mil{cython}, \mil{scipy}, \mil{numpy}, and \mil{scikit-image}. Theses are easily installed by using PyPI. If the user has PyPI installed on their system, then these packages can be installed using
\begin{minted}{bash}
pip install cython, scipy, numpy, scikit-image, matplotlib
\end{minted}
\mil{hazma} can be installed in the same way, using:
\begin{minted}{bash}
pip install hazma
\end{minted}
This will download a tarball from the PyPI repository, compile all the c-code and install \mil{hazma} on the system. Alternatively, the user can install \mil{hazma} by downloading the package from \url{https://github.com/LoganAMorrison/Hazma.git}. Once downloaded, navigate to the package directory using the command line and run either
\begin{minted}{bash}
pip install .
\end{minted}
or
\begin{minted}{bash}
python setup.py install
\end{minted}

Note that since \mil{hazma} makes extensive usage of the package \mil{cython}, the user will need to have a \mil{c} and \mil{c++} compiler installed on their system (for example \mil{gcc} and \mil{g++} on unix-like systems or Microsoft Visual Studios 2015 or later on Windows). For more information, see the \mil{cython} installation guide: \url{https://cython.readthedocs.io/en/latest/src/quickstart/install.html}

\section{Basic Usage}
\label{sec:basic_usage}

\subsection{Using the Built-In Models}
\label{sub:basic_usage_simp_models}

Throughout the text, we gave usage examples of how to use the models built into \mil{hazma}. Here we give a compact review of how to use these models in \mil{hazma}. All the models built into \mil{hazma} have identical interfaces. The only difference in the interfaces is the parameters which need to be specified for the particular model being used. Thus, we will only show the usage of one of the models. The others can be used in an identical fashion. The example we will use is the \mil{KineticMixing} model. To create a \mil{KineticMixing} model, we use:
\begin{minted}{python}
# Import the model
>>> from hazma.vector_mediator import KineticMixing
# Specify the parameters
>>> params = {'mx': 250.0, 'mv': 1e6, 'gvxx': 1.0, 'eps': 1e-3}
# Create KineticMixing object
>>> km = KineticMixing(**params)
\end{minted}
Here we have created a model with a dark matter fermion of mass $250$ MeV, a vector mediator which mixes with the standard model with a mass of $1$ TeV. We set the coupling of the vector mediator to the dark matter, $g_{V\chi} = 1$ and set the kinetic mixing parameter $\epsilon = 10^{-3}$. To list all of the available final states for which the dark matter can annihilate into, we use:
\begin{minted}{python}
>>> km.list_annihilation_final_states()
['mu mu', 'e e', 'pi pi', 'pi0 g', 'pi0 v', 'v v']
\end{minted}
This tells us that we can potentially annihilate through: $\chi\bar{\chi}\to\mu^{+}\mu^{-},e^{+}e^{-},\pi^{+}\pi^{-},\pi^{0}\gamma, \pi^0 V$ or $V^{\mu}V^{\mu}$. However, which of these final states is actually available depends on the center of mass energy. We can see this fact by looking at the annihilation cross sections or branching fractions, which can be computed using:
\begin{minted}{python}
>>> cme = 2.0 * km.mx * (1.0 + 0.5 * 1e-6)
>>> km.annihilation_cross_sections(cme)
{'mu mu': 8.998735387276086e-25,
 'e e': 9.115024836416874e-25,
 'pi pi': 1.3013263970304e-25,
 'pi0 g': 1.7451483984993156e-29,
 'pi0 v': 0.0,
 'v v': 0.0,
 'total': 1.941526113556321e-24}
>>> km.annihilation_branching_fractions(cme)
{'mu mu': 0.4634877339245762,
 'e e': 0.46947732367713324,
 'pi pi': 0.06702595385888176,
 'pi0 g': 8.988539408840104e-06,
 'pi0 v': 0.0,
 'v v': 0.0}
\end{minted}
Here we have chosen a realistic center of mass energy for dark matter in our galaxy, which as a velocity dispersion of $\delta v\sim 10^{-3}$. We can that the $V^{\mu}V^{\mu}$ final state is unavailable, as it should be since the vector mediator mass is too heavy. In this theory, the vector mediator can decay. If we would like to know the decay width and the partial widths, we can use:
\begin{minted}{python}
>>> km.partial_widths()
{'pi pi': 0.0006080948890354345,
 'pi0 g': 0.23374917012731816,
 'x x': 26525.823848648604,
 'e e': 0.0024323798404358825,
 'mu mu': 0.0024323798404358803,
 'total': 26526.0630706733}
\end{minted}
If we would like to know the gamma-ray spectrum from dark matter annihilations, we can use 
\begin{minted}{python}
# Specify photon energies to compute spectrum at
>>> photon_energies = np.array([cme / 4])
>>> km.spectra(photon_energies, cme)
{'mu mu': array([2.8227189e-05]),
 'e e': array([0.00013172]),
 'pi pi': array([2.1464018e-06]),
 'pi0 g': array([7.66452627e-08]),
 'pi0 v': array([0.]),
 'v v': array([0.]),
 'total': array([0.00016217])}
\end{minted}
Note that we only used a single photon energy because of display purposes, but in general the user can specify any number of photon energies. If the user would like access to the underlying spectrum functions, they can use:
\begin{minted}{python}
>>> spec_funs = km.spectrum_functions()
>>> spec_funs['mu mu'](photon_energies, cme)
array([6.09016958e-05])
>>> mumu_bf = km.annihilation_branching_fractions(cme)['mu mu']
>>> mumu_bf * spec_funs['mu mu'](photon_energies, cme)
array([2.8227189e-05])
\end{minted}
Notice that the direct call to the spectrum function for $\chi\bar{\chi}\to\mu^{+}\mu^{-}$ doesn't given the same result as \mil{km.spectra(photon_energies, cme)['mu mu']}. This is because the branching fractions are not applied for the \mil{spec_funs = km.spectrum_functions()}. If the user doesn't care about the underlying components of the gamma-ray spectra, the can simply call
\begin{minted}{python}
>>> km.total_spectrum(photon_energies, cme)
array([0.00016217])
\end{minted}
to get the total gamma-ray spectrum. The reader may have caught the fact that there is a gamma-ray line in the spectrum for $\chi\bar{\chi}\to\pi^{0}\gamma$. To get the location of this line, the user can use:
\begin{minted}{python}
>>> km.gamma_ray_lines(cme)
{'pi0 g': {'energy': 231.78145156177675, 'bf': 8.988539408840104e-06}}
\end{minted}
This tells us the process which produces the line, the location of the line and the branching fraction for the process. We do not include the line in the total spectrum since the line produces a Dirac-delta function. In order to get a realistic spectrum including the line, we need to convolve the gamma-ray spectrum with an energy resolution. This can be achieved using:
\begin{minted}{python}
>>> min_photon_energy = 1e-3
>>> max_photon_energy = cme
>>> energy_resolution = lambda photon_energy : 1.0
>>> number_points = 1000
>>> spec = km.total_conv_spectrum_fn(min_photon_energy, max_photon_energy, 
...                                  cme, energy_resolution, number_points)
>>> spec(cme / 4)  # compute the spectrum at a photon energy of `cme/4`
array(0.00167605)
\end{minted}
The \mil{km.total_conv_spectrum_fn} computes and returns an interpolating function of the convolved function. An important thing to note here is that the \mil{km.total_conv_spectrum_fn} takes in a \emph{function} for the energy resolution. This allows the user to define the energy resolution to depend on the specific photon energy, which is often the case for gamma-ray telescopes.

Next we present the positron spectra. These have an identical interface to the gamma-ray spectra, so we only show how to call the functions and we suppress the output
\begin{minted}{python}
>>> from hazma.parameters import electron_mass as me
>>> positron_energies = np.logspace(np.log10(me), np.log10(cme), num=100)
>>> km.positron_spectra(positron_energies, cme)
>>> km.positron_lines(cme)
>>> km.total_positron_spectrum(positron_energies, cme)
>>> dnde_pos = km.total_conv_positron_spectrum_fn(min(positron_energies),
...                                           max(positron_energies),
...                                           cme, 
...                                           energy_resolution,
...                                           number_points)
\end{minted}
The last thing that we would like to demonstrate is how to compute limits. In order to compute the limits on the annihilation cross section of a model from a gamma-ray telescope, say EGRET, we can use:
\begin{minted}{python}
>>> from hazma.gamma_ray_parameters import egret_diffuse
# Choose DM masses from half the electron mass to 250 MeV
>>> mxs = np.linspace(me/2., 250., num=100)
# Compute limits from e-ASTROGAM
>>> limits = np.zeros(len(mxs), dtype=float)
>>> for i, mx in enumerate(mxs):
...     km.mx = mx
...     limits[i] = km.binned_limit(egret_diffuse)
\end{minted}
Similarly, if we would like to set constraints using e-ASTROGAM, one can use:
\begin{minted}{python}
>>> # Import target and background model for the e-ASTROGAM telescope
>>> from hazma.gamma_ray_parameters import gc_target, gc_bg_model
>>> # Import telescope parameters
>>> from hazma.gamma_ray_parameters import A_eff_e_astrogam
>>> from hazma.gamma_ray_parameters import energy_res_e_astrogam
>>> from hazma.gamma_ray_parameters import T_obs_e_astrogam
>>> # Choose DM masses from half the electron mass to 250 MeV
>>> mxs = np.linspace(me/2., 250., num=100)
>>> # Compute limits
>>> limits = np.zeros(len(mxs), dtype=float)
>>> for i, mx in enumerate(mxs):
...     km.mx = mx
...     limits[i] = km.unbinned_limit(A_eff_e_astrogam,
...                                   energy_res_e_astrogam,
...                                   T_obs_e_astrogam,
...                                   gc_target,
...                                   gc_bg_model)
\end{minted}

\subsection{Subclassing the Built-In Models}
\label{sub:basic_usage_sub_class}

As mentioned in Sec.~(\ref{sec:structure_workflow}), the user may not be interested in the generic models built into \mil{hazma}, but instead a more specialized model. In this case, it makes sense for the user to subclass one of the models (i.e. create a class which inherits from one of the models.) As and example, we illustrate how to do this with the Higgs-portal model (of course this model is already built into \mil{hazma}, but it works nicely as an example.) Recall that the full set of parameters for the scalar mediator model are:
\begin{enumerate}
    \item $m_{\chi}$: dark matter mass,
    \item $m_{S}$: scalar mediator mass,
    \item $g_{S\chi}$: coupling of scalar mediator to dark matter,
    \item $g_{Sf}$: coupling of scalar mediator to standard model fermions,
    \item $g_{SG}$: effective coupling of scalar mediator to gluons,
    \item $g_{SF}$: effective coupling of scalar mediator to photons and
    \item $\Lambda$: cut-off scale for the effective interactions.
\end{enumerate}
In the case of the Higgs-portal model, the scalar mediator talks to the standard model only through the Higgs boson, i.e. it mixes with the Higgs. Therefore, the scalar mediator inherits its interactions with the standard model fermions, gluons and photon through the Higgs. In Sec.~(\ref{sec:models_scalar_mediator}), we gave the relationships between the couplings $g_{Sf}, g_{SG}, g_{SF}, \Lambda$ and $\sin\theta, v_{h}$, where $\sin\theta$ is the mixing angle for the scalar and Higgs and $v_{h}$ is the Higgs vacuum expectation value. These relationships are:
\begin{align}
    g_{Sf} &= \sin\theta, & g_{SG} &= 3\sin\theta, & g_{SF}& = -\frac{5}{6}\sin\theta, & \Lambda &= v_{h}.
\end{align}
Therefore, the Higgs-portal model doesn't need to know about all of the couplings of the generic scalar mediator model. It only needs to know about $m_{\chi}, m_{S}, g_{S\chi}$ and $\sin\theta$. Below, we construct a class which subclasses the scalar mediator class to implement the Higgs-portal model.
\begin{minted}{python}
from hazma.scalar_mediator import ScalarMediator
from hazma.parameters import vh

class HiggsPortal(ScalarMediator):
    def __init__(self, mx, ms, gsxx, stheta):
        self._lam = vh
        self._stheta = stheta
        super(HiggsPortal, self).__init__(mx, ms, gsxx, stheta, 3.*stheta,
                                          -5.*stheta/6., vh)

    @property
    def stheta(self):
        return self._stheta

    @stheta.setter
    def stheta(self, stheta):
        self._stheta = stheta
        self.gsff = stheta
        self.gsGG = 3. * stheta
        self.gsFF = - 5. * stheta / 6.

    # Hide underlying properties' setters
    @ScalarMediator.gsff.setter
    def gsff(self, gsff):
        raise AttributeError("Cannot set gsff")

    @ScalarMediator.gsGG.setter
    def gsGG(self, gsGG):
        raise AttributeError("Cannot set gsGG")

    @ScalarMediator.gsFF.setter
    def gsFF(self, gsFF):
        raise AttributeError("Cannot set gsFF")
\end{minted}
There are a couple things to note about our above implementation. First, our model only takes in $m_{\chi}, m_{S}, g_{S\chi}$ and $\sin\theta$, as desired. But the underlying model, i.e. the \mil{ScalarMediator} model only knows about $m_{\chi}, m_{S}, g_{S\chi}, g_{Sf}, g_{SG}, g_{SF}$ and $\Lambda$. So if we update $\sin\theta$, we additionally need to update the underlying parameters, $g_{Sf}, g_{SG}, g_{SF}$ and $\Lambda$. The easiest way to do this is using getters and setters by defining $\sin\theta$ to be a \mil{property} through the \mil{@property} decorator. Then every time we update $\sin\theta$, we can also update the underlying parameters. The second thing to note is that we want to make sure we don't accidentally change the underlying parameters directly, since in this model, they are only defined through $\sin\theta$. We an ensure that we cannot change the underlying parameters directly by overriding the getters and setters for \mil{gsff}, \mil{gsGG} and \mil{gsGG} and raising an error if we try to change them. This isn't strictly necessary (as long as the user is careful), but can help avoid confusing behavior. 

\section{Advanced Usage}
\label{sec:usage}

\subsection{Adding New Gamma-Ray Experiments}
\label{app:gamma_ray_limits}

Currently \mil{hazma} only includes information for producing projected unbinned limits with e-ASTROGAM, using the dwarf Draco or inner $10^\circ\times10^\circ$ region of the Milky Way as a target. Adding new detectors and target regions is straightforward. As described in Sec.~(\ref{sec:gr_limits}), a detector is characterized by the effective area $A_\u{eff}(E)$, the energy resolution $\epsilon(E)$ and observation time $T_\u{obs}$. In \mil{hazma}, the first two can be any callables (functions) and the third must be a float. The region of interest is defined by a \mil{TargetParams} object, which can be instantiated with:
\begin{minted}{python}
>>> from hazma.gamma_ray_parameters import TargetParams
>>> tp = TargetParams(J=1e29, dOmega=0.1)
\end{minted}
The background model should be packaged in an object of type \mil{BackgroundModel}. This light-weight class has a function \mil{dPhi_dEdOmega()} for computing the differential photon flux per solid angle (in $\u{MeV}^{-1}\us{sr}$) and an attribute \mil{e_range} specifying the energy range over which the model is valid (in MeV). New background models are defined by passing these two the initializer:
\begin{minted}{python}
>>> from hazma.background_model import BackgroundModel
>>> bg = BackgroundModel(e_range=[0.5, 1e4],
...                      dPhi_dEdOmega=lambda e: 2.7e-3 / e**2)
\end{minted}

As explained in Sec.~(\ref{sec:gr_limits}), gamma-ray observation information from Fermi-LAT, EGRET and COMPTEL is included with \mil{hazma}, and other observations can be added using the container class \mil{FluxMeasurement}. The initializer requires:
\begin{itemize}
    \item The name of a CSV file containing gamma-ray observations. The file's columns must contain:
    \begin{enumerate}
        \item Lower bin edge (MeV)
        \item Upper bin edge (MeV)
        \item $E^n d^2\Phi / dE\, d\Omega$ (in $\u{MeV}^{n-1} \us{cm}^{-2} \us{s}^{-1} \us{sr}^{-1}$)
        \item Upper error bar (in $\u{MeV}^{n-1} \us{cm}^{-2} \us{s}^{-1} \us{sr}^{-1}$)
        \item Lower error bar (in $\u{MeV}^{n-1} \us{cm}^{-2} \us{s}^{-1} \us{sr}^{-1}$)
    \end{enumerate}
    Note that the error bar values are their $y$-coordinates, not their relative distances from the central flux.
    \item The detector's energy resolution function.
    \item A \mil{TargetParams} object for the target region.
\end{itemize}
For example, a CSV file \mil{obs.csv} containing observations
\begin{minted}{text}
150.0,275.0,0.0040,0.0043,0.0038
650.0,900.0,0.0035,0.0043,0.003
\end{minted}
with $n=2$ for an instrument with energy resolution $\epsilon(E) = 0.05$ observing the target region \mil{tp} defined above can be loaded using\footnote{If the CSV containing the observations uses a different power of $E$ than $n=2$, this can be specified using the \mil{power} keyword argument to the initializer for \mil{FluxMeasurement}.}
\begin{minted}{python}
>>> from hazma.flux_measurement import FluxMeasurement
>>> obs = FluxMeasurement("obs.dat", lambda e: 0.05, tp)
\end{minted}
The attributes of the \mil{FluxMeasurement} store all of the provide information, with the $E^n$ prefactor removed from the flux and error bars, and the errors converted from the positions of the error bars to their sizes. These are used internally by the \mil{Theory.binned_limit()} method, and can be accessed as follows:
\begin{minted}{python}
>>> obs.e_lows, obs.e_highs
(array([150., 650.]), array([275., 900.]))
>>> obs.target
<hazma.gamma_ray_parameters.TargetParams at 0x1c1bbbafd0>
>>> obs.fluxes
array([8.85813149e-08, 5.82726327e-09])
>>> obs.upper_errors
array([6.64359862e-09, 1.33194589e-09])
>>> obs.lower_errors
array([4.42906574e-09, 8.32466181e-10])
>>> obs.energy_res(10.)
0.05
\end{minted}

\subsection{User-Defined Models}
\label{app:user_defined_models}

In this subsection, we demonstrate how to implement new models in Hazma. A notebook containing all the code in this appendix can be downloaded from GitHub \href{https://github.com/LoganAMorrison/Hazma/blob/b525ac3482ce355f4109c9ad6622470c14d0931f/notebooks/hazma_paper/hazma_example.ipynb}{\faBook}. The model we will consider is an effective field theory with a Dirac fermion DM particle which talks to neutral and charged pions through gauge-invariant dimension-5 operators. The Lagrangian for this model is:
\begin{align}
    \mathcal{L} &\supset \dfrac{c_1}{\Lambda}\overline{\chi}\chi\pi^{+}\pi^{-} + \dfrac{c_2}{\Lambda}\overline{\chi}\chi\pi^{0}\pi^{0}
\end{align}
where \(c_{1}, c_{2}\) are dimensionless Wilson coefficients and \(\Lambda\) is the cut-off scale of the theory. In order to implement this model in Hazma, we need to compute the annihilation cross sections and the FSR spectra. The annihilation channels for this model are simply \(\bar{\chi}\chi\to\pi^{0}\pi^{0}\) and \(\bar{\chi}\chi\to\pi^{+}\pi^{-}\). The computations for the cross sections are straight forward and yield:
\begin{align}
    \sigma(\bar{\chi}\chi\to\pi^{+}\pi^{-}) &= \frac{c_1^2 \sqrt{1-4 \mu _{\pi }^2} \sqrt{1-4 \mu _{\chi }^2}}{32 \pi \Lambda^2}\\
    \sigma(\bar{\chi}\chi\to\pi^{0}\pi^{0}) &= \frac{c_2^2 \sqrt{1-4 \mu_{\pi^{0}}^2} \sqrt{1-4 \mu_{\chi}^2}}{8 \pi \Lambda^2}
\end{align}
where \(Q\) is the center of mass energy, \(\mu_{\chi} = m_{\chi}/Q\), \(\mu_{\pi} = m_{\pi^{\pm}}/Q\) and \(\mu_{\pi^{0}} = m_{\pi^{0}}/Q\). In addition to the cross sections, we need the FSR spectrum for \(\overline{\chi}\chi\to\pi^{+}\pi^{-}\gamma\). This is:
\begin{align}
    \dfrac{dN(\bar{\chi}\chi\to\pi^{+}\pi^{-}\gamma)}{dE_{\gamma}} &= \frac{\alpha  \left(2 f(x)-2\left(1-x-2 \mu_{\pi} ^2\right)
   \log \left(\frac{1-x-f(x)}{1-x+f(x)}\right)\right)}{\pi  \sqrt{1-4 \mu_{\pi} ^2} x}
\end{align}
where 
\begin{align}
    f(x) &= \sqrt{1-x} \sqrt{1-x-4 \mu_{\pi} ^2}
\end{align}
We are now ready to set up the Hazma model. For \mil{hazma} to work properly, we will need to define the following functions in our model:
\begin{itemize}

    \item \mil{annihilation_cross_section_funcs()}: A function returning a dictionary of the annihilation cross sections functions, each of which take a center of mass energy.
    \item \mil{spectrum_funcs()}: A function returning a dictionary of functions which take photon energies and a center of mass energy and return the gamma-ray spectrum contribution from each final state.
    \item \mil{gamma_ray_lines(e_cm)}: A function returning a dictionary of the gamma-ray lines for a given center of mass energy.
    \item \mil{positron_spectrum_funcs()}: Like \mil{spectrum_funcs()}, but for positron spectra.
    \item \mil{positron_lines(e_cm)}: A function returning a dictionary of the electron/positron lines for a center of mass energy.
\end{itemize}
In the interests of compartmentalization, we find it easiest place each of these components in its own class (a mixin in python terminology) which are then combined into a master class representing our model. Before we begin writing the classes, we need to import a few helper functions and constants from \mil{hazma}:
\begin{minted}{python}
import numpy as np # NumPy is heavily used
import matplotlib.pyplot as plt # Plotting utilities
# neutral and charged pion masses
from hazma.parameters import neutral_pion_mass as mpi0
from hazma.parameters import charged_pion_mass as mpi
from hazma.parameters import qe # Electric charge
# Positron spectra for neutral and charged pions
from hazma.positron_spectra import charged_pion as pspec_charged_pion
# Deay spectra for neutral and charged pions
from hazma.decay import neutral_pion, charged_pion
# The `Theory` class which we will ultimately inherit from
from hazma.theory import Theory
\end{minted}
Now, we implement a cross section class:
\begin{minted}{python}
class HazmaExampleCrossSection:
    def sigma_xx_to_pipi(self, Q):
        mupi = mpi / Q
        mux = self.mx / Q

        if Q > 2 * self.mx and Q > 2 * mpi:
            sigma = (self.c1**2 * np.sqrt(1 - 4 * mupi**2) * 
                     np.sqrt(1 - 4 * mux**2)**2 / 
                     (32.0 * self.lam**2 * np.pi))
        else:
            sigma = 0.0

        return sigma

    def sigma_xx_to_pi0pi0(self, Q):
        mupi0 = mpi0 / Q
        mux = self.mx / Q

        if Q > 2 * self.mx and Q > 2 * mpi0:
            sigma = (self.c2**2 * np.sqrt(1 - 4 * mux**2) *
                     np.sqrt(1 - 4 * mupi0**2) /
                     (8.0 * self.lam**2 * np.pi))
        else:
            sigma = 0.0

        return sigma

    def annihilation_cross_section_funcs(self):
        return {'pi0 pi0': self.sigma_xx_to_pi0pi0,
                'pi pi': self.sigma_xx_to_pipi}
\end{minted}
The key function is \mil{annihilation_cross_section_funcs}, which is \textbf{required} to be implemented. Next, we implement the spectrum functions which will produce the FSR and decay spectra:
\begin{minted}{python}
class HazmaExampleSpectra:
    def dnde_pi0pi0(self, e_gams, e_cm):
        # Decay spectra for the neutral pions
        return 2.0 * neutral_pion(e_gams, e_cm / 2.0)
    
    def __dnde_xx_to_pipig(self, e_gam, Q):
        # Unvectorized function for computing FSR spectrum
        mupi = mpi / Q
        mux = self.mx / Q
        x = 2.0 * e_gam / Q
        if 0.0 < x and x < 1. - 4. * mupi**2:
            dnde = (
                qe ** 2
                * (
                    2 * np.sqrt(1 - x) * np.sqrt(1 - 4 * mupi ** 2 - x)
                    + (-1 + 2 * mupi ** 2 + x)
                    * np.log(
                        (-1 + np.sqrt(1 - x) *
                         np.sqrt(1 - 4 * mupi ** 2 - x) + x) ** 2
                        / (1 + np.sqrt(1 - x) *
                           np.sqrt(1 - 4 * mupi ** 2 - x) - x) ** 2
                    )
                )
            ) / (Q * 2.0 * np.sqrt(1 - 4 * mupi ** 2) * np.pi ** 2 * x)
        else:
            dnde = 0
        
        return dnde

    def dnde_pipi(self, e_gams, e_cm):
        # Sum the FSR and decay spectra for the charged pions
        return (np.vectorize(self.__dnde_xx_to_pipig)(e_gams, e_cm) +
                2. * charged_pion(e_gams, e_cm / 2.0))
    
    def spectrum_funcs(self):
        return {'pi0 pi0':  self.dnde_pi0pi0,
                'pi pi':  self.dnde_pipi}

    def gamma_ray_lines(self, e_cm):
        return {}
\end{minted}
The required functions here are \mil{spetrum_funcs} and \mil{gamma_ray_lines}. Note the the second \mil{__dnde_xx_to_pipig} is an unvectorized helper function, which is not to be used directly. Next we implement the positron spectra:
\begin{minted}{python}
class HazmaExamplePositronSpectra:
    def dnde_pos_pipi(self, e_ps, e_cm):
        return pspec_charged_pion(e_ps, e_cm / 2.)
    
    def positron_spectrum_funcs(self):
        return {"pi pi": self.dnde_pos_pipi}

    def positron_lines(self, e_cm):
        return {}
\end{minted}
Lastly, we group all of these classes into a master class and we're done:
\begin{minted}{python}
class HazmaExample(HazmaExampleCrossSection,
                   HazmaExamplePositronSpectra,
                   HazmaExampleSpectra,
                   Theory):
    # Model parameters are DM mass: mx, 
    # Wilson coefficients: c1, c2 and
    # cutoff scale: lam
    def __init__(self, mx, c1, c2, lam):
        self.mx = mx
        self.c1 = c1
        self.c2 = c2
        self.lam = lam
    
    @staticmethod
    def list_annihilation_final_states():
        return ['pi pi', 'pi0 pi0']
\end{minted}
Now we can easily compute gamma-ray spectra, positron spectra and limit on our new model from gamma-ray telescopes. To implement our new model with \(m_{\chi} = 200~\mathrm{MeV}, c_{1} = c_{2} = 1\) and \(\Lambda = 100~\mathrm{GeV}\), we can use:
\begin{minted}{python}
>>> model = HazmaExample(200.0, 1.0, 1.0, 100e3)
\end{minted}
To compute a gamma-ray spectrum:
\begin{minted}{python}
# Photon energies from 1 keV to 1 GeV
>>> e_gams = np.logspace(-3.0, 3.0, num=150)
# Assume the DM is moving with a velocity of 10^-3
>>> vdm = 1e-3
# Compute CM energy assuming the above velocity
>>> Q = 2.0 * model.mx * (1 + 0.5 * vdm**2)
# Compute spectra
>>> spectra = model.spectra(e_gams, Q)
\end{minted}
Then we can plot the spectra using:
\begin{minted}{python}
>>> plt.figure(dpi=100)
>>> for key, val in spectra.items():
...     plt.plot(e_gams, val, label=key)
>>> plt.xlabel(r'$E_{\gamma} (\mathrm{MeV})$', fontsize=16)
>>> plt.ylabel(r'$\frac{dN}{dE_{\gamma}} (\mathrm{MeV}^{-1})$', fontsize=16)
>>> plt.xscale('log')
>>> plt.yscale('log')
>>> plt.legend()
\end{minted}
Additionally, we can compute limits on the thermally-averaged annihilation cross section of our model for various DM masses using 
\begin{minted}{python}
# Import target and background model for the E-Astrogam telescope
>>> from hazma.gamma_ray_parameters import gc_target, gc_bg_model
# Choose DM masses from half the pion mass to 250 MeV
>>> mxs = np.linspace(mpi/2., 250., num=100)
# Compute limits from E-Astrogam
>>> limits = np.zeros(len(mxs), dtype=float)
>>> for i, mx in enumerate(mxs):
...     model.mx = mx
...     limits[i] = model.unbinned_limit(target_params=gc_target,
...                                      bg_model=gc_bg_model)
\end{minted}

\section{Multi-particle phase space integration: \texorpdfstring{\mil{rambo}}{rambo}}
\label{sec:rambo}

The \mil{rambo} module implements the RAMBO algorithm, developed by Kleiss, Stirling and Ellis \cite{Kleiss:1985gy}, for generating phase-space points using a Monte Carlo procedure. We advise the interested reader to consult the original paper for details on how this algorithm works. We will only described how the algorithm works at a high-level.

The RAMBO algorithm works by first democratically generations random, independent, isotropic, massless four-momenta, $q_{i}^{\mu}$, with $i$ running from $1$ to $N$, $N$ being the number of final state particles. The momenta $q_{i}^{\mu}$ are generated with energies distributed according to $q_{i}^{0}\exp(-q_{i}^{0})dq_{i}^{0}$. These four momenta are then transformed into four-momenta $p_{i}^{\mu}$ by performing a particular Lorentz boost such that $\sum_i p_{i}^{\mu} = P^{\mu}$ where $P^{\mu}$ is the total four-momenta of the given process in the center-of-mass frame (i.e. $P^{0} = E_{\mathrm{CM}}$ and $\vec{P} = \vec{0}$.) At this stage, a corresponding weight $W_{0}(p_{i})$ is computed to give the probability of the event with the particular four-momenta $p_{i}^{\mu}$. Lastly, the four-momenta $p_{i}^{\mu}$ are transformed into four-momenta $k_{i}^{\mu}$ such that the $k_{i}^{\mu}$ have the correct masses (i.e. $k_{i}^{\mu}k_{i,\mu} = m_{i}^2$) and a new weight $W(k_{i})$ is computed. This completes the algorithm. Below we give pseudo-code for the algorithm:
\begin{enumerate}
    \item Generate the $q_{i}^{\mu}$'s with energies distributed according to $q^{0}_{i}\exp(-q^{0}_{i})dq^{0}_{i}$ using:
    \begin{align}
        q_{i}^{0} &= -\log(\rho_{i}^{(3)}\rho_{i}^{(4)}), & q_{i}^{x} &= q_{i}^{0}\sqrt{1-c_{i}^2}\cos{\phi_{i}}, & q_{i}^{y} &= q_{i}^{0}\sqrt{1-c_{i}^2}\sin{\phi_{i}}, & q_{i}^{z} &= q_{i}^{0}c_{i}.
    \end{align}
    where $c_{i} = 2\rho_{i}^{(1)} - 1$, $\phi_{i}=2\pi\rho_{i}^{(2)}$ and $\rho_{i}^{(1,2,3,4)}$ are uniform random numbers on $(0,1)$. See \cite{Kleiss:1985gy} for a proof that these are indeed distributed according to $q^{0}_{i}\exp(-q^{0}_{i})dq^{0}_{i}$.
    \item Transform the $q_{i}^{\mu}$'s into $p_{i}^{\mu}$'s such that the $\sum_{i}p_{i}^{\mu} = P^{\mu}$ with $P^{0} = E_{\mathrm{CM}}$ and $\vec{P} = \vec{0}$. This is done using:
    \begin{align}
        p_{i}^{0} &= x(\gamma q^{0}_{i}+\vec{b}\cdot\vec{q}_{i}), & \vec{p}_{i} &= x(\vec{q}_{i}+ \vec{b}q_{i}^{0} + a (\vec{b}\cdot\vec{q}_{i})\vec{b}).
    \end{align}
    where 
    \begin{align}
        \vec{b} &= -\vec{Q}/M, & x &= E_{\mathrm{CM}}/M, & \gamma &= Q^{0}/M, & a &= 1/\sqrt{1+\gamma},\notag\\ 
        Q^{\mu} &= \sum_{i}q_{i}^{\mu}, & M &= \sqrt{Q_{\mu}Q^{\mu}}
    \end{align}
    Assign the following weight to the event:
    \begin{align}
        W_{0} &= \left(\dfrac{\pi}{2}\right)^{N-1}E_{\mathrm{CM}}^{2N-4}\dfrac{(2\pi)^{4-3N}}{\Gamma(N)\Gamma(N-1)}
    \end{align}
    \item Next we transform the $p_{i}^{\mu}$'s to $k_{i}^{\mu}$'s such that $k_{i}^{\mu}k_{i,\mu} = m_{i}$. This is done via:
    \begin{align}
        \vec{k}_{i} &= \xi\vec{p}_{i}, & k_{i}^{0} &= \sqrt{m_{i} +(\xi p_{i}^{0})^2}
    \end{align}
    where $\xi$ is the solution to 
    \begin{align}
        E_{\mathrm{CM}} &= \sum_{i=1}^{N}\sqrt{m_{i}^2 + (\xi p_{i}^{0})^2}
    \end{align}
    This equation can be solved efficiently using Newton's method with the initial guess $\xi_{0} = \sqrt{1 - \left(\sum_{i=1}^{N}m_{i}/E_{\mathrm{CM}}\right)^2}$. Lastly, we re-weight the probability of the event using:
    \begin{align}
        W = W_{0} E_{\mathrm{CM}}\left(\sum_{i=1}^{N}\dfrac{|\vec{k}_{i}|}{E_{\mathrm{CM}}}\right)^{2N-3}
        \left(\sum_{i=1}^{N}\dfrac{|\vec{k}_{i}|^2}{k_{i}^{0}}\right)^{-1}
        \left(\prod_{i=1}^{N}\dfrac{|\vec{k}_{i}|}{k_{i}^{0}}\right)
    \end{align}
    We note that this fixes a typo from the original paper, where the overall factor of $E_{\mathrm{CM}}$ was missing. Additionally, we note that the weights clearly don't take into account the matrix element of the particle physics process. The matrix elements are included by simply multiplying the weights by the corresponding matrix element. 
\end{enumerate}
This completes the pseudo-code for the RAMBO algorithm. 

In \mil{hazma}, the \mil{rambo} module can be used to perform various tasks. At the highest level, the function \mil{compute_annihilation_cross_section} (\mil{compute_decay_width}) is provided for computing cross-sections (decay widths) for $2\to N$ ($1\to N$) processes. For example, consider computing the partial decay width of $\mu^- \to e^- \overline{\nu}_e \nu_\mu$. One first declares a function that takes a list of the final state particles' four-momenta and muon's four-momentum, and returns the matrix element squared for the reaction:
\begin{minted}{python}
# Import the fermi constant
>>> from hazma.parameters import GF
# Import a helper function for scalar products of four-vectors
>>> from hazma.field_theory_helper_functions.common_functions import \
...    minkowski_dot as MDot
# Declare the matrix element
>>> def msqrd_mu_to_enunu(momenta):
...    pe = momenta[0] # electron four-momentum
...    pve = momenta[1] # electron-neutrino four-momentum
...    pvmu = momenta[2] # muon-neutrino four-momentum
...    pmu = sum(momenta) # muon four-momentum
       # squared matrix element
...    return 64. * GF**2 * MDot(pe, pvmu) * MDot(pmu, pve)
\end{minted}
Then, the partial decay width can be computed using:
\begin{minted}{python}
# Import function to compute decay width
>>> from hazma.rambo import compute_decay_width
# import masses of muon and electron
>>> from hazma.parameters import muon_mass as mmu
>>> from hazma.parameters import electron_mass as me
# Specify the masses of the electron and neutrinos
>>> fsp_masses = np.array([me, 0., 0.])
# compute the partial width
>>> partial_width = compute_decay_width(fsp_masses, mmu, num_ps_pts=50000,
...                                     mat_elem_sqrd=msqrd_mu_to_enunu)
\end{minted}
Using 50000 phase-space points gives a result within $5\%$ of the analytical value.

In addition, \mil{rambo} includes a function for performing integrations over all variables except the energy of one of the final-state particles called \mil{generate_energy_histogram}. This is useful for computing energy spectra for FSR from final states with large numbers of particles. The following demonstrates how it can be used:
\begin{minted}{python}
>>> from hazma.rambo import generate_energy_histogram
>>> import numpy as np
>>> num_ps_pts = 100000 #number of phase-space points to use
# masses of final-state particles
>>> masses = np.array([100., 100., 0.0, 0.0])
>>> cme = 1000. # center-of-mass energy
>>> num_bins = 100 # number of energy bins to use
# computing energy histograms
>>> eng_hist = generate_energy_histogram(num_ps_pts, masses, cme,
...                                  num_bins=num_bins)
# plot the results
>>> import matplotlib as plt
>>> for i in range(len(masses)):
...     # pts[i, 0] are the energies of particle i
...     # pts[i, 1] are the probabilities
...     plt.loglog(pts[i, 0], pts[i, 1])                                   
\end{minted}

At the lowest level, \mil{rambo} includes a function for generating phase-space points called \mil{generate_phase_space}. We also include a function \mil{integrate_over_phase_space} which will perform the integral
\begin{align}
    \int \left(\prod_{i=1}^{N}\dfrac{d^{3}\vec{p}_{i}}{(2\pi)^3}\dfrac{1}{2E_{i}}\right)(2\pi)^4\delta^{4}\left(P-\sum_{i=1}^{N}p_{i}\right)\left|\mathcal{M}\right|^2
\end{align}
where $P^{\mu}$ is the total four-momentum, $p_{i}^{\mu}$ are the individual four-momenta for each of the $N$ final-state particles and $\mathcal{M}$ is the matrix element. 

\section{Gamma-ray Spectra from many final state particles: \texorpdfstring{\mil{gamma_ray}}{gamma\_ray}}
\label{sec:gamma_ray}

Since computing gamma-ray spectra is a model-dependent process, we include in \mil{hazma} tools for computing gamma-ray spectra from both FSR and the decay of final state-particles. The \mil{gamma_ray} module contains two functions called \mil{gamma_ray_decay} and \mil{gamma_ray_fsr}. The \mil{gamma_ray_decay} function accepts a list of the final-state particles, the center-of-mass energy, the gamma-ray energies to compute the spectrum at and optionally the matrix element squared. Currently, the final-state particles can be $\pi^{0}, \pi^{\pm}, \mu^{\pm}, e^{\pm}, K^{\pm}, K_{L}$ and $K_{S}$ where $K$ stands for kaon. We caution that when including many final-state mesons, one needs to take care to supply a matrix element squared that is valid at the reaction's center-of-mass energy (see Sec.~(\ref{sec:introduction}) for a discussion on the validity of ChPT for large energies).

The \mil{gamma_ray_decay} function works by first computing the energies distributions of all the final-state particles and convolving these with the decay spectra of the final-state particles. It can be used as follows:
\begin{minted}{python}
>>> from hazma.gamma_ray_decay import gamma_ray_decay
>>> import numpy as np
# specify the final-state particles
>>> particles = np.array(['muon', 'charged_kaon', 'long_kaon'])
# choose the center of mass energy
>>> e_cm = 5000.
# choose list of the gamma-ray energies to compute spectra at
>>> e_gams = np.logspace(0., np.log10(e_cm), num=200, dtype=np.float64)
# compute the gamma-ray spectra assuming a constant matrix element
>>> spec = gamma_ray_decay(particles, e_cm, e_gams)
\end{minted}

The \mil{gamma_ray_fsr} function computes the gamma-ray spectrum from $X\to Y\gamma$, i.e.:
\begin{align}
    \dfrac{dN(X\to Y\gamma)}{dE_{\gamma}} = \dfrac{1}{\sigma(X\to Y)}\dfrac{d\sigma(X\to Y\gamma)}{dE_{\gamma}}
\end{align}
where $X$ and $Y$ are any particles excluding the photon. As input this function takes a list of the initial state particle masses (either 1 or 2 particles), the final state particle masses, the center-of-mass energy, a function for the tree-level matrix element squared (for $X\to Y$) and a function for the radiative matrix element squared ($X\to Y\gamma$). The functions for the squared matrix elements must take is a single argument which is a list of the four-momenta for the final state particles. As an example, we consider the process of two dark-matter particles annihilating into charged pions, $\chi\bar{\chi}\to \pi^{+}\pi^{-}(\gamma)$ using the model from App.~(\ref{app:user_defined_models}). In App~(\ref{app:user_defined_models}), we gave the analytic expressions for the gamma-ray spectra. The tree-level and radiative matrix elements squared for this process are:
\begin{align}
    |\mathcal{M}(\chi\bar{\chi}\to \pi^{+}\pi^{-})|^2 &= \frac{c_1^2 \left(s-4 m_{\chi}^2\right)}{2 \Lambda^2}\\
    |\mathcal{M}(\chi\bar{\chi}\to \pi^{+}\pi^{-}\gamma)|^2 &= \dfrac{2 c_1^2 \left(4 \mu_{\chi}^2-1\right) Q^2 e^2}{\Lambda^2 \left(t-\mu_{\pi}^2 Q^2\right)^2
   \left(u-\mu_{\pi}^2 Q^2\right)^2}\\
   &\qquad \times \left(\left(\mu_{\pi} Q (t+u)-2 \mu_{\pi}^3 Q^3\right)^2+s
   \left(t-\mu_{\pi}^2 Q^2\right) \left(\mu_{\pi}^2
   Q^2-u\right)\right)\notag
\end{align}
where $Q$ is the center-of-mass energy, $e$ is the electromagnetic coupling, $\mu_{\pi,\chi} = m_{\pi,\chi}/Q$ and 
\begin{align}
    s &= (p_{\pi,1} + p_{\pi,2})^2, &t &= (p_{\pi,1} + k)^2, & u &= (p_{\pi,2} + k)^2
\end{align}
with $p_{\pi,1,2}$ are the four-momenta of the two final-state pions and $k$ is the four-momenta of the final-state photon. Below, we create a class to implement functions for the tree and radiative squared matrix elements. Note that these functions take in an array of four-momenta. 
\begin{minted}{python}
from hazma.field_theory_helper_functions.common_functions import \
    minkowski_dot as MDot
    
class Msqrd(object):
    def __init__(self, mx, c1, lam):
        self.mx = mx # DM mass
        self.c1 = c1 # effective coupling of DM to charged pions
        self.lam = lam # cut off scale for effective theory

    def tree(self, momenta):
        ppi1 = momenta[0] # first charged pion four-momentum
        ppi2 = momenta[1] # second charged pion four-momentum
        Q = ppi1[0] + ppi2[0] # center-of-mass energy
        return -((self.c1**2 * (4 * self.mx**2-Q**2)) / (2 * self.lam**2))

    def radiative(self, momenta):
        ppi1 = momenta[0] # first charged pion four-momentum
        ppi2 = momenta[1] # second charged pion four-momentum
        k = momenta[2] # photon four-momentum
        Q = ppi1[0] + ppi2[0] + k[0] # center-of-mass energy

        mux = self.mx / Q
        mupi = mpi / Q

        s = MDot(ppi1 + ppi2, ppi1 + ppi2)
        t = MDot(ppi1 + k, ppi1 + k)
        u = MDot(ppi2 + k, ppi2 + k)

        return ((2*self.c1**2*(-1 + 4*mux**2)*Q**2*qe**2 *
                 (s*(-(mupi**2*Q**2) + t)*(mupi**2*Q**2 - u) +
                  (-2*mupi**3*Q**3 + mupi*Q*(t + u))**2)) /
                (self.lam**2*(-(mupi**2*Q**2) + t)**2*
                (-(mupi**2*Q**2) + u)**2))
\end{minted}
Next, we can compute the gamma-ray spectrum for $\chi\bar{\chi}\to\pi^{+}\pi^{-}\gamma$ using:
\begin{minted}{python}
>>> from hazma.gamma_ray import gamma_ray_fsr
# specify the parameters of the model
>>> params = {'mx': 200.0, 'c1':1.0, 'lam':1e4}
# create instance of our Msqrd class
>>> msqrds = Msqrd(**params)
# specify the initial and final state masses
>>> isp_masses = np.array([msqrds.mx, msqrds.mx])
>>> fsp_masses = np.array([mpi, mpi, 0.0])
# choose the center-of-mass energy
>>> e_cm = 4.0 * msqrds.mx
# compute the gamma-ray spectrum
>>> spec = gamma_ray_fsr(isp_masses, fsp_masses, e_cm, msqrds.tree,
                         msqrds.radiative, num_ps_pts=500000, num_bins=50)
# plot the spectrum
>>> import matplotlib.pyplot as plt 
>>> plt.figure(dpi=100)
>>> plt.plot(spec[0], spec[1])
>>> plt.yscale('log')
>>> plt.xscale('log')
>>> plt.ylabel(r'$dN/dE_{\gamma} \ (\mathrm{MeV}^{-1})$', fontsize=16)
>>> plt.xlabel(r'$E_{\gamma} \ (\mathrm{MeV})$', fontsize=16)
\end{minted}

\bibliographystyle{apsrev}
\bibliography{hazma}

\begin{thebibliography}{74}
\expandafter\ifx\csname natexlab\endcsname\relax\def\natexlab#1{#1}\fi
\expandafter\ifx\csname bibnamefont\endcsname\relax
  \def\bibnamefont#1{#1}\fi
\expandafter\ifx\csname bibfnamefont\endcsname\relax
  \def\bibfnamefont#1{#1}\fi
\expandafter\ifx\csname citenamefont\endcsname\relax
  \def\citenamefont#1{#1}\fi
\expandafter\ifx\csname url\endcsname\relax
  \def\url#1{\texttt{#1}}\fi
\expandafter\ifx\csname urlprefix\endcsname\relax\def\urlprefix{URL }\fi
\providecommand{\bibinfo}[2]{#2}
\providecommand{\eprint}[2][]{\url{#2}}

\bibitem[{\citenamefont{Lee and Weinberg}(1977)}]{Lee:1977ua}
\bibinfo{author}{\bibfnamefont{B.~W.} \bibnamefont{Lee}} \bibnamefont{and}
  \bibinfo{author}{\bibfnamefont{S.}~\bibnamefont{Weinberg}},
  \bibinfo{journal}{Phys. Rev. Lett.} \textbf{\bibinfo{volume}{39}},
  \bibinfo{pages}{165} (\bibinfo{year}{1977}), \bibinfo{note}{[,183(1977)]}.

\bibitem[{\citenamefont{Profumo}(2008)}]{Profumo:2008yg}
\bibinfo{author}{\bibfnamefont{S.}~\bibnamefont{Profumo}},
  \bibinfo{journal}{Phys. Rev.} \textbf{\bibinfo{volume}{D78}},
  \bibinfo{pages}{023507} (\bibinfo{year}{2008}), \eprint{0806.2150}.

\bibitem[{\citenamefont{Atwood et~al.}(2009{\natexlab{a}})}]{fermilat}
\bibinfo{author}{\bibfnamefont{W.~B.} \bibnamefont{Atwood}}
  \bibnamefont{et~al.} (\bibinfo{collaboration}{Fermi-LAT}),
  \bibinfo{journal}{Astrophys. J.} \textbf{\bibinfo{volume}{697}},
  \bibinfo{pages}{1071} (\bibinfo{year}{2009}{\natexlab{a}}),
  \eprint{0902.1089}.

\bibitem[{\citenamefont{Aleksi{\'c} et~al.}(2016)}]{magic}
\bibinfo{author}{\bibfnamefont{J.}~\bibnamefont{Aleksi{\'c}}}
  \bibnamefont{et~al.} (\bibinfo{collaboration}{MAGIC}),
  \bibinfo{journal}{Astropart. Phys.} \textbf{\bibinfo{volume}{72}},
  \bibinfo{pages}{76} (\bibinfo{year}{2016}), \eprint{1409.5594}.

\bibitem[{\citenamefont{Abramowski et~al.}(2014)}]{hess}
\bibinfo{author}{\bibfnamefont{A.}~\bibnamefont{Abramowski}}
  \bibnamefont{et~al.} (\bibinfo{collaboration}{H.E.S.S.}),
  \bibinfo{journal}{Phys. Rev.} \textbf{\bibinfo{volume}{D90}},
  \bibinfo{pages}{112012} (\bibinfo{year}{2014}), \eprint{1410.2589}.

\bibitem[{\citenamefont{Holder et~al.}(2006)\citenamefont{Holder, Atkins,
  Badran, Blaylock, Bradbury, Buckley, Byrum, Carter-Lewis, Celik, Chow
  et~al.}}]{holder2006first}
\bibinfo{author}{\bibfnamefont{J.}~\bibnamefont{Holder}},
  \bibinfo{author}{\bibfnamefont{R.}~\bibnamefont{Atkins}},
  \bibinfo{author}{\bibfnamefont{H.}~\bibnamefont{Badran}},
  \bibinfo{author}{\bibfnamefont{G.}~\bibnamefont{Blaylock}},
  \bibinfo{author}{\bibfnamefont{S.}~\bibnamefont{Bradbury}},
  \bibinfo{author}{\bibfnamefont{J.}~\bibnamefont{Buckley}},
  \bibinfo{author}{\bibfnamefont{K.}~\bibnamefont{Byrum}},
  \bibinfo{author}{\bibfnamefont{D.}~\bibnamefont{Carter-Lewis}},
  \bibinfo{author}{\bibfnamefont{O.}~\bibnamefont{Celik}},
  \bibinfo{author}{\bibfnamefont{Y.}~\bibnamefont{Chow}}, \bibnamefont{et~al.},
  \bibinfo{journal}{Astroparticle Physics} \textbf{\bibinfo{volume}{25}},
  \bibinfo{pages}{391} (\bibinfo{year}{2006}).

\bibitem[{\citenamefont{Aguilar et~al.}(2016)}]{AMS}
\bibinfo{author}{\bibfnamefont{M.}~\bibnamefont{Aguilar}} \bibnamefont{et~al.}
  (\bibinfo{collaboration}{AMS}), \bibinfo{journal}{Phys. Rev. Lett.}
  \textbf{\bibinfo{volume}{117}}, \bibinfo{pages}{231102}
  (\bibinfo{year}{2016}).

\bibitem[{\citenamefont{Gondolo et~al.}(2004)\citenamefont{Gondolo, Edsjo,
  Ullio, Bergstrom, Schelke, and Baltz}}]{Gondolo:2004sc}
\bibinfo{author}{\bibfnamefont{P.}~\bibnamefont{Gondolo}},
  \bibinfo{author}{\bibfnamefont{J.}~\bibnamefont{Edsjo}},
  \bibinfo{author}{\bibfnamefont{P.}~\bibnamefont{Ullio}},
  \bibinfo{author}{\bibfnamefont{L.}~\bibnamefont{Bergstrom}},
  \bibinfo{author}{\bibfnamefont{M.}~\bibnamefont{Schelke}}, \bibnamefont{and}
  \bibinfo{author}{\bibfnamefont{E.~A.} \bibnamefont{Baltz}},
  \bibinfo{journal}{JCAP} \textbf{\bibinfo{volume}{0407}}, \bibinfo{pages}{008}
  (\bibinfo{year}{2004}), \eprint{astro-ph/0406204}.

\bibitem[{\citenamefont{Belanger et~al.}(2014)\citenamefont{Belanger, Boudjema,
  Pukhov, and Semenov}}]{Belanger:2013oya}
\bibinfo{author}{\bibfnamefont{G.}~\bibnamefont{Belanger}},
  \bibinfo{author}{\bibfnamefont{F.}~\bibnamefont{Boudjema}},
  \bibinfo{author}{\bibfnamefont{A.}~\bibnamefont{Pukhov}}, \bibnamefont{and}
  \bibinfo{author}{\bibfnamefont{A.}~\bibnamefont{Semenov}},
  \bibinfo{journal}{Comput. Phys. Commun.} \textbf{\bibinfo{volume}{185}},
  \bibinfo{pages}{960} (\bibinfo{year}{2014}), \eprint{1305.0237}.

\bibitem[{\citenamefont{Cirelli et~al.}(2011)\citenamefont{Cirelli, Corcella,
  Hektor, Hutsi, Kadastik, Panci, Raidal, Sala, and Strumia}}]{PPPC}
\bibinfo{author}{\bibfnamefont{M.}~\bibnamefont{Cirelli}},
  \bibinfo{author}{\bibfnamefont{G.}~\bibnamefont{Corcella}},
  \bibinfo{author}{\bibfnamefont{A.}~\bibnamefont{Hektor}},
  \bibinfo{author}{\bibfnamefont{G.}~\bibnamefont{Hutsi}},
  \bibinfo{author}{\bibfnamefont{M.}~\bibnamefont{Kadastik}},
  \bibinfo{author}{\bibfnamefont{P.}~\bibnamefont{Panci}},
  \bibinfo{author}{\bibfnamefont{M.}~\bibnamefont{Raidal}},
  \bibinfo{author}{\bibfnamefont{F.}~\bibnamefont{Sala}}, \bibnamefont{and}
  \bibinfo{author}{\bibfnamefont{A.}~\bibnamefont{Strumia}},
  \bibinfo{journal}{JCAP} \textbf{\bibinfo{volume}{1103}}, \bibinfo{pages}{051}
  (\bibinfo{year}{2011}), \bibinfo{note}{[Erratum: JCAP1210,E01(2012)]},
  \eprint{1012.4515}.

\bibitem[{\citenamefont{Sjostrand et~al.}(2008)\citenamefont{Sjostrand, Mrenna,
  and Skands}}]{pythia}
\bibinfo{author}{\bibfnamefont{T.}~\bibnamefont{Sjostrand}},
  \bibinfo{author}{\bibfnamefont{S.}~\bibnamefont{Mrenna}}, \bibnamefont{and}
  \bibinfo{author}{\bibfnamefont{P.~Z.} \bibnamefont{Skands}},
  \bibinfo{journal}{Comput. Phys. Commun.} \textbf{\bibinfo{volume}{178}},
  \bibinfo{pages}{852} (\bibinfo{year}{2008}), \eprint{0710.3820}.

\bibitem[{\citenamefont{Rando et~al.}(2019)\citenamefont{Rando, De~Angelis, and
  Mallamaci}}]{eastrogam}
\bibinfo{author}{\bibfnamefont{R.}~\bibnamefont{Rando}},
  \bibinfo{author}{\bibfnamefont{A.}~\bibnamefont{De~Angelis}},
  \bibnamefont{and} \bibinfo{author}{\bibfnamefont{M.}~\bibnamefont{Mallamaci}}
  (\bibinfo{collaboration}{thee-ASTROGAM}), \bibinfo{journal}{J. Phys. Conf.
  Ser.} \textbf{\bibinfo{volume}{1181}}, \bibinfo{pages}{012044}
  (\bibinfo{year}{2019}).

\bibitem[{\citenamefont{Tavani et~al.}(2018)}]{DeAngelis:2017gra}
\bibinfo{author}{\bibfnamefont{M.}~\bibnamefont{Tavani}} \bibnamefont{et~al.}
  (\bibinfo{collaboration}{e-ASTROGAM}), \bibinfo{journal}{JHEAp}
  \textbf{\bibinfo{volume}{19}}, \bibinfo{pages}{1} (\bibinfo{year}{2018}),
  \eprint{1711.01265}.

\bibitem[{\citenamefont{McEnery et~al.}(2019)\citenamefont{McEnery, Barrio,
  Agudo, Ajello, {\'A}lvarez, Ansoldi, Anton, Auricchio, Stephen, Baldini
  et~al.}}]{mcenery2019all}
\bibinfo{author}{\bibfnamefont{J.}~\bibnamefont{McEnery}},
  \bibinfo{author}{\bibfnamefont{J.~A.} \bibnamefont{Barrio}},
  \bibinfo{author}{\bibfnamefont{I.}~\bibnamefont{Agudo}},
  \bibinfo{author}{\bibfnamefont{M.}~\bibnamefont{Ajello}},
  \bibinfo{author}{\bibfnamefont{J.-M.} \bibnamefont{{\'A}lvarez}},
  \bibinfo{author}{\bibfnamefont{S.}~\bibnamefont{Ansoldi}},
  \bibinfo{author}{\bibfnamefont{S.}~\bibnamefont{Anton}},
  \bibinfo{author}{\bibfnamefont{N.}~\bibnamefont{Auricchio}},
  \bibinfo{author}{\bibfnamefont{J.~B.} \bibnamefont{Stephen}},
  \bibinfo{author}{\bibfnamefont{L.}~\bibnamefont{Baldini}},
  \bibnamefont{et~al.}, \bibinfo{journal}{arXiv preprint arXiv:1907.07558}
  (\bibinfo{year}{2019}).

\bibitem[{\citenamefont{{Hunter} et~al.}(2010)\citenamefont{{Hunter}, {Bloser},
  {Dion}, {McConnell}, {de Nolfo}, {Son}, {Ryan}, and {Stecker}}}]{adept}
\bibinfo{author}{\bibfnamefont{S.~D.} \bibnamefont{{Hunter}}},
  \bibinfo{author}{\bibfnamefont{P.~F.} \bibnamefont{{Bloser}}},
  \bibinfo{author}{\bibfnamefont{M.~P.} \bibnamefont{{Dion}}},
  \bibinfo{author}{\bibfnamefont{M.~L.} \bibnamefont{{McConnell}}},
  \bibinfo{author}{\bibfnamefont{G.~A.} \bibnamefont{{de Nolfo}}},
  \bibinfo{author}{\bibfnamefont{S.}~\bibnamefont{{Son}}},
  \bibinfo{author}{\bibfnamefont{J.~M.} \bibnamefont{{Ryan}}},
  \bibnamefont{and} \bibinfo{author}{\bibfnamefont{F.~W.}
  \bibnamefont{{Stecker}}}, in \emph{\bibinfo{booktitle}{Space Telescopes and
  Instrumentation 2010: Ultraviolet to Gamma Ray}} (\bibinfo{year}{2010}), vol.
  \bibinfo{volume}{7732} of \emph{\bibinfo{series}{Proceedings of the SPIE}},
  p. \bibinfo{pages}{773221}.

\bibitem[{\citenamefont{Wu}(2016)}]{pangu}
\bibinfo{author}{\bibfnamefont{X.}~\bibnamefont{Wu}}, \bibinfo{journal}{PoS}
  \textbf{\bibinfo{volume}{ICRC2015}}, \bibinfo{pages}{964}
  (\bibinfo{year}{2016}).

\bibitem[{\citenamefont{Duncan et~al.}(2016)}]{grips}
\bibinfo{author}{\bibfnamefont{N.}~\bibnamefont{Duncan}} \bibnamefont{et~al.},
  \bibinfo{journal}{Proc. SPIE Int. Soc. Opt. Eng.}
  \textbf{\bibinfo{volume}{9905}}, \bibinfo{pages}{99052Q}
  (\bibinfo{year}{2016}), \eprint{1609.08558}.

\bibitem[{\citenamefont{Contributors}(2019)}]{hazma}
\bibinfo{author}{\bibfnamefont{P.~U.~W.} \bibnamefont{Contributors}},
  \emph{\bibinfo{title}{Hazma}} (\bibinfo{year}{2019}),
  \urlprefix\url{https://pokemon-uranium.fandom.com/wiki/Hazma}.

\bibitem[{\citenamefont{Scherer}(2003)}]{Scherer2003}
\bibinfo{author}{\bibfnamefont{S.}~\bibnamefont{Scherer}},
  \emph{\bibinfo{title}{Introduction to Chiral Perturbation Theory}}
  (\bibinfo{publisher}{Springer US}, \bibinfo{address}{Boston, MA},
  \bibinfo{year}{2003}), pp. \bibinfo{pages}{277--538}, ISBN
  \bibinfo{isbn}{978-0-306-47916-8}, \eprint{hep-ph/0210398}.

\bibitem[{\citenamefont{{Coogan, Adam and Morrison, Logan and Profumo,
  Stefano}}(to appear)}]{companion}
\bibinfo{author}{\bibnamefont{{Coogan, Adam and Morrison, Logan and Profumo,
  Stefano}}} (\bibinfo{year}{to appear}).

\bibitem[{\citenamefont{D'Eramo and Profumo}(2018)}]{DEramo:2018khz}
\bibinfo{author}{\bibfnamefont{F.}~\bibnamefont{D'Eramo}} \bibnamefont{and}
  \bibinfo{author}{\bibfnamefont{S.}~\bibnamefont{Profumo}},
  \bibinfo{journal}{Phys. Rev. Lett.} \textbf{\bibinfo{volume}{121}},
  \bibinfo{pages}{071101} (\bibinfo{year}{2018}), \eprint{1806.04745}.

\bibitem[{\citenamefont{Alloul et~al.}(2014)\citenamefont{Alloul, Christensen,
  Degrande, Duhr, and Fuks}}]{Alloul:2013bka}
\bibinfo{author}{\bibfnamefont{A.}~\bibnamefont{Alloul}},
  \bibinfo{author}{\bibfnamefont{N.~D.} \bibnamefont{Christensen}},
  \bibinfo{author}{\bibfnamefont{C.}~\bibnamefont{Degrande}},
  \bibinfo{author}{\bibfnamefont{C.}~\bibnamefont{Duhr}}, \bibnamefont{and}
  \bibinfo{author}{\bibfnamefont{B.}~\bibnamefont{Fuks}},
  \bibinfo{journal}{Comput. Phys. Commun.} \textbf{\bibinfo{volume}{185}},
  \bibinfo{pages}{2250} (\bibinfo{year}{2014}), \eprint{1310.1921}.

\bibitem[{\citenamefont{Hahn}(2001)}]{Hahn:2000kx}
\bibinfo{author}{\bibfnamefont{T.}~\bibnamefont{Hahn}},
  \bibinfo{journal}{Comput. Phys. Commun.} \textbf{\bibinfo{volume}{140}},
  \bibinfo{pages}{418} (\bibinfo{year}{2001}), \eprint{hep-ph/0012260}.

\bibitem[{\citenamefont{Shtabovenko et~al.}(2016)\citenamefont{Shtabovenko,
  Mertig, and Orellana}}]{Shtabovenko:2016sxi}
\bibinfo{author}{\bibfnamefont{V.}~\bibnamefont{Shtabovenko}},
  \bibinfo{author}{\bibfnamefont{R.}~\bibnamefont{Mertig}}, \bibnamefont{and}
  \bibinfo{author}{\bibfnamefont{F.}~\bibnamefont{Orellana}},
  \bibinfo{journal}{Comput. Phys. Commun.} \textbf{\bibinfo{volume}{207}},
  \bibinfo{pages}{432} (\bibinfo{year}{2016}), \eprint{1601.01167}.

\bibitem[{\citenamefont{Oliphant}(2006--)}]{numpy}
\bibinfo{author}{\bibfnamefont{T.}~\bibnamefont{Oliphant}},
  \emph{\bibinfo{title}{{NumPy}: A guide to {NumPy}}},
  \bibinfo{howpublished}{USA: Trelgol Publishing} (\bibinfo{year}{2006--}),
  \bibinfo{note}{online; accessed 2019-07-23},
  \urlprefix\url{http://www.numpy.org/}.

\bibitem[{\citenamefont{Behnel et~al.}(2011)\citenamefont{Behnel, Bradshaw,
  Citro, Dalcin, Seljebotn, and Smith}}]{behnel2011cython}
\bibinfo{author}{\bibfnamefont{S.}~\bibnamefont{Behnel}},
  \bibinfo{author}{\bibfnamefont{R.}~\bibnamefont{Bradshaw}},
  \bibinfo{author}{\bibfnamefont{C.}~\bibnamefont{Citro}},
  \bibinfo{author}{\bibfnamefont{L.}~\bibnamefont{Dalcin}},
  \bibinfo{author}{\bibfnamefont{D.~S.} \bibnamefont{Seljebotn}},
  \bibnamefont{and} \bibinfo{author}{\bibfnamefont{K.}~\bibnamefont{Smith}},
  \bibinfo{journal}{Computing in Science \& Engineering}
  \textbf{\bibinfo{volume}{13}}, \bibinfo{pages}{31} (\bibinfo{year}{2011}).

\bibitem[{\citenamefont{Weinberg}(1979)}]{WEINBERG1979327}
\bibinfo{author}{\bibfnamefont{S.}~\bibnamefont{Weinberg}},
  \bibinfo{journal}{Physica A: Statistical Mechanics and its Applications}
  \textbf{\bibinfo{volume}{96}}, \bibinfo{pages}{327 } (\bibinfo{year}{1979}),
  ISSN \bibinfo{issn}{0378-4371},
  \urlprefix\url{http://www.sciencedirect.com/science/article/pii/0378437179902231}.

\bibitem[{\citenamefont{Gasser and Leutwyler}(1984)}]{Gasser:1983yg}
\bibinfo{author}{\bibfnamefont{J.}~\bibnamefont{Gasser}} \bibnamefont{and}
  \bibinfo{author}{\bibfnamefont{H.}~\bibnamefont{Leutwyler}},
  \bibinfo{journal}{Annals Phys.} \textbf{\bibinfo{volume}{158}},
  \bibinfo{pages}{142} (\bibinfo{year}{1984}).

\bibitem[{\citenamefont{Gasser and Leutwyler}(1985)}]{GASSER1985465}
\bibinfo{author}{\bibfnamefont{J.}~\bibnamefont{Gasser}} \bibnamefont{and}
  \bibinfo{author}{\bibfnamefont{H.}~\bibnamefont{Leutwyler}},
  \bibinfo{journal}{Nuclear Physics B} \textbf{\bibinfo{volume}{250}},
  \bibinfo{pages}{465 } (\bibinfo{year}{1985}), ISSN \bibinfo{issn}{0550-3213},
  \urlprefix\url{http://www.sciencedirect.com/science/article/pii/0550321385904924}.

\bibitem[{\citenamefont{Ecker}(1995)}]{ecker1995chiral}
\bibinfo{author}{\bibfnamefont{G.}~\bibnamefont{Ecker}},
  \bibinfo{journal}{Progress in Particle and Nuclear Physics}
  \textbf{\bibinfo{volume}{35}}, \bibinfo{pages}{1} (\bibinfo{year}{1995}).

\bibitem[{\citenamefont{Pich}(1995)}]{Pich:1995bw}
\bibinfo{author}{\bibfnamefont{A.}~\bibnamefont{Pich}}, \bibinfo{journal}{Rept.
  Prog. Phys.} \textbf{\bibinfo{volume}{58}}, \bibinfo{pages}{563}
  (\bibinfo{year}{1995}), \eprint{hep-ph/9502366}.

\bibitem[{\citenamefont{Meissner}(1993)}]{Meissner:1993ah}
\bibinfo{author}{\bibfnamefont{U.~G.} \bibnamefont{Meissner}},
  \bibinfo{journal}{Rept. Prog. Phys.} \textbf{\bibinfo{volume}{56}},
  \bibinfo{pages}{903} (\bibinfo{year}{1993}), \eprint{hep-ph/9302247}.

\bibitem[{\citenamefont{Pelaez}(2016)}]{f0500_review}
\bibinfo{author}{\bibfnamefont{J.~R.} \bibnamefont{Pelaez}},
  \bibinfo{journal}{Phys. Rept.} \textbf{\bibinfo{volume}{658}},
  \bibinfo{pages}{1} (\bibinfo{year}{2016}), \eprint{1510.00653}.

\bibitem[{\citenamefont{Truong}(1988)}]{PhysRevLett.61.2526}
\bibinfo{author}{\bibfnamefont{T.~N.} \bibnamefont{Truong}},
  \bibinfo{journal}{Phys. Rev. Lett.} \textbf{\bibinfo{volume}{61}},
  \bibinfo{pages}{2526} (\bibinfo{year}{1988}),
  \urlprefix\url{https://link.aps.org/doi/10.1103/PhysRevLett.61.2526}.

\bibitem[{\citenamefont{Ecker et~al.}(1989)\citenamefont{Ecker, Gasser, Pich,
  and Rafael}}]{ECKER1989311}
\bibinfo{author}{\bibfnamefont{G.}~\bibnamefont{Ecker}},
  \bibinfo{author}{\bibfnamefont{J.}~\bibnamefont{Gasser}},
  \bibinfo{author}{\bibfnamefont{A.}~\bibnamefont{Pich}}, \bibnamefont{and}
  \bibinfo{author}{\bibfnamefont{E.~D.} \bibnamefont{Rafael}},
  \bibinfo{journal}{Nuclear Physics B} \textbf{\bibinfo{volume}{321}},
  \bibinfo{pages}{311 } (\bibinfo{year}{1989}), ISSN \bibinfo{issn}{0550-3213},
  \urlprefix\url{http://www.sciencedirect.com/science/article/pii/0550321389903465}.

\bibitem[{\citenamefont{Soto et~al.}(2013)\citenamefont{Soto, Talavera, and
  Tarrus}}]{Soto:2011ap}
\bibinfo{author}{\bibfnamefont{J.}~\bibnamefont{Soto}},
  \bibinfo{author}{\bibfnamefont{P.}~\bibnamefont{Talavera}}, \bibnamefont{and}
  \bibinfo{author}{\bibfnamefont{J.}~\bibnamefont{Tarrus}},
  \bibinfo{journal}{Nucl. Phys.} \textbf{\bibinfo{volume}{B866}},
  \bibinfo{pages}{270} (\bibinfo{year}{2013}), \eprint{1110.6156}.

\bibitem[{\citenamefont{Krnjaic}(2016)}]{Krnjaic:2015mbs}
\bibinfo{author}{\bibfnamefont{G.}~\bibnamefont{Krnjaic}},
  \bibinfo{journal}{Phys. Rev.} \textbf{\bibinfo{volume}{D94}},
  \bibinfo{pages}{073009} (\bibinfo{year}{2016}), \eprint{1512.04119}.

\bibitem[{\citenamefont{Marciano et~al.}(2012)\citenamefont{Marciano, Zhang,
  and Willenbrock}}]{Marciano:2011gm}
\bibinfo{author}{\bibfnamefont{W.~J.} \bibnamefont{Marciano}},
  \bibinfo{author}{\bibfnamefont{C.}~\bibnamefont{Zhang}}, \bibnamefont{and}
  \bibinfo{author}{\bibfnamefont{S.}~\bibnamefont{Willenbrock}},
  \bibinfo{journal}{Phys. Rev.} \textbf{\bibinfo{volume}{D85}},
  \bibinfo{pages}{013002} (\bibinfo{year}{2012}), \eprint{1109.5304}.

\bibitem[{\citenamefont{Witten}(1983)}]{Witten:1983tw}
\bibinfo{author}{\bibfnamefont{E.}~\bibnamefont{Witten}},
  \bibinfo{journal}{Nucl. Phys.} \textbf{\bibinfo{volume}{B223}},
  \bibinfo{pages}{422} (\bibinfo{year}{1983}).

\bibitem[{\citenamefont{Wess and Zumino}(1971)}]{Wess:1971yu}
\bibinfo{author}{\bibfnamefont{J.}~\bibnamefont{Wess}} \bibnamefont{and}
  \bibinfo{author}{\bibfnamefont{B.}~\bibnamefont{Zumino}},
  \bibinfo{journal}{Phys. Lett.} \textbf{\bibinfo{volume}{37B}},
  \bibinfo{pages}{95} (\bibinfo{year}{1971}).

\bibitem[{\citenamefont{Low}(1958)}]{Low1958}
\bibinfo{author}{\bibfnamefont{F.~E.} \bibnamefont{Low}},
  \bibinfo{journal}{Phys. Rev.} \textbf{\bibinfo{volume}{110}}
  (\bibinfo{year}{1958}).

\bibitem[{\citenamefont{Burnett and Kroll}(1968)}]{BurnettKroll1968}
\bibinfo{author}{\bibfnamefont{T.~H.} \bibnamefont{Burnett}} \bibnamefont{and}
  \bibinfo{author}{\bibfnamefont{N.~M.} \bibnamefont{Kroll}},
  \bibinfo{journal}{Phys. Rev. Lett.} \textbf{\bibinfo{volume}{20}}
  (\bibinfo{year}{1968}).

\bibitem[{\citenamefont{Collins et~al.}(1989)\citenamefont{Collins, Soper, and
  Sterman}}]{collins1989factorization}
\bibinfo{author}{\bibfnamefont{J.~C.} \bibnamefont{Collins}},
  \bibinfo{author}{\bibfnamefont{D.~E.} \bibnamefont{Soper}}, \bibnamefont{and}
  \bibinfo{author}{\bibfnamefont{G.}~\bibnamefont{Sterman}}, in
  \emph{\bibinfo{booktitle}{Perturbative QCD}} (\bibinfo{publisher}{World
  Scientific}, \bibinfo{year}{1989}), pp. \bibinfo{pages}{1--91}.

\bibitem[{\citenamefont{Chen et~al.}(2017)\citenamefont{Chen, Han, and
  Tweedie}}]{Chen:2016wkt}
\bibinfo{author}{\bibfnamefont{J.}~\bibnamefont{Chen}},
  \bibinfo{author}{\bibfnamefont{T.}~\bibnamefont{Han}}, \bibnamefont{and}
  \bibinfo{author}{\bibfnamefont{B.}~\bibnamefont{Tweedie}},
  \bibinfo{journal}{JHEP} \textbf{\bibinfo{volume}{11}}, \bibinfo{pages}{093}
  (\bibinfo{year}{2017}), \eprint{1611.00788}.

\bibitem[{\citenamefont{Birkedal et~al.}(2005)\citenamefont{Birkedal, Matchev,
  Perelstein, and Spray}}]{Birkedal:2005ep}
\bibinfo{author}{\bibfnamefont{A.}~\bibnamefont{Birkedal}},
  \bibinfo{author}{\bibfnamefont{K.~T.} \bibnamefont{Matchev}},
  \bibinfo{author}{\bibfnamefont{M.}~\bibnamefont{Perelstein}},
  \bibnamefont{and} \bibinfo{author}{\bibfnamefont{A.}~\bibnamefont{Spray}}
  (\bibinfo{year}{2005}), \eprint{hep-ph/0507194}.

\bibitem[{\citenamefont{Bartels et~al.}(2017)\citenamefont{Bartels, Gaggero,
  and Weniger}}]{Bartels2017}
\bibinfo{author}{\bibfnamefont{R.}~\bibnamefont{Bartels}},
  \bibinfo{author}{\bibfnamefont{D.}~\bibnamefont{Gaggero}}, \bibnamefont{and}
  \bibinfo{author}{\bibfnamefont{C.}~\bibnamefont{Weniger}},
  \bibinfo{journal}{Journal of Cosmology and Astroparticle Physics}
  \textbf{\bibinfo{volume}{2017}}, \bibinfo{pages}{001} (\bibinfo{year}{2017}),
  \eprint{1703.02546 [astro-ph.HE]}.

\bibitem[{\citenamefont{Beacom et~al.}(2005)\citenamefont{Beacom, Bell, and
  Bertone}}]{PhysRevLett.94.171301}
\bibinfo{author}{\bibfnamefont{J.~F.} \bibnamefont{Beacom}},
  \bibinfo{author}{\bibfnamefont{N.~F.} \bibnamefont{Bell}}, \bibnamefont{and}
  \bibinfo{author}{\bibfnamefont{G.}~\bibnamefont{Bertone}},
  \bibinfo{journal}{Phys. Rev. Lett.} \textbf{\bibinfo{volume}{94}},
  \bibinfo{pages}{171301} (\bibinfo{year}{2005}), \eprint{astro-ph/0409403},
  \urlprefix\url{https://link.aps.org/doi/10.1103/PhysRevLett.94.171301}.

\bibitem[{\citenamefont{Mardon et~al.}(2009)\citenamefont{Mardon, Nomura,
  Stolarski, and Thaler}}]{cascadia}
\bibinfo{author}{\bibfnamefont{J.}~\bibnamefont{Mardon}},
  \bibinfo{author}{\bibfnamefont{Y.}~\bibnamefont{Nomura}},
  \bibinfo{author}{\bibfnamefont{D.}~\bibnamefont{Stolarski}},
  \bibnamefont{and} \bibinfo{author}{\bibfnamefont{J.}~\bibnamefont{Thaler}},
  \bibinfo{journal}{Journal of Cosmology and Astroparticle Physics}
  \textbf{\bibinfo{volume}{2009}}, \bibinfo{pages}{016} (\bibinfo{year}{2009}),
  \eprint{0901.2926 [hep-ph]},
  \urlprefix\url{http://stacks.iop.org/1475-7516/2009/i=05/a=016}.

\bibitem[{\citenamefont{Bringmann et~al.}(2008)\citenamefont{Bringmann,
  Bergstrom, and Edsjo}}]{1126-6708-2008-01-049}
\bibinfo{author}{\bibfnamefont{T.}~\bibnamefont{Bringmann}},
  \bibinfo{author}{\bibfnamefont{L.}~\bibnamefont{Bergstrom}},
  \bibnamefont{and} \bibinfo{author}{\bibfnamefont{J.}~\bibnamefont{Edsjo}},
  \bibinfo{journal}{Journal of High Energy Physics}
  \textbf{\bibinfo{volume}{2008}}, \bibinfo{pages}{049} (\bibinfo{year}{2008}),
  \eprint{0710.3169 [hep-ph]},
  \urlprefix\url{http://stacks.iop.org/1126-6708/2008/i=01/a=049}.

\bibitem[{\citenamefont{Essig et~al.}(2009)\citenamefont{Essig, Sehgal, and
  Strigari}}]{PhysRevD.80.023506}
\bibinfo{author}{\bibfnamefont{R.}~\bibnamefont{Essig}},
  \bibinfo{author}{\bibfnamefont{N.}~\bibnamefont{Sehgal}}, \bibnamefont{and}
  \bibinfo{author}{\bibfnamefont{L.~E.} \bibnamefont{Strigari}},
  \bibinfo{journal}{Phys. Rev. D} \textbf{\bibinfo{volume}{80}},
  \bibinfo{pages}{023506} (\bibinfo{year}{2009}), \eprint{0902.4750 [hep-ph]}.

\bibitem[{\citenamefont{Schwartz}(2017)}]{schwartz_2017}
\bibinfo{author}{\bibfnamefont{M.~D.} \bibnamefont{Schwartz}},
  \emph{\bibinfo{title}{Quantum field theory and the standard model}}
  (\bibinfo{publisher}{Cambridge University Press}, \bibinfo{year}{2017}).

\bibitem[{\citenamefont{Kuno and Okada}(2001)}]{RevModPhys.73.151}
\bibinfo{author}{\bibfnamefont{Y.}~\bibnamefont{Kuno}} \bibnamefont{and}
  \bibinfo{author}{\bibfnamefont{Y.}~\bibnamefont{Okada}},
  \bibinfo{journal}{Rev. Mod. Phys.} \textbf{\bibinfo{volume}{73}},
  \bibinfo{pages}{151} (\bibinfo{year}{2001}),
  \urlprefix\url{https://link.aps.org/doi/10.1103/RevModPhys.73.151}.

\bibitem[{\citenamefont{Bryman et~al.}(1982)\citenamefont{Bryman, Depommier,
  and Leroy}}]{bryman1982pi}
\bibinfo{author}{\bibfnamefont{D.}~\bibnamefont{Bryman}},
  \bibinfo{author}{\bibfnamefont{P.}~\bibnamefont{Depommier}},
  \bibnamefont{and} \bibinfo{author}{\bibfnamefont{C.}~\bibnamefont{Leroy}},
  \bibinfo{journal}{Physics Reports} \textbf{\bibinfo{volume}{88}},
  \bibinfo{pages}{151} (\bibinfo{year}{1982}).

\bibitem[{\citenamefont{Tanabashi et~al.}(2018)\citenamefont{Tanabashi,
  Hagiwara, Hikasa, Nakamura, Sumino, Takahashi, Tanaka, Agashe, Aielli, Amsler
  et~al.}}]{PhysRevD.98.030001}
\bibinfo{author}{\bibfnamefont{M.}~\bibnamefont{Tanabashi}},
  \bibinfo{author}{\bibfnamefont{K.}~\bibnamefont{Hagiwara}},
  \bibinfo{author}{\bibfnamefont{K.}~\bibnamefont{Hikasa}},
  \bibinfo{author}{\bibfnamefont{K.}~\bibnamefont{Nakamura}},
  \bibinfo{author}{\bibfnamefont{Y.}~\bibnamefont{Sumino}},
  \bibinfo{author}{\bibfnamefont{F.}~\bibnamefont{Takahashi}},
  \bibinfo{author}{\bibfnamefont{J.}~\bibnamefont{Tanaka}},
  \bibinfo{author}{\bibfnamefont{K.}~\bibnamefont{Agashe}},
  \bibinfo{author}{\bibfnamefont{G.}~\bibnamefont{Aielli}},
  \bibinfo{author}{\bibfnamefont{C.}~\bibnamefont{Amsler}},
  \bibnamefont{et~al.} (\bibinfo{collaboration}{Particle Data Group}),
  \bibinfo{journal}{Phys. Rev. D} \textbf{\bibinfo{volume}{98}},
  \bibinfo{pages}{030001} (\bibinfo{year}{2018}),
  \urlprefix\url{https://link.aps.org/doi/10.1103/PhysRevD.98.030001}.

\bibitem[{\citenamefont{Kappadath}(1993)}]{Kappadath:1993}
\bibinfo{author}{\bibfnamefont{S.~C.} \bibnamefont{Kappadath}}, Ph.D. thesis,
  \bibinfo{school}{University of New Hampshire} (\bibinfo{year}{1993}).

\bibitem[{\citenamefont{Michel}(1950)}]{ku}
\bibinfo{author}{\bibfnamefont{L.}~\bibnamefont{Michel}},
  \bibinfo{journal}{Proc. Phys. Soc.} \textbf{\bibinfo{volume}{A63}},
  \bibinfo{pages}{514} (\bibinfo{year}{1950}), \bibinfo{note}{[,45(1949)]}.

\bibitem[{\citenamefont{Bringmann et~al.}(2009)\citenamefont{Bringmann, Doro,
  and Fornasa}}]{Bringmann2009}
\bibinfo{author}{\bibfnamefont{T.}~\bibnamefont{Bringmann}},
  \bibinfo{author}{\bibfnamefont{M.}~\bibnamefont{Doro}}, \bibnamefont{and}
  \bibinfo{author}{\bibfnamefont{M.}~\bibnamefont{Fornasa}},
  \bibinfo{journal}{Journal of Cosmology and Astroparticle Physics}
  \textbf{\bibinfo{volume}{2009}}, \bibinfo{pages}{016} (\bibinfo{year}{2009}),
  \eprint{0809.2269 [astro-ph]}.

\bibitem[{\citenamefont{Essig et~al.}(2013{\natexlab{a}})\citenamefont{Essig,
  Kuflik, McDermott, Volansky, and Zurek}}]{Essig:2013goa}
\bibinfo{author}{\bibfnamefont{R.}~\bibnamefont{Essig}},
  \bibinfo{author}{\bibfnamefont{E.}~\bibnamefont{Kuflik}},
  \bibinfo{author}{\bibfnamefont{S.~D.} \bibnamefont{McDermott}},
  \bibinfo{author}{\bibfnamefont{T.}~\bibnamefont{Volansky}}, \bibnamefont{and}
  \bibinfo{author}{\bibfnamefont{K.~M.} \bibnamefont{Zurek}},
  \bibinfo{journal}{JHEP} \textbf{\bibinfo{volume}{11}}, \bibinfo{pages}{193}
  (\bibinfo{year}{2013}{\natexlab{a}}), \eprint{1309.4091}.

\bibitem[{\citenamefont{Boddy and Kumar}(2015)}]{Boddy2015}
\bibinfo{author}{\bibfnamefont{K.~K.} \bibnamefont{Boddy}} \bibnamefont{and}
  \bibinfo{author}{\bibfnamefont{J.}~\bibnamefont{Kumar}},
  \bibinfo{journal}{Phys. Rev. D} \textbf{\bibinfo{volume}{92}},
  \bibinfo{pages}{023533} (\bibinfo{year}{2015}), \eprint{1504.04024
  [astro-ph.CO]}.

\bibitem[{\citenamefont{Thompson et~al.}(1993)}]{Thompson:1993zz}
\bibinfo{author}{\bibfnamefont{D.~J.} \bibnamefont{Thompson}}
  \bibnamefont{et~al.}, \bibinfo{journal}{Astrophys. J. Suppl.}
  \textbf{\bibinfo{volume}{86}}, \bibinfo{pages}{629} (\bibinfo{year}{1993}).

\bibitem[{\citenamefont{Atwood et~al.}(2009{\natexlab{b}})}]{FermiInstrument}
\bibinfo{author}{\bibfnamefont{W.~B.} \bibnamefont{Atwood}}
  \bibnamefont{et~al.}, \bibinfo{journal}{The Astrophysical Journal}
  \textbf{\bibinfo{volume}{697}}, \bibinfo{pages}{1071}
  (\bibinfo{year}{2009}{\natexlab{b}}),
  \urlprefix\url{http://stacks.iop.org/0004-637X/697/i=2/a=1071}.

\bibitem[{\citenamefont{Angelis et~al.}(2018)}]{eASTROGAMWhitebook}
\bibinfo{author}{\bibfnamefont{A.~D.} \bibnamefont{Angelis}}
  \bibnamefont{et~al.}, \bibinfo{type}{Tech. Rep.} (\bibinfo{year}{2018}),
  \eprint{astro-ph/0406254}.

\bibitem[{\citenamefont{Strong et~al.}(2004)\citenamefont{Strong, Moskalenko,
  and Reimer}}]{Strong:2004de}
\bibinfo{author}{\bibfnamefont{A.~W.} \bibnamefont{Strong}},
  \bibinfo{author}{\bibfnamefont{I.~V.} \bibnamefont{Moskalenko}},
  \bibnamefont{and} \bibinfo{author}{\bibfnamefont{O.}~\bibnamefont{Reimer}},
  \bibinfo{journal}{Astrophys. J.} \textbf{\bibinfo{volume}{613}},
  \bibinfo{pages}{962} (\bibinfo{year}{2004}), \eprint{astro-ph/0406254}.

\bibitem[{\citenamefont{Ackermann et~al.}(2012)\citenamefont{Ackermann, Ajello,
  Atwood, Baldini, Ballet, Barbiellini, Bastieri, Bechtol, Bellazzini, Berenji
  et~al.}}]{Ackermann2012}
\bibinfo{author}{\bibfnamefont{M.}~\bibnamefont{Ackermann}},
  \bibinfo{author}{\bibfnamefont{M.}~\bibnamefont{Ajello}},
  \bibinfo{author}{\bibfnamefont{W.~B.} \bibnamefont{Atwood}},
  \bibinfo{author}{\bibfnamefont{L.}~\bibnamefont{Baldini}},
  \bibinfo{author}{\bibfnamefont{J.}~\bibnamefont{Ballet}},
  \bibinfo{author}{\bibfnamefont{G.}~\bibnamefont{Barbiellini}},
  \bibinfo{author}{\bibfnamefont{D.}~\bibnamefont{Bastieri}},
  \bibinfo{author}{\bibfnamefont{K.}~\bibnamefont{Bechtol}},
  \bibinfo{author}{\bibfnamefont{R.}~\bibnamefont{Bellazzini}},
  \bibinfo{author}{\bibfnamefont{B.}~\bibnamefont{Berenji}},
  \bibnamefont{et~al.}, \bibinfo{journal}{The Astrophysical Journal}
  \textbf{\bibinfo{volume}{750}}, \bibinfo{pages}{3} (\bibinfo{year}{2012}),
  \urlprefix\url{https://doi.org/10.1088%2F0004-637x%2F750%2F1%2F3}.

\bibitem[{\citenamefont{Essig et~al.}(2013{\natexlab{b}})\citenamefont{Essig,
  Kuflik, McDermott, Volansky, and Zurek}}]{Essig2013}
\bibinfo{author}{\bibfnamefont{R.}~\bibnamefont{Essig}},
  \bibinfo{author}{\bibfnamefont{E.}~\bibnamefont{Kuflik}},
  \bibinfo{author}{\bibfnamefont{S.~D.} \bibnamefont{McDermott}},
  \bibinfo{author}{\bibfnamefont{T.}~\bibnamefont{Volansky}}, \bibnamefont{and}
  \bibinfo{author}{\bibfnamefont{K.~M.} \bibnamefont{Zurek}},
  \bibinfo{journal}{Journal of High Energy Physics}
  \textbf{\bibinfo{volume}{2013}}, \bibinfo{pages}{193}
  (\bibinfo{year}{2013}{\natexlab{b}}), ISSN \bibinfo{issn}{1029-8479},
  \eprint{hep-ph/1309.4091},
  \urlprefix\url{https://doi.org/10.1007/JHEP11(2013)193}.

\bibitem[{\citenamefont{Colafrancesco et~al.}(2006)\citenamefont{Colafrancesco,
  Profumo, and Ullio}}]{Colafrancesco:2005ji}
\bibinfo{author}{\bibfnamefont{S.}~\bibnamefont{Colafrancesco}},
  \bibinfo{author}{\bibfnamefont{S.}~\bibnamefont{Profumo}}, \bibnamefont{and}
  \bibinfo{author}{\bibfnamefont{P.}~\bibnamefont{Ullio}},
  \bibinfo{journal}{Astron. Astrophys.} \textbf{\bibinfo{volume}{455}},
  \bibinfo{pages}{21} (\bibinfo{year}{2006}), \eprint{astro-ph/0507575}.

\bibitem[{\citenamefont{Colafrancesco et~al.}(2007)\citenamefont{Colafrancesco,
  Profumo, and Ullio}}]{Colafrancesco:2006he}
\bibinfo{author}{\bibfnamefont{S.}~\bibnamefont{Colafrancesco}},
  \bibinfo{author}{\bibfnamefont{S.}~\bibnamefont{Profumo}}, \bibnamefont{and}
  \bibinfo{author}{\bibfnamefont{P.}~\bibnamefont{Ullio}},
  \bibinfo{journal}{Phys. Rev.} \textbf{\bibinfo{volume}{D75}},
  \bibinfo{pages}{023513} (\bibinfo{year}{2007}), \eprint{astro-ph/0607073}.

\bibitem[{\citenamefont{Chen and Kamionkowski}(2004)}]{Chen:2003gz}
\bibinfo{author}{\bibfnamefont{X.-L.} \bibnamefont{Chen}} \bibnamefont{and}
  \bibinfo{author}{\bibfnamefont{M.}~\bibnamefont{Kamionkowski}},
  \bibinfo{journal}{Phys. Rev.} \textbf{\bibinfo{volume}{D70}},
  \bibinfo{pages}{043502} (\bibinfo{year}{2004}), \eprint{astro-ph/0310473}.

\bibitem[{\citenamefont{Padmanabhan and Finkbeiner}(2005)}]{Padmanabhan:2005es}
\bibinfo{author}{\bibfnamefont{N.}~\bibnamefont{Padmanabhan}} \bibnamefont{and}
  \bibinfo{author}{\bibfnamefont{D.~P.} \bibnamefont{Finkbeiner}},
  \bibinfo{journal}{Phys. Rev.} \textbf{\bibinfo{volume}{D72}},
  \bibinfo{pages}{023508} (\bibinfo{year}{2005}), \eprint{astro-ph/0503486}.

\bibitem[{\citenamefont{Galli et~al.}(2009)\citenamefont{Galli, Iocco, Bertone,
  and Melchiorri}}]{Galli:2009zc}
\bibinfo{author}{\bibfnamefont{S.}~\bibnamefont{Galli}},
  \bibinfo{author}{\bibfnamefont{F.}~\bibnamefont{Iocco}},
  \bibinfo{author}{\bibfnamefont{G.}~\bibnamefont{Bertone}}, \bibnamefont{and}
  \bibinfo{author}{\bibfnamefont{A.}~\bibnamefont{Melchiorri}},
  \bibinfo{journal}{Phys. Rev.} \textbf{\bibinfo{volume}{D80}},
  \bibinfo{pages}{023505} (\bibinfo{year}{2009}), \eprint{0905.0003}.

\bibitem[{\citenamefont{Slatyer et~al.}(2009)\citenamefont{Slatyer,
  Padmanabhan, and Finkbeiner}}]{Slatyer:2009yq}
\bibinfo{author}{\bibfnamefont{T.~R.} \bibnamefont{Slatyer}},
  \bibinfo{author}{\bibfnamefont{N.}~\bibnamefont{Padmanabhan}},
  \bibnamefont{and} \bibinfo{author}{\bibfnamefont{D.~P.}
  \bibnamefont{Finkbeiner}}, \bibinfo{journal}{Phys. Rev.}
  \textbf{\bibinfo{volume}{D80}}, \bibinfo{pages}{043526}
  (\bibinfo{year}{2009}), \eprint{0906.1197}.

\bibitem[{\citenamefont{Slatyer}(2016)}]{PhysRevD.93.023527}
\bibinfo{author}{\bibfnamefont{T.~R.} \bibnamefont{Slatyer}},
  \bibinfo{journal}{Phys. Rev. D} \textbf{\bibinfo{volume}{93}},
  \bibinfo{pages}{023527} (\bibinfo{year}{2016}), \eprint{1506.03811 [hep-ph]},
  \urlprefix\url{https://link.aps.org/doi/10.1103/PhysRevD.93.023527}.

\bibitem[{\citenamefont{Aghanim et~al.}(2018)}]{planck2018}
\bibinfo{author}{\bibfnamefont{N.}~\bibnamefont{Aghanim}} \bibnamefont{et~al.}
  (\bibinfo{collaboration}{Planck}) (\bibinfo{year}{2018}),
  \eprint{1807.06209}.

\bibitem[{\citenamefont{Kleiss et~al.}(1986)\citenamefont{Kleiss, Stirling, and
  Ellis}}]{Kleiss:1985gy}
\bibinfo{author}{\bibfnamefont{R.}~\bibnamefont{Kleiss}},
  \bibinfo{author}{\bibfnamefont{W.~J.} \bibnamefont{Stirling}},
  \bibnamefont{and} \bibinfo{author}{\bibfnamefont{S.~D.} \bibnamefont{Ellis}},
  \bibinfo{journal}{Comput. Phys. Commun.} \textbf{\bibinfo{volume}{40}},
  \bibinfo{pages}{359} (\bibinfo{year}{1986}).

\end{thebibliography}
\end{document}